\documentclass[twoside,12pt]{article}
\usepackage{CJK}
\usepackage{indentfirst}
\usepackage{bm}
\usepackage{epsfig}
\usepackage{graphicx}
\usepackage{epstopdf}
\usepackage{color}
\usepackage{mdsymbol}
\usepackage{booktabs}
\usepackage[numbers,sort&compress]{natbib}

\begin{document}

\begin{CJK*}{GBK}{song}

\thispagestyle{empty} \vspace*{0.8cm}\hbox
to\textwidth{\vbox{\hfill\huge\sf Commun. Theor. Phys.\hfill}}
\par\noindent\rule[3mm]{\textwidth}{0.2pt}\hspace*{-\textwidth}\noindent
\rule[2.5mm]{\textwidth}{0.2pt}


\begin{center}
\LARGE\bf Discrete Boltzmann model with split collision for nonequilibrium reactive flows$^{*}$
\end{center}

\footnotetext{\hspace*{-.45cm}\footnotesize $^*$ This work is supported by National Natural Science Foundation of China (under Grant Nos. U2242214 and 91441120), Guangdong Basic and Applied Basic Research Foundation (under Grant No. 2022A1515012116), Natural Science Foundation of Fujian Province (under Grant Nos. 2021J01652, 2021J01655), and China Scholarship Council (No. 202306380288). Support from the UK Engineering and Physical Sciences Research Council under the project ``UK Consortium on Mesoscale Engineering Sciences (UKCOMES)" (Grant No. EP/X035875/1) is gratefully acknowledged. This work made use of computational support by CoSeC, the Computational Science Centre for Research Communities, through UKCOMES.}
\footnotetext{\hspace*{-.45cm}\footnotesize $^\dag$Corresponding author, E-mail: hllai@fjnu.edu.cn}

\begin{center}
\rm Chuandong Lin$^{\rm a,b,c)}$, \ \ Kai H. Luo$^{\rm d)}$, \ and  \ Huilin Lai$^{\rm e)\dagger}$
\end{center}

\begin{center}
\begin{footnotesize} \sl
${}^{\rm a)}$ Sino-French Institute of Nuclear Engineering and Technology, Sun Yat-sen University, Zhuhai 519082, China \\
${}^{\rm b)}$ Key Laboratory for Thermal Science and Power Engineering of Ministry of Education, Department of Energy and Power Engineering, Tsinghua University, Beijing 100084, China \\
${}^{\rm c)}$ Department of Mechanical Engineering, National University of Singapore, 10 Kent Ridge Crescent, 119260, Singapore \\
${}^{\rm d)}$ Department of Mechanical Engineering, University College London, Torrington Place, London WC1E 7JE, United Kingdom \\
${}^{\rm e)}$ School of Mathematics and Statistics, Key Laboratory of Analytical Mathematics and Applications (Ministry of Education), Fujian Key Laboratory of Analytical Mathematics and Applications (FJKLAMA), Center for Applied Mathematics of Fujian Province (FJNU), Fujian Normal University, 350117 Fuzhou, China \\
\end{footnotesize}
\end{center}

\begin{center}
\footnotesize (Received XXXX; revised manuscript received XXXX)

\end{center}

\vspace*{2mm}

\begin{center}
\begin{minipage}{15.5cm}
\parindent 20pt\footnotesize
A multi-relaxation-time discrete Boltzmann model (DBM) with split collision is proposed for both subsonic and supersonic compressible reacting flows, where chemical reactions take place among various components. The physical model is based on a unified set of discrete Boltzmann equations that describes the evolution of each chemical species with adjustable acceleration, specific heat ratio, and Prandtl number. On the righ-hand side of discrete Boltzmann equations, the collision, force, and reaction terms denote the change rates of distribution functions due to self- and cross-collisions, external forces, and chemical reactions, respectively. The source terms can be calculated in three ways, among which the matrix inversion method possesses the highest physical accuracy and computational efficiency. Through Chapman-Enskog analysis, it is proved that the DBM is consistent with the reactive Navier-Stokes equations, Fick's law and Stefan-Maxwell diffusion equation in the hydrodynamic limit. Compared with the one-step-relaxation model, the split collision model offers a detailed and precise measurement of hydrodynamic, thermodynamic, and chemical nonequilibrium effects. Finally, the model is validated by six benchmarks, including multicomponent diffusion, mixture in the force field, Kelvin-Helmholtz instability, flame at constant pressure, opposing chemical reaction, and steady detonation.
\end{minipage}
\end{center}

\begin{center}
\begin{minipage}{15.5cm}
\begin{minipage}[t]{2.3cm}{\bf Keywords:}\end{minipage}
\begin{minipage}[t]{13.1cm}
discrete Boltzmann method, reactive flow, detonation, nonequilibrium effect
\end{minipage}\par\vglue8pt

\end{minipage}
\end{center}

\section{Introduction}
Reactive flows are a complex physicochemical phenomenon where different chemical species collide randomly and react violently, various interfacial and/or mechanical structures coexist, the chemical, hydrodynamic and thermodynamic nonequilibrium effects play significant roles \cite{Law2006,Williams2018,Fickett2000,Nagnibeda2009}. Due to its practical importance in both nature and society, reacting flows have been widely studied in human history. With the rapid development of computer hardware and computational science in recent decades, numerical simulation has become indispensable for academic research. Roughly speaking, there are three levels of physical description of reactive fluids, i.e., the macroscopic, mesoscopic and microscopic models. The most commonly used method is the macroscopic description based upon the continuous models, such as the reactive Euler or Navier-Stokes (NS) equations, where hydrodynamic quantities such as the density, velocity, temperature, and pressure are utilized to characterize the reactive system \cite{Law2006,Williams2018,Fickett2000}. Despite their great success, the traditional hydrodynamic governing equations ignore detailed thermodynamic nonequilibrium effects that often play essential roles, especially in a microscopic system or a local structure with sharp physical gradients \cite{Fickett2000,Nagnibeda2009}.
By contrast, the microscopic description is generally based on molecular dynamics (MD) where the interaction potential between molecules is considered \cite{Mao2023PECS}. Although the exact position and velocity of each molecule can be obtained dynamically, the MD is not capable of mimicking a relatively large system due to its excessive computational cost.

To solve the aforementioned issues, one way is to resort to the mesoscopic description that bridges the microscopic molecular and macroscopic continuous models. As a kinetic methodology, the discrete Boltzmann method (DBM) is regarded as a promising tool for mimicking multi-physics flow phenomena \cite{Gan2013EPL,Yan2013FOP,Lin2015PRE,Lin2016CNF,Zhang2016CNF,Lin2017SR,Lin2018CNF,Lin2019PRE,Lin2020CESW,Ji2021AIP,Ji2022JCP,Su2022CTP,Su2023CTP}. The DBM is a variant version of the standard lattice Boltzmann method (LBM) that has achieved great success in simulating complex reactive or nonreactive flows \cite{SucciBook,GuoBook2013,Succi1997JSC,ChiavazzoCNF2010,Kang2014PRE,Chen2015IJHMT,Hosseini2018PA,Liu2020CNF,Tayyab2020CNF,Dugast2020JCP,Chai2021PRE,Karlin2021JFM,Huang2019JCP,Boivin2021POF,Lei2021FP,Jiang2022ICHMT,Lei2023PCI,Li2016PECS,Hosseini2024PECS}. Early in 1997, Succi et al. proposed the pioneering simple extension of the LBM for numerical combustion where reactive flow dynamics is handled in the limit of fast chemistry \cite{Succi1997JSC} . In 2010, Chiavazzo et al. presented the modelling of reactive flows based upon a coupling between accurate reduced reaction mechanism and the LBM simulation of flow phenomena \cite{ChiavazzoCNF2010}. In 2014, Kang et al. proposed a thermal multicomponent LBM for catalytic reactive flows where the Bhatnagar-Gross-Krook (BGK) relaxation process is split into two parts: the first part is characterized by the relaxation toward an auxiliary state and the second part describes the relaxation toward the thermodynamic equilibrium \cite{Kang2014PRE}. In 2020, Dugast et al. proposed a topology optimization algorithm based on a multi-relaxation-time (MRT) LBM coupled with a level-set method for reactive fluid flows, allowing higher Reynolds numbers flow simulations compared to the ordinary single-relaxation-time model \cite{Dugast2020JCP}. In 2021, Lei and Luo developed a sophisticated LBM for reactive flows in porous media, where separate equations describe the evolution of multicomponent flows and chemical species \cite{Lei2021FP}. In 2022, Jiang et al.  proposed an immersed boundary LBM for particle combustion with varying thermodynamic and transport properties, and conducted the hydrodynamics-resolved simulation of a char particle combustion \cite{Jiang2022ICHMT}.

Although a series of LBM studies have been performed for reactive flows over the past two decades, they have not been developed to simulate complex compressible reacting flows where significant hydrodynamic, thermodynamic and chemical nonequilibrium effects coexist and interact with each other. In fact, both the DBM and basic LBM are intended as solvers for a truncated version of the Boltzmann equation, and they have the following same advantages:
(i) Simple scheme. It is easy to code the algorithm for the discrete Boltzmann equation with a simplified collision operator.
(ii) Parallel computation. It is convenient for massively parallel computing as all the information transfer is local in time and space \cite{Lin2019CTP}.
(iii) Boundary condition. It is straightforward to deal with complex geometry by changing the discrete distribution functions \cite{Liu2020CNF,Zhang2019CPC}.
Moreover, the differences between the DBM and basic LBM are as follows:
(i) Phase-space discretization.
Classical LBM solvers rely on a specific stencil to discretize the phase space and a Lagrangian approach is used for the time-space discretization (integration along characteristic lines) resulting in the streaming-collision algorithm; The DBM is a classical Eulerian solver for the discrete Boltzmann equation where the discretization is performed using a moment-matching approach.
(ii) Physical model.
Traditional LBMs mainly serve as the solver of NS equations or other partial differential equations, while the DBM is fairly equivalent to a modified hydrodynamic model plus a coarse-grained model of thermodynamic nonequilibrium behaviors. Namely, the DBM is strictly inherited by the Boltzmann equation and can describe nonequilibrium system beyond macroscopic governing equations, while the standard LBM is a numerical scheme characterized by the streaming-collision steps and can be used to model a large family of advection equations beyond the topic of the Boltzmann equation.

It should be pointed out that the last aforementioned difference is the key reason why the DBM has been developed. The DBM is derived from the nonequilibrium statistical physics and has been successfully applied to investigate compressible reacting flows  \cite{Yan2013FOP,Lin2015PRE,Lin2016CNF,Zhang2016CNF,Lin2017SR,Lin2018CNF,Lin2019PRE,Lin2020CESW,Ji2021AIP,Ji2022JCP,Su2022CTP,Su2023CTP}, multiphase \cite{Gan2022JFM,Zhang2022AIP,Wang2023CAF,Sun2024POF}, and fluid instabilities \cite{Lin2014PRE,Lin2017PRE,Lai2016PRE,Chen2018POF,Gan2019FOP,Lai2021FOP,Lin2021PRE,Li2022FOP,lai2023CAF}, etc. In 2013, the pioneering DBM for combustion was developed by using a hybrid scheme, and the hydrodynamic and thermodynamic nonequilibrium effects were investigated around the detonation wave \cite{Yan2013FOP}. In 2016, the DBM was employed to probe detonation with negative temperature coefficient from three aspects: hydrodynamic quantities, nonequilibrium quantities and entropy productions \cite{Zhang2016CNF}. In the same year, a DBM was constructed where one (another) set of distribution function describes chemical reactant (product) \cite{Lin2016CNF}. In 2017, a BGK DBM for multicomponent reactive flows was presented, where the relaxation time has the same value for one species and the Prandtl number is fixed to $\Pr = 1$ \cite{Lin2017SR}.
In 2019, an efficient MRT DBM was developed to tackle steady or unsteady supersonic reactive flows \cite{Lin2019PRE}. Since 2021, the DBM has been extended to three-dimensional steady and unsteady detonation \cite{Ji2021AIP,Ji2022JCP}. In 2022, the DBM was used to study the quantitative discrepancy between equilibrium and nonequilibrium distribution functions around the detonation wave \cite{Su2022CTP}. In 2023, via the DBM and the fast Fourier transform, the deviations of the velocity distribution function from the equilibrium state have been investigated and the kinetic moments of reaction terms have been discussed in the evolution of unsteady detonation \cite{Su2023CTP}.

Base on previous works \cite{Lin2017SR,Lin2019PRE,Lin2020CESW,Lin2021PRE}, we develop an MRT DBM with a split collision term for reacting flows with both hydrodynamic and thermodynamic nonequilibrium effects. Compared with the BGK model \cite{Lin2017SR}, the MRT DBM has various relaxation times for different nonequilibrium processes and a flexible $\Pr$. In contrast to the single-distribution-function DBM \cite{Lin2019PRE,Lin2020CESW}, the current DBM takes account of collisions and reactions among different chemical species. Different from the MRT DBM for multicomponent mixtures \cite{Lin2021PRE}, this model involves the effects of chemical reaction and external force, and a splitting technique is applied to the collision term. 
In historical context, the splitting technique originated within the realm of plasma physics for electron-ion systems, where the distinct impacts of self-collision and cross-collision on each species were recognized \cite{Gross1956PR,Sofonea2001PA}. 
Physically, complex systems often undergo multiple evolution stages characterized by varying time scales, and their components, such as electrons or ions, may exhibit divergent temperatures \cite{Gross1956PR,Sofonea2001PA}. Consequently, employing an appropriate methodology becomes imperative to investigate such intricate physical systems, especially those involving reactive flows with numerous chemical species. This serves as the primary motivation for this study. 
Specifically, the MRT DBM with split collision is developted for a mixture system to delineate the effects of self-collision and cross-collision on individual species. In contrast, the DBM without split collision represents a simplified version of the current model, applicable under the condition where relaxation frequencies during self-collision match those in cross-collision scenarios.

The rest of the paper is organized as follows. First, we introduce details of the MRT DBM with the split collision term for compressible reacting flows. The model is subsequently validated by six benchmarks, i.e., the multicomponent diffusion, homogeneous mixture in the force field, Kelvin-Helmholtz (KH) instability, flame at constant pressure, opposing chemical reaction, and steady detonation. Finally, conclusions are drawn.

\section{Discrete Bolzmann method} \label{SecII}

The coarse-grained physical modeling from the Boltzmann equation to the DBM mainly involves three key steps \cite{Lin2017PRE,Gan2018PRE}: (i) Simplification of the collision term. It is difficult to solve the original Boltzmann equation where the collision term is too complex in the integral form. Hence, it is necessary to simplify the collision term in order to utilize the Boltzmann equation. In this paper, we employ the widely used MRT collision model \cite{Lin2015PRE,Lin2019PRE,Chen2018POF}. (ii) Discretization of the particle velocity. For the purpose of physical accuracy and numerical efficiency, the particle velocity space is discretized with the matrix inversion method \cite{Gan2013EPL,Lin2019PRE}. (iii) Description of nonequilibrium effects. The main purpose of the DBM is to probe and extract essential nonequilibrium information beyond traditional hydrodynamic models \cite{Yan2013FOP,Lin2016CNF,Lin2014PRE}. Note that the first two steps are for coarse-grained modeling, and the third one is the core of DBM. The last step is not only an extension to the first two, but also puts forward stricter physical requirements for the construction. To be specific, all kinetic moment relations in the DBM should be consistent with those in the nonequilibrium statistical physics, and multi-physics fields (including density, temperature, velocity) should be coupled naturally as various physical quantities are obtained from the kinetic moments of the same distribution function.

\begin{figure}
	\begin{center}
		\includegraphics[bbllx=0pt,bblly=0pt,bburx=143pt,bbury=143pt,width=0.4\textwidth]{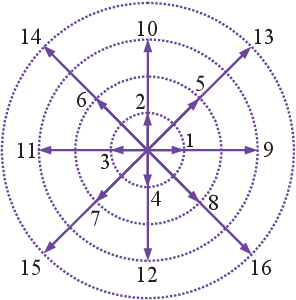}
	\end{center}
	\caption{Sketch of discrete velocities.}
	\label{Fig01}
\end{figure}

In this work, the MRT discrete Boltzmann equations, which describe the spatio-temporal evolution of reacting flows, take the form,
\begin{equation}
	\dfrac{\partial {{\mathbf{f}}^{\sigma }}}{\partial t}+{{\mathbf{v}}^{\sigma }}\cdot \nabla {{\mathbf{f}}^{\sigma }}={{\mathbf{\Omega }}^{\sigma }}+{{\mathbf{F}}^{\sigma }}+{{\mathbf{R}}^{\sigma }}
	\label{DBEquation}
	\tt{.}
\end{equation}
Here $t$ represents the time. The superscript $\sigma = 1$, $2$, $\dots$, $N_s$ indicates chemical species with the number $N_s$ in total. The column matrices
${{\mathbf{f}}^{\sigma }}={{\left( \begin{matrix} {f_{1}^{\sigma }} \ f_{2}^{\sigma } \ \cdots  \ f_{N}^{\sigma } \end{matrix} \right)}^{\rm{T}}}$,
${{\mathbf{\Omega }}^{\sigma }}={{\left( \begin{matrix} {\Omega _{1}^{\sigma }} \ \Omega _{2}^{\sigma } \ \cdots  \ \Omega _{N}^{\sigma } \end{matrix} \right)}^{\rm{T}}}$,  ${{\mathbf{F}}^{\sigma }}={{\left( \begin{matrix} {F_{1}^{\sigma }} \ F_{2}^{\sigma } \ \cdots  \ F_{N}^{\sigma } \end{matrix} \right)}^{\rm{T}}}$,
and ${{\mathbf{R}}^{\sigma }}={{\left( \begin{matrix} {R_{1}^{\sigma }} \ R_{2}^{\sigma } \ \cdots  \ R_{N}^{\sigma } \end{matrix} \right)}^{\rm{T}}}$
denote the discrete distribution functions, collision terms, force terms, and reaction terms, respectively. The diagonal matrix ${{\mathbf{v}}^{\sigma }} = \rm{diag}\left( \begin{matrix}
	{{\mathbf{v}}^{\sigma }_{1}} \ {{\mathbf{v}}^{\sigma }_{2}} \ \cdots  \ {{\mathbf{v}}^{\sigma }_{N}}
\end{matrix} \right)$ stands for the discrete velocities, and the subscript $i = 1$, $2$, $\dots$, $N$ is the index of discrete velocities, with $N = 16$ in this paper.
As shown in Fig. \ref{Fig01}, a discrete velocity set reads
\begin{equation}
	{{\mathbf{v}}^{\sigma }_{i}}=\left\{ \begin{array}{*{35}{l}}
		{\rm{cyc}}:{{v}_{a}^{\sigma}}\left( \pm 1,0 \right), & 1\le i\le 4,  \\
		{\rm{cyc}}:{{v}_{b}^{\sigma}}\left( \pm 1,\pm 1 \right), & 5\le i\le 8,  \\
		{\rm{cyc}}:{{v}_{c}^{\sigma}}\left( \pm 1,0 \right), & 9\le i\le 12,  \\
		{\rm{cyc}}:{{v}_{d}^{\sigma}}\left( \pm 1,\pm 1 \right), & 13\le i\le 16,
	\end{array} \right.
	\label{DVM}
\end{equation}
where (${{v}_{a}^{\sigma}}$, ${{v}_{b}^{\sigma}}$, ${{v}_{c}^{\sigma}}$, ${{v}_{d}^{\sigma}}$) are flexible parameters.
Besides, to take account of the part of internal energies due to molecular rotation and/or vibration, we introduce the symbol ${\eta }_{i}^{\sigma} = {\eta }_{a }^{\sigma}$, ${\eta }_{b}^{\sigma}$, ${\eta }_{c}^{\sigma}$, and ${\eta }_{d}^{\sigma}$ for $1\le i\le 4$, $5\le i\le 8$, $9\le i\le 12$, and $13\le i\le 16$, respectively, where the parameters (${\eta }_{a }^{\sigma}$, ${\eta }_{b}^{\sigma}$, ${\eta }_{c}^{\sigma}$, ${\eta }_{d}^{\sigma}$) are tunable as well.

It is worth emphasizing that the values of ${{\mathbf{v}}^{\sigma }_{i}}$ and ${\eta }_{i}^{\sigma}$ can be adjusted to optimize the DBM robustness and accuracy. On the one hand, the parameters (${{v}_{a}^{\sigma}}$, ${{v}_{b}^{\sigma}}$, ${{v}_{c}^{\sigma}}$, ${{v}_{d}^{\sigma}}$) can be chosen around the values of flow velocity $\mathbf{u}^{\sigma }$ and sound speed $v_s^{\sigma } = \sqrt{{\gamma^{\sigma}} {T^{\sigma}}/{m^{\sigma }}}$, where ${T^{\sigma}}$ stands for the temperature, ${m^{\sigma }}$ the molar mass, $\gamma^{\sigma} = (D + I^{\sigma} + 2) / (D + I^{\sigma})$ the specific heat ratio, $D = 2$ the spatial dimension in this paper, and $I^{\sigma}$ extra degrees of freedom due to molecular rotation and/or vibration. On the other hand, among (${\eta }_{a }^{\sigma}$, ${\eta }_{b}^{\sigma}$, ${\eta }_{c}^{\sigma}$, ${\eta }_{d}^{\sigma}$), one (another) parameter should be less (greater) than $\bar{\eta} ^ {\sigma} =\sqrt{I^{\sigma} {T^{\sigma}} / m^{\sigma}}$, because the extra internal energy is about $\dfrac{1}{2} m^{\sigma } {\bar{\eta}}^{2}=\dfrac{1}{2} I^{\sigma} {T^{\sigma}}$ according to the equipartition of energy theorem.

\subsection{Macroscopic quantities}

The individual molar concentration ${n}^{\sigma }$, mass density ${\rho}^{\sigma }$, velocity ${\mathbf{u}}^{\sigma }$, energy ${{E}^{\sigma }}$, and temperature ${{T}^{\sigma }}$ are given by,
\begin{equation}
	{{n}^{\sigma }}=\sum\nolimits_{i}{f_{i}^{\sigma }}
	\label{n_sigma}
	\tt{,}
\end{equation}
\begin{equation}
	{{\rho }^{\sigma }}={{m}^{\sigma }}{{n}^{\sigma }}
	\tt{,}
\end{equation}
\begin{equation}
	{{n }^{\sigma }} \mathbf{u}^{\sigma }=\sum\nolimits_{i}{f_{i}^{\sigma } \mathbf{v}_{i }^{\sigma }}
	\label{Momentum}
	\tt{,}
\end{equation}
\begin{equation}
	{{E}^{\sigma }}=\dfrac{m^{\sigma}}{2}\sum\nolimits_{i}{f_{i}^{\sigma eq}\left( |\mathbf{v}_{i}^{\sigma}|^{2}+\eta _{i}^{\sigma 2} \right)}
	\label{Energy}
	\tt{,}
\end{equation}
\begin{equation}
	{{T}^{\sigma }}=\dfrac{2{{E}^{\sigma }}-{{\rho }^{\sigma }}{{u}^{\sigma 2}}}{\left( D+{{I}^{\sigma }} \right){{n}^{\sigma }}}
	\label{TemperatureI}
	\tt{,}
\end{equation}
respectively. The mixing number density ${n}$, mass density ${\rho}$, velocity $\mathbf{u}$, energy $E$, internal energy ${{E}_{int}}$, and temperature $T$ are obtained from
\begin{equation}
	n=\sum\nolimits_{\sigma }{{{n}^{\sigma }}}
	\tt{,}
\end{equation}
\begin{equation}
	\rho =\sum\nolimits_{\sigma }{{{\rho }^{\sigma }}}
	\tt{,}
\end{equation}
\begin{equation}
	\rho \mathbf{u}=\sum\nolimits_{\sigma }{{{\rho }^{\sigma }} \mathbf{u}^{\sigma }}
	\tt{,}
\end{equation}
\begin{equation}
	E=\sum\nolimits_{\sigma }{{{E}^{\sigma }}}
	\label{Energy_Mixing}
	\tt{,}
\end{equation}
\begin{equation}
	{{E}_{{int}}}=E-\dfrac{1}{2}\rho {{\left| \mathbf{u} \right|}^{2}}
	\tt{,}
\end{equation}
\begin{equation}
	T=\dfrac{2E_{int}}{\sum\nolimits_{\sigma }{\left( D+{{I}^{\sigma }} \right){{n}^{\sigma }}}}
	\label{T_expression}
	\tt{,}
\end{equation}
respectively. Actually, energies ${{E}^{\sigma }}$ in Eq. (\ref{Energy}) and $E$ in Eq. (\ref{Energy_Mixing}) are not conserved during chemical reaction and may be called the
``sensible" or ``total nonchemical" energies.

It should be mentioned that, with the substitution of the equilibrium discrete distribution functions ${f}^{\sigma eq}$ for the discrete distribution functions ${f}^{\sigma}$, the formulas (\ref{n_sigma}) - (\ref{Energy}) still holds. In fact, the aforementioned physical quantities in Eqs. (\ref{n_sigma}) - (\ref{T_expression}) are macroscopic parameters that are statistical results of particles with random motion, and can be conveniently measured by traditional numerical or experimental methods.

\subsection{Split collision term}

In fact, the discrete Boltzmann equation (\ref{DBEquation}) is a simplified form of the original Boltzmann equation, and the collision term is a reduced expression of its original nonlinear integral term. To be specific, the collision term is composed of three parts, i.e.,
\begin{equation}
	{{\mathbf{\Omega }}^{\sigma }}={{\mathbf{\Omega }}^{1 \sigma }} + {{\mathbf{\Omega }}^{2 \sigma }} + {{\mathbf{\Omega }}^{3 \sigma }}
	\tt{,}
	\label{Omega_sum}
\end{equation}
in terms of
\begin{equation}
	{{\mathbf{\Omega }}^{1 \sigma }} = -{({{\mathbf{M}}^{\sigma}})^{-1}}{{\mathbf{S}}^{1\sigma }}\left( {{{\mathbf{\hat{f}}}}^{\sigma }}-{{{\mathbf{\hat{f}}}}^{\sigma seq}} \right)
	\tt{,}
	\label{Omega_1}
\end{equation}
\begin{equation}
	{{\mathbf{\Omega }}^{2 \sigma }} = -{({{{\mathbf{M}}^{\sigma}}})^{-1}}{{\mathbf{S}}^{2\sigma }}\left( {{{\mathbf{\hat{f}}}}^{\sigma seq}}-{{{\mathbf{\hat{f}}}}^{\sigma eq}} \right)
	\tt{,}
	\label{Omega_2}
\end{equation}
and
\begin{equation}
	{{\mathbf{\Omega }}^{3 \sigma }} = {({{{\mathbf{M}}^{\sigma}}})^{-1}}{{\mathbf{\hat{A}}}^{\sigma }}
	\tt{,}
	\label{Omega_3}
\end{equation}
where the diagonal matrix ${{\mathbf{S}}^{1 \sigma }} = {\rm{diag}} \left( \begin{matrix} {{S}^{1 \sigma }_{1}} \ {{S}^{1 \sigma }_{2}} \ \cdots  \ {{S}^{1 \sigma }_{N}} \end{matrix} \right)$ indicates the relaxation frequencies that control the relaxation speed of kinetic moments ${{\mathbf{\hat{f}}}}^{\sigma } = {{\left( \begin{matrix} {\hat{f}_{1}^{\sigma }} \ \hat{f}_{2}^{\sigma } \ \cdots  \ \hat{f}_{N}^{\sigma } \end{matrix} \right)}^{\rm{T}}}$ approaching their individual intermediate equilibrium counterparts ${{\mathbf{\hat{f}}}^{\sigma seq}} = {{\left( \begin{matrix} {\hat{f}_{1}^{\sigma seq}} \ \hat{f}_{2}^{\sigma seq} \ \cdots  \ \hat{f}_{N}^{\sigma seq} \end{matrix} \right)}^{\rm{T}}}$, and
${{\mathbf{S}}^{2 \sigma }} = {\rm{diag}}\left( \begin{matrix} {{S}^{2 \sigma }_{1}} \ {{S}^{2 \sigma }_{2}} \ \cdots  \ {{S}^{2 \sigma }_{N}} \end{matrix} \right)$ denotes the relaxation frequencies which govern the relaxation speed of  ${{\mathbf{\hat{f}}}^{\sigma seq}} = {{\left( \begin{matrix} {\hat{f}_{1}^{\sigma seq}} \ \hat{f}_{2}^{\sigma seq} \ \cdots  \ \hat{f}_{N}^{\sigma seq} \end{matrix} \right)}^{\rm{T}}}$ approaching the ultimate equilibrium counterparts ${{\mathbf{\hat{f}}}^{\sigma eq}} = {{\left( \begin{matrix} {\hat{f}_{1}^{\sigma eq}} \ \hat{f}_{2}^{\sigma eq} \ \cdots  \ \hat{f}_{N}^{\sigma eq} \end{matrix} \right)}^{\rm{T}}}$.
Here ${{\mathbf{\hat{f}}}^{\sigma eq}}$ is the function of ($n^{\sigma}$, ${u}_{\alpha}$, $T$), and
${{\mathbf{\hat{f}}}^{\sigma seq}}$ is expressed by substituting
(${u}^{\sigma }_{\alpha}$, $T^{\sigma }$) for (${u}_{\alpha}$, $T$) in the formula of ${{\mathbf{\hat{f}}}^{\sigma eq}}$, see Appendix \ref{APPENDIXA}. The square matrix $\mathbf{M}^{\sigma} = \left( M^{\sigma}_{il} \right)$ and its inverse ${({\mathbf{M}^{\sigma}})^{-1}} = \left ( {\left({M^{\sigma}_{il}}\right)^{-1}} \right)$, both of which have $N \times N$ elements, act as the link between the velocity and moment spaces. To be specific,
\begin{equation}
	{{\mathbf{\hat{f}}}^{\sigma eq}}={{\mathbf{M}}^{\sigma }}{{\mathbf{f}}^{\sigma eq}}
	\label{Moment_feq}
	\tt{,}
\end{equation}
\begin{equation}
	{{\mathbf{\hat{f}}}^{\sigma seq}}={{\mathbf{M}}^{\sigma }}{{\mathbf{f}}^{\sigma seq}}
	\label{Moment_fseq}
	\tt{,}
\end{equation}
\begin{equation}
	{{\mathbf{\hat{f}}}^{\sigma }}={{\mathbf{M}}^{\sigma }}{{\mathbf{f}}^{\sigma }}
	\label{Moment_f}
	\tt{,}
\end{equation}
with ${{\mathbf{f}}^{\sigma eq}}={{\left( \begin{matrix} {f_{1}^{\sigma eq}} \ f_{2}^{\sigma eq} \ \cdots  \ f_{N}^{\sigma eq}  \end{matrix} \right)}^{\rm{T}}}$ and ${{\mathbf{f}}^{\sigma seq}}={{\left( \begin{matrix} {f_{1}^{\sigma seq}} \ f_{2}^{\sigma seq} \ \cdots  \ f_{N}^{\sigma seq}  \end{matrix} \right)}^{\rm{T}}}$. Here (${{\mathbf{\hat{f}}}^{\sigma eq}}$, ${{\mathbf{\hat{f}}}^{\sigma seq}}$, ${{\mathbf{\hat{f}}}^{\sigma }}$) and (${{\mathbf{f}}^{\sigma eq}}$, ${{\mathbf{f}}^{\sigma seq}}$, ${{\mathbf{f}}^{\sigma }}$) correspond to the moment and velocity spaces, respectively. In fact, Eq. (\ref{Moment_feq}) is equivalent to the following relationship,
\begin{equation}
	\iint{{{f}^{\sigma eq}}\Psi d\mathbf{v}d\eta }=\sum\nolimits_{i}{f_{i}^{\sigma eq}}{{\Psi }_{i}}
	\label{Moment_feq0}
	\tt{,}
\end{equation}
in terms of $\Psi = 1$, $\mathbf{v}$, $\left( |\mathbf{v}{{|}^{2}}+{{\eta }^{2}} \right)$, $\mathbf{vv}$, $\left( |\mathbf{v}{{|}^{2}}+{{\eta }^{2}} \right)\mathbf{v}$, $\mathbf{vvv}$, $\left( |\mathbf{v}{{|}^{2}}+{{\eta }^{2}} \right)\mathbf{vv}$,
and their corresponding discrete counterparts
$\Psi_{i} = 1$, $\mathbf{v}_{i}^{\sigma }$, $\left( |\mathbf{v}_{i}^{\sigma }{{|}^{2}}+\eta _{i}^{2} \right)$, $\mathbf{v}_{i}^{\sigma }\mathbf{v}_{i}^{\sigma }$, $\left( |\mathbf{v}_{i}^{\sigma }{{|}^{2}}+\eta _{i}^{\sigma 2} \right)\mathbf{v}_{i}^{\sigma }$, $\mathbf{v}_{i}^{\sigma }\mathbf{v}_{i}^{\sigma }\mathbf{v}_{i}^{\sigma }$, $\left( |\mathbf{v}_{i}^{\sigma }{{|}^{2}}+\eta _{i}^{\sigma 2} \right)\mathbf{v}_{i}^{\sigma }\mathbf{v}_{i}^{\sigma }$. In this work, the theoretical equilibrium distribution function is expressed by \cite{Lin2017SR}
\begin{equation}
	{{f}^{\sigma eq}}={{n}^{\sigma }}{{\left( \dfrac{{{m}^{\sigma }}}{2\pi T} \right)}^{D/2}}{{\left( \dfrac{{{m}^{\sigma }}}{2\pi {{I}^{\sigma }}T} \right)}^{1/2}}\exp \left[ -\dfrac{{{m}^{\sigma }}{{\left| \mathbf{v}-\mathbf{u} \right|}^{2}}}{2T}-\dfrac{{{m}^{\sigma }}{{\eta }^{2}}}{2{{I}^{\sigma }}T} \right]
	\label{Formula_Feq}
	\tt{.}
\end{equation}
Actually, the formula (\ref{Moment_feq0}) is a necessary condition to recover the NS equations in the hydrodynamic limit.

Note that in the simplification process from the original Boltzmann equation to the DBM, the physical quantities (such as density, momentum, energy, and lower-order kinetic moments) under consideration remain unchanged, while some other physical information (such as higher-order kinetic moments and the interactions between them) may be lost. The loss of relevant information constrains the applications of the physical model, and may lead to inaccuracy for particular real situations. To rectify this, Chapman-Enskog (CE) multiscale analysis can be employed to identify and amend the deficiencies in the physical model. For this purpose, an additional term ${{\mathbf{\hat{A}}}^{\sigma }}$ is incorporated into the collision term to make up for the missing relation between physical quantities $\hat{f}_{5}^{\sigma }$,  $\hat{f}_{6}^{\sigma }$, $\hat{f}_{7}^{\sigma }$, $\hat{f}_{8}^{\sigma }$, and $\hat{f}_{9}^{\sigma }$.
Specifically, the term ${{\mathbf{\hat{A}}}^{\sigma }} = {{\left( \begin{matrix} 0 \ \cdots  \ 0 \ \hat{A}_{8}^{\sigma } \ \hat{A}_{9}^{\sigma } \ 0 \ \cdots  \ 0  \end{matrix} \right)}^{\rm{T}}}$ depends upon
\begin{equation}
	\hat{A}_{8}^{\sigma }=2\left( S_{8}^{1 \sigma }-S_{5}^{1 \sigma } \right)u_{x}^{\sigma }\Delta _{5}^{\sigma }+2\left( S_{8}^{1 \sigma }-S_{6}^{1 \sigma } \right)u_{y}^{\sigma }\Delta _{6}^{\sigma }
	\label{A_8}
	\tt{,}
\end{equation}
\begin{equation}
	\hat{A}_{9}^{\sigma }=2\left( S_{9}^{1 \sigma }-S_{7}^{1 \sigma } \right)u_{y}^{\sigma }\Delta _{7}^{\sigma }+2\left( S_{9}^{1 \sigma }-S_{6}^{1 \sigma } \right)u_{x}^{\sigma }\Delta _{6}^{\sigma }
	\label{A_9}
	\tt{,}
\end{equation}
with
\begin{eqnarray}
	\Delta _{5}^{\sigma }=\dfrac{2{{n }^{\sigma }}{{T}^{\sigma }}}{S_{5}^{1\sigma }{{m}^{\sigma }}}\left( \dfrac{1-D-{{I}^{\sigma }}}{D+{{I}^{\sigma }}}\dfrac{\partial u_{x}^{\sigma }}{\partial x}+\dfrac{1}{D+{{I}^{\sigma }}}\dfrac{\partial u_{y}^{\sigma }}{\partial y} \right)
	\tt{,}
	\label{Delta5}
\end{eqnarray}
\begin{eqnarray}
	\Delta _{6}^{\sigma }=-\dfrac{{{n }^{\sigma }}{{T}^{\sigma }}}{S_{6}^{1\sigma }{{m}^{\sigma }}}\left( \dfrac{\partial u_{x}^{\sigma }}{\partial y}+\dfrac{\partial u_{y}^{\sigma }}{\partial x} \right)
	\tt{,}
	\label{Delta6}
\end{eqnarray}
\begin{eqnarray}
	\Delta _{7}^{\sigma }=\dfrac{2{{n }^{\sigma }}{{T}^{\sigma }}}{S_{7}^{1\sigma }{{m}^{\sigma }}}\left( \dfrac{1}{D+{{I}^{\sigma }}}\dfrac{\partial u_{x}^{\sigma }}{\partial x}+\dfrac{1-D-{{I}^{\sigma }}}{D+{{I}^{\sigma }}}\dfrac{\partial u_{y}^{\sigma }}{\partial y} \right)
	\tt{.}
	\label{Delta7}
\end{eqnarray}

It should be further explained that the physical meaning of the first two parts in Eq. (\ref{Omega_sum}) is as follows: there are two split steps during the thermodynamic relaxation process. The distribution function ${{{\mathbf{\hat{f}}}}^{\sigma }}$ firstly approachs its temporary equilibrium state ${{{\mathbf{\hat{f}}}}^{\sigma seq}}$ under the control of relaxation frequency ${{\mathbf{S}}^{1\sigma }}$, then tends toward the local ultimate equilibrium state ${{{\mathbf{\hat{f}}}}^{\sigma eq}}$ with relaxation frequency ${{\mathbf{S}}^{2\sigma }}$. It is worth mentioning that, there are more flexible parameters in the two-step-relaxation collision term, which is suitable for a wider application range of physical systems. Through the CE expansion, the relations can be determined between the relaxation parameters and other physical quantities, such as the nonequilibrium quantities (\ref{Moment_f_feq}) - (\ref{hatf7_sneq}), diffusivity (\ref{Diffusion_expression}), thermal conductivity (\ref{Thermal_conductivity}),
dynamic viscosity (\ref{Dynamic_viscosity}). Consequently, compared with the one-step-relaxation MRT or BGK model, the two-step-relaxation collision term presents a more detailed relationship between the thermodynamic relaxation process and nonequilibrium effects.

Moreover, substituting Eqs. (\ref{Omega_1}) - (\ref{Omega_3}) into (\ref{Omega_sum}) leads to the following expression
\begin{equation}
	{{\mathbf{\Omega }}^{\sigma }}=-{({{\mathbf{M}}^{\sigma}})^{-1}}
	\left[
	{{\mathbf{S}}^{1\sigma }}\left( {{{\mathbf{\hat{f}}}}^{\sigma }}-{{{\mathbf{\hat{f}}}}^{\sigma seq}} \right)
	+{{\mathbf{S}}^{2\sigma }}\left( {{{\mathbf{\hat{f}}}}^{\sigma seq}}-{{{\mathbf{\hat{f}}}}^{\sigma eq}} \right)
	-{{\mathbf{\hat{A}}}^{\sigma }} \right]
	\label{Collision_term}
	\tt{.}
\end{equation}
Clearly, in the case of ${{\mathbf{S}}^{1 \sigma }} = {{\mathbf{S}}^{2 \sigma }} = {{\mathbf{S}}^{\sigma }}$, the split collision model (\ref{Collision_term}) (called two-step-relaxation collision model) reduces to the popular MRT model (named a one-step-relaxation collision model)
\begin{equation}
	{{\mathbf{\Omega }}^{\sigma }}=-{({{{\mathbf{M}}^{\sigma}}})^{-1}}
	\left[
	{{\mathbf{S}}^{\sigma }}\left( {{{\mathbf{\hat{f}}}}^{\sigma }}-{{{\mathbf{\hat{f}}}}^{\sigma eq}} \right)
	-{{\mathbf{\hat{A}}}^{\sigma }} \right]
	\tt{,}
\end{equation}
which further reduces to the single-relaxation model
\begin{equation}
	{{\mathbf{\Omega }}^{\sigma }}=-\dfrac{1}{{\tau}^{\sigma }}\left( {{\mathbf{f}}^{\sigma }}-{{\mathbf{f}}^{\sigma eq}} \right)
	\tt{,}
	\label{single-relaxation model}
\end{equation}
if ${{\mathbf{S}}^{\sigma }} = \mathbf{I} / {{\tau}^{\sigma }}$, and $\mathbf{I}$ denotes the unit tensor. Clearly, $\hat{A}_{8}^{\sigma }=\hat{A}_{9}^{\sigma }=0$ when $S_{5}^{1 \sigma }=S_{6}^{1 \sigma }=S_{7}^{1 \sigma }=S_{8}^{1 \sigma }=S_{9}^{1 \sigma }$, as seen in Eqs. (\ref{A_8}) and (\ref{A_9}). Namely, the additional term ${{\mathbf{\hat{A}}}^{\sigma }}$ disappears in the single-relaxation case. A widely used single-relaxation model is the BGK model. In fact, both MRT and BGK models are also suitable for a fluid system if there is only one component or the average mixing effect is under consideration. Besides, in order to achieve local momentum conservation for a two-component system, the relaxation times of the two components should be equal: ${\tau}^{\sigma} = {\tau}^{\bar{\sigma}}= {\tau}$, which is a constraint of the single-relaxation BGK model \cite{Sofonea2001PA}.

In addition, the collision term in Eq. (\ref{Omega_sum}) can be rewritten as,
\begin{equation}
	{{\mathbf{\Omega }}^{\sigma }}={{\mathbf{\Omega }}^{1 \sigma *}} + {{\mathbf{\Omega }}^{2 \sigma *}} + {{\mathbf{\Omega }}^{3 \sigma }}
	\tt{,}
	\label{Omega*_sum}
\end{equation}
with
\begin{equation}
	{{\mathbf{\Omega }}^{1 \sigma *}} = -{({{\mathbf{M}}^{\sigma}})^{-1}}{{\mathbf{S}}^{1\sigma * }}\left( {{{\mathbf{\hat{f}}}}^{\sigma }}-{{{\mathbf{\hat{f}}}}^{\sigma seq}} \right)
	\tt{,}
	\label{Omega*_1}
\end{equation}
\begin{equation}
	{{\mathbf{\Omega }}^{2 \sigma *}} = -{({{{\mathbf{M}}^{\sigma}}})^{-1}}{{\mathbf{S}}^{2\sigma * }}\left( {{{\mathbf{\hat{f}}}}^{\sigma }}-{{{\mathbf{\hat{f}}}}^{\sigma eq}} \right)
	\tt{,}
	\label{Omega*_2}
\end{equation}
where ${{\mathbf{S}}^{1\sigma *}}={{\mathbf{S}}^{1\sigma }}-{{\mathbf{S}}^{2\sigma }}$ and ${{\mathbf{S}}^{2\sigma *}}={{\mathbf{S}}^{2\sigma }}$. The terms ${{\mathbf{\Omega }}^{1 \sigma *}}$ and ${{\mathbf{\Omega }}^{2 \sigma *}}$ are related to the self-collision and cross-collision among various particles, respectively.
In fact, both self-collision and cross-collision affect the evolution of the discrete distribution functions. In other words, the effects of self-collision and cross-collision on the evolution of physical systems are taken into consideration.

\subsection{Force term}

Physically, the force term denotes the change rate of distribution function due to the external force. How to calculate the force term is a key to an accurate physical model. In this part, we introduce three ways to obtain the mathematical expression of the force term. The first two methods, which were actually proposed for two-component fluids in Ref. \cite{Lin2017PRE}, are extended to multicomponent systems in this work. The last method that is named the matrix inversion method \cite{Gan2013EPL,Lin2019PRE} is developed for multicomponent systems for the first time as well.

\textbf{Method I}

Via the Taylor expansion, it can be found that the main part of the distribution function is the equilibrium distribution function in a system not too far from equilibrium \cite{Lin2017PRE}. Theoretically, $f^{\sigma}$ is close to $f^{\sigma seq}( n^{\sigma}, \mathbf{u}^{\sigma}, T^{\sigma})$ rather than $f^{\sigma eq}( n^{\sigma}, \mathbf{u}, T)$, especially in a non-premixed or partially premixed system. Hence the approximation $f^{\sigma }\approx f^{\sigma seq}( n^{\sigma}, \mathbf{u}^{\sigma}, T^{\sigma})$ can be used as follows,
\begin{equation}
	{{F}^{\sigma }}
	= - {\mathbf{a}^{\sigma}} \cdot \frac{\partial {f}^{\sigma}}{\partial \mathbf{v}}
	\approx
	- {\mathbf{a}^{\sigma}} \cdot \frac{\partial {f}^{\sigma seq}}{\partial \mathbf{v}}
	= {\mathbf{a}^{\sigma}} \cdot \left( \mathbf{v}-{{\mathbf{u}}^{\sigma }} \right)\frac{{{m}^{\sigma }}}{{{T}^{\sigma }}}{{f}^{\sigma seq}}
	\tt{,}
\end{equation}
where ${{\mathbf{a}}^{\sigma }}=a_{\alpha }^{\sigma }{{\mathbf{e}}_{\alpha }}$ stands for the body acceleration of species $\sigma$, and $\mathbf{e}_{\alpha }$ the unit vector in the ${\alpha }$ direction. Then, the force terms are obtained in the discretization form directly
\begin{equation}
	F_{i}^{\sigma } = {{\mathbf{a}}^{\sigma }}\cdot \left( \mathbf{v}_{i}^{\sigma }-{{\mathbf{u}}^{\sigma }} \right)\frac{{{m}^{\sigma }}}{{{T}^{\sigma }}}{{f}^{\sigma seq}_{i}}
	\tt{.}
	\label{ForceTerm1}
\end{equation}
In fact, Eq. (\ref{ForceTerm1}) is a conventional way to calculate the force terms \cite{Lai2016PRE,Chen2018POF}.

\textbf{Method II}

The force terms are used to incorporate forcing effects into the Boltzmann equation. According to its physical meaning, the force terms can be expressed by the change of discrete distribution functions $\delta f_{i}^{\sigma}$ due to the external force over a small time interval $\delta t$, i.e.,
\begin{equation}
	F_{i}^{\sigma }
	={{\left. \frac{\partial f_{i}^{\sigma }}{\partial t} \right|}_{\text{Force}}}
	=\underset{\delta t\to 0}{\mathop{\lim }}\,{{\left. \frac{\delta f_{i}^{\sigma }}{\delta t} \right|}_{\text{Force}}}
	\approx {{\left. \frac{\Delta f_{i}^{\sigma seq}}{\Delta t} \right|}_{\text{Force}}}
	\label{ForceTerm_2}
	\tt{,}
\end{equation}
where $\Delta f_{i}^{\sigma seq}$ represents the corresponding change of the equilibrium distribution function within a time step $\Delta t$ and is a function of the concentration, velocity and temperature.

In classical physics, the impulse (work) done by an external force changes the momentum (kinetic energy) of a system directly, while the mass, internal energy, and temperature remain constant in the force field.
In other words, the force changes the velocity and energy of the fluid components, but does not have a direct influence on the density or temperature. Consequently, $\Delta f_{i}^{\sigma seq}$ in Eq. (\ref{ForceTerm_2}) can be written as
\begin{equation}
	\Delta f_{i}^{\sigma seq} = f_{i}^{\sigma seq\dagger }-f_{i}^{\sigma seq}
	\tt{,}
	\label{ForceTerm2}
\end{equation}
where the equilibrium distribution functions change from $f_{i}^{\sigma seq} = f_{i}^{\sigma seq}\left( {{n}^{\sigma }}, {{\mathbf{u}}^{\sigma }}, {{T}^{\sigma }} \right)$ to $f_{i}^{\sigma seq\dagger } = f_{i}^{\sigma seq}\left( {{n}^{\sigma }}, {{\mathbf{u}}^{\sigma \dagger }}, {{T}^{\sigma }} \right)$, and the flow velocity changes from ${{\mathbf{u}}^{\sigma }}$ to ${{\mathbf{u}}^{\sigma \dagger }} = {{\mathbf{u}}^{\sigma }}+{{\mathbf{a}}^{\sigma }} \Delta t$ within a time step.

Theoretically, because of the external force, the energy of component $\sigma$ change from  ${{E}^{\sigma }}$ into ${{E}^{\sigma \dagger }} = {{E}^{\sigma }}+{{\rho }^{\sigma }}{{\mathbf{u}}^{\sigma }}\cdot {{\mathbf{a}}^{\sigma }}\Delta t$, then it can be derived from Eq. (\ref{TemperatureI}) that the temperature of component $\sigma$ is
\begin{equation}
	{{T}^{\sigma \dagger }}
	= {{T}^{\sigma }}-\frac{{{m}^{\sigma }}{{\left| {{\mathbf{a}}^{\sigma }} \right|}^{2}}}{D+{{I}^{\sigma }}}{{\left( \Delta t \right)}^{2}}
	\approx {{T}^{\sigma }}
	\label{Error_T}
	\tt{.}
\end{equation}
Clearly, ${{T}^{\sigma \dagger }}$ equals ${{T}^{\sigma }}$ as $\Delta t$ approaches zero. In other words, the temperature is not changed by the external force. It can be found from Eq. (\ref{Error_T}) that the expression in Eq. (\ref{ForceTerm2}) is of the second order accuracy.

\textbf{Method III}

Let us consider the following relation \cite{Lin2019PRE},
\begin{equation}
	\iint{{{F}^{\sigma }}\Psi d\mathbf{v}d\eta } = \sum\nolimits_{i}{F_{i}^{\sigma }}{{\Psi }_{i}}
	\label{MomentF0}
	\tt{,}
\end{equation}
where $\Psi$ and $\Psi_{i}$ are the same as those in Eq. (\ref{Moment_feq0}), and ${F}^{\sigma }$ and $F_{i}^{\sigma }$ denote the force terms in the continuous and discrete velocity spaces, respectively.

In fact, the formula (\ref{MomentF0}) is equivalent to the following matrix form
\begin{equation}
	{{\mathbf{\hat{F}}}^{\sigma}}={{\mathbf{M}}^{\sigma }}{{\mathbf{F}}^{\sigma}}
	\label{Moment_F}
	\tt{,}
\end{equation}
where the elements of ${{\mathbf{\hat{F}}}^{\sigma}}$ are given in Appendix \ref{APPENDIXA}. From Eq. (\ref{Moment_F}), the force terms can be expressed by
\begin{equation}
	{{\mathbf{F}}^{\sigma}} = {({{\mathbf{M}}^{\sigma}})^{-1}} {{\mathbf{\hat{F}}}^{\sigma}}
	\label{ForceTerm3}
	\tt{.}
\end{equation}

It is noteworthy that there are two similarities among above three methods.
(I) Based upon the approximation $f^{\sigma }\approx f^{\sigma seq}(n^{\sigma}, \mathbf{u}^{\sigma}, T^{\sigma})$, the force terms are expressed with the discrete equilibrium distribution functions;
(II) The relations satisfied by the force terms are sufficient to recover the NS equations, see Appendix \ref{APPENDIXB}.
Besides, the differences among above three methods are as follows.
(I) As for Method I, there are nine relations satisfied by the force terms, which are the necessary and sufficient conditions to recover the NS equations. In contrast, besides the nine relations, another seven relations are satisfied by the force terms in the last two methods as well. That is to say, there are sixteen relationships in Methods II and III, respectively.
(II) As shown in Eqs. (\ref{ForceTerm1}) and (\ref{ForceTerm2}), the force terms are a function of the equilibrium discrete distribution functions. Consequently, it is necessary to calculate the equilibrium discrete distribution functions in the program for Method I or II. In contrast, as shown in Eq. (\ref{ForceTerm3}), the force terms are computed by using an  inverse matrix in the last method, which has a higher computational efficiency. In other words, the matrix inversion method has the characteristic of high physical accuracy and computational efficiency, see Table \ref{TableII}. Consequently, this methodology is utilized to calculate the reaction terms in the next subsection as well.

\subsection{Reaction term}

The reaction term, which represents the change rate of distribution function because of the chemical reaction, satisfies the following relationship \cite{Lin2019PRE},
\begin{equation}
	\iint{{{R}^{\sigma }}\Psi d\mathbf{v}d\eta } = \sum\nolimits_{i}{R_{i}^{\sigma }}{{\Psi }_{i}}
	\label{MomentR0}
	\tt{,}
\end{equation}
where $\Psi$ and $\Psi_{i}$ are the same as those in Eq. (\ref{Moment_feq0}), and ${R}^{\sigma }$ and $R_{i}^{\sigma }$ are the reaction terms in the continuous and discrete velocity spaces, respectively. The expression of ${R}^{\sigma }$ reads \cite{Lin2015PRE},
\begin{equation}
	{{R}^{\sigma }}={{f}^{\sigma eq}}\dfrac{{{n}^{\sigma }}^{\prime }}{{{n}^{\sigma }}}+{{f}^{\sigma eq}}\dfrac{-\left( 1+D \right){{I}^{\sigma }}T+{{m}^{\sigma }}{{I}^{\sigma }}{{\left| \mathbf{v}-\mathbf{u} \right|}^{2}}+{{m}^{\sigma }}{{\eta }^{2}}}{2{{I}^{\sigma }}{{T}^{2}}}{T}'
	\tt{,}
\end{equation}
where ${n}^{\sigma \prime }$ stands for the concentration variation rate of species $\sigma$,
\begin{equation}
	{T}'=\dfrac{2\left[ {{E}'}\cdot \sum\nolimits_{\sigma }{{{n}^{\sigma }}\left( D+{{I}^{\sigma }} \right)}-{{E}_{int}}\cdot \sum\nolimits_{\sigma }{{{n}^{\sigma }}^{\prime }\left( D+{{I}^{\sigma }} \right)} \right]}{{{\left[ \sum\nolimits_{\sigma }{{{n}^{\sigma }}\left( D+{{I}^{\sigma }} \right)} \right]}^{2}}}
	\tt{,}
\end{equation}
is the temperature variation rate, and
\begin{equation}
	{E}' = {\omega_{\rm{ov}}} {Q}
	\label{E_ratio}
	\tt{,}
\end{equation}
indicates the release rate of chemical heat that equals the energy variation rate due to the chemical reaction. In Eq. (\ref{E_ratio}), $\omega_{\rm{ov}}$ stands for the chemical reaction rate, and $Q$ stands for the chemical heat release of reactant per unit mole.

In addition, the formula (\ref{MomentR0}) is equivalent to the following matrix form
\begin{equation}
	{{\mathbf{\hat{R}}}^{\sigma}}={{\mathbf{M}}^{\sigma }}{{\mathbf{R}}^{\sigma}}
	\label{Moment_R}
	\tt{,}
\end{equation}
where the elements of ${{\mathbf{R}}^{\sigma}}$ are given in Appendix \ref{APPENDIXA}. From Eq. (\ref{MomentR0}), the expressions of reaction terms can be obtained
\begin{equation}
	{{\mathbf{R}}^{\sigma}} = {({{\mathbf{M}}^{\sigma}})^{-1}} {{\mathbf{\hat{R}}}^{\sigma}}
	\label{ReactionTerm}
	\tt{.}
\end{equation}

As for the description of chemical reactions, we can adopt the one-step reaction, two-step reaction, detailed or reduced multi-step chemical kinetics. For example, without loss of generality, three simple chemical reaction models are adopted in this manuscript.

It should be emphasized that, via the reaction terms on the right-hand side of the discrete Boltzmann equation (\ref{DBEquation}), the multi-physics and chemical reactions are naturally coupled. The matrix inversion method is a precise and efficient calculation approach to compute the reaction terms in Eq. (\ref{ReactionTerm}), because the sixteen moment relations of reaction terms are satisfied in an elegant way.

\subsection{Nonequilibrium effects}

In fact, the difference between Eqs. (\ref{Moment_feq}) and (\ref{Moment_f}) indicates the nonequilibrium departure degree of the physical system,
\begin{equation}
	{{\mathbf{\hat{f}}}^{\sigma neq}}
	={{\mathbf{\hat{f}}}^{\sigma }}-{{\mathbf{\hat{f}}}^{\sigma eq}}
	={{\mathbf{M}}^{\sigma }}\left( {{\mathbf{f}}^{\sigma }}-{{\mathbf{f}}^{\sigma eq}} \right)
	={{\mathbf{M}}^{\sigma }}{{\mathbf{f}}^{\sigma neq}}
	\label{Moment_f_feq}
	\tt{,}
\end{equation}
with
${{\mathbf{f}}^{\sigma neq}}={{\left( \begin{matrix}
			{f_{1}^{\sigma neq}} \ f_{2}^{\sigma neq} \ \cdots  \ f_{N}^{\sigma neq}  \end{matrix} \right)}^{\rm{T}}}$ and
${{\mathbf{\hat{f}}}^{\sigma neq}}={{\left( \begin{matrix}
			{\hat{f}_{1}^{\sigma neq}} \ \hat{f}_{2}^{\sigma neq} \ \cdots  \ \hat{f}_{N}^{\sigma neq}  \end{matrix} \right)}^{\rm{T}}}$.
Similarly, we can define
\begin{equation}
	{{\mathbf{\hat{f}}}^{\sigma sneq}}
	={{\mathbf{\hat{f}}}^{\sigma }}-{{\mathbf{\hat{f}}}^{\sigma seq}}
	={{\mathbf{M}}^{\sigma }}\left( {{\mathbf{f}}^{\sigma }}-{{\mathbf{f}}^{\sigma seq}} \right)
	={{\mathbf{M}}^{\sigma }}{{\mathbf{f}}^{\sigma sneq}}
	\label{Moment_f_fseq}
	\tt{,}
\end{equation}
with 
${{\mathbf{f}}^{\sigma sneq}} = {{\left( \begin{matrix}
	{f_{1}^{\sigma sneq}} \ f_{2}^{\sigma sneq} \ \cdots  \ f_{N}^{\sigma sneq}  \end{matrix} \right)}^{\rm{T}}}$ 
and
${{\mathbf{\hat{f}}}^{\sigma sneq}} = {{\left( \begin{matrix}
	{\hat{f}_{1}^{\sigma sneq}} \ \hat{f}_{2}^{\sigma sneq} \ \cdots  \ \hat{f}_{N}^{\sigma sneq}  \end{matrix} \right)}^{\rm{T}}}$. 
Physically, Eq. (\ref{Moment_f_fseq}) means the kinetic moment deviation of a fluid component from its individual temporary equilibrium state, and Eq. (\ref{Moment_f_feq}) denotes the departure of a chemical species from the local mixing eventual equilibrium.

It is noteworthy that the physical quantities ${{\mathbf{f}}^{\sigma neq}}$ and ${{\mathbf{f}}^{\sigma sneq}}$ represent the nonequilibrium effects from various aspects. To be specific, $\hat{f}_{1}^{\sigma neq} = \hat{f}_{1}^{\sigma sneq} = 0$ in line with the mass conservation; $\hat{f}_{2}^{\sigma sneq} = 0$ and $\hat{f}_{3}^{\sigma sneq} = 0$ due to the momentum conservation; $m^{\sigma} \hat{f}_{2}^{\sigma neq} = \rho ^{\sigma} (u^{\sigma}_{x} - u_{x})$ and $m^{\sigma} \hat{f}_{3}^{\sigma neq} = \rho ^{\sigma} (u^{\sigma}_{y} - u_{y})$ denote the individual mass diffusion fluxes in the $x$ and $y$ directions, respectively; $\hat{f}_{4}^{\sigma sneq} = 0$ on account of the energy conservation; Neither $\hat{f}_{i}^{\sigma sneq}$ nor $\hat{f}_{i}^{\sigma neq}$ may be zero for $i \ge 5$ when a system is in the nonequilibrium state.

Furthermore, Eqs. (\ref{Delta5})-(\ref{Delta7}) are parts of the following nonequilibrium quantities,
\begin{eqnarray}
	& \hat{f}_{5}^{\sigma sneq}=\Delta _{5}^{\sigma }-\dfrac{{{\rho }^{\sigma \prime }}}{S_{5}^{1\sigma }{{m}^{\sigma }}}\dfrac{{{\left( u_{x}^{\sigma }-{{u}_{x}} \right)}^{2}}+{{\left( u_{y}^{\sigma }-{{u}_{y}} \right)}^{2}}}{D+{{I}^{\sigma }}}+\dfrac{{{\rho }^{\sigma \prime }}}{S_{5}^{1\sigma }{{m}^{\sigma }}}{{\left( u_{x}^{\sigma }-{{u}_{x}} \right)}^{2}} \nonumber \\
	& +\dfrac{S_{4}^{2\sigma }{{\rho }^{\sigma }}}{S_{5}^{1\sigma }{{m}^{\sigma }}}\dfrac{{{u}^{\sigma 2}}-{{u}^{2}}}{D+{{I}^{\sigma }}}-\dfrac{S_{5}^{2\sigma }{{\rho }^{\sigma }}}{S_{5}^{1\sigma }{{m}^{\sigma }}}\left( u_{x}^{\sigma 2}-u_{x}^{2} \right)+\dfrac{S_{4}^{2\sigma }-S_{5}^{2\sigma }}{S_{5}^{1\sigma }}\dfrac{{{\rho }^{\sigma }}}{{{m}^{\sigma 2}}}\left( {{T}^{\sigma }}-T \right) \nonumber \\
	& +\dfrac{S_{2}^{2\sigma }{{\rho }^{\sigma }}}{S_{5}^{1\sigma }{{m}^{\sigma }}}\dfrac{D+{{I}^{\sigma }}-1}{D+{{I}^{\sigma }}}2u_{x}^{\sigma }\left( u_{x}^{\sigma }-{{u}_{x}} \right)-\dfrac{S_{3}^{2\sigma }{{\rho }^{\sigma }}}{S_{5}^{1\sigma }{{m}^{\sigma }}}\dfrac{2u_{y}^{\sigma }\left( u_{y}^{\sigma }-{{u}_{y}} \right)}{D+{{I}^{\sigma }}}
	\label{hatf5_sneq}
	\tt{,}
\end{eqnarray}
\begin{eqnarray}
	& \hat{f}_{6}^{\sigma sneq}=\Delta _{6}^{\sigma }+\dfrac{S_{2}^{2\sigma }{{\rho }^{\sigma }}}{S_{6}^{1\sigma }{{m}^{\sigma }}}\left( u_{x}^{\sigma }-{{u}_{x}} \right)u_{y}^{\sigma }+\dfrac{S_{3}^{2\sigma }{{\rho }^{\sigma }}}{S_{6}^{1\sigma }{{m}^{\sigma }}}u_{x}^{\sigma }\left( u_{y}^{\sigma }-{{u}_{y}} \right) \nonumber \\
	& -\dfrac{S_{6}^{2\sigma }{{\rho }^{\sigma }}}{S_{6}^{1\sigma }{{m}^{\sigma }}}\left( u_{x}^{\sigma }u_{y}^{\sigma }-{{u}_{x}}{{u}_{y}} \right)+\dfrac{{{\rho }^{\sigma \prime }}}{S_{6}^{1\sigma }{{m}^{\sigma }}}\left( {{u}_{x}}{{u}_{y}}+u_{x}^{\sigma }u_{y}^{\sigma }-{{u}_{x}}u_{y}^{\sigma }-u_{x}^{\sigma }{{u}_{y}} \right)
	\label{hatf6_sneq}
	\tt{,}
\end{eqnarray}
\begin{eqnarray}
	& \hat{f}_{7}^{\sigma sneq}=\Delta _{7}^{\sigma }-\dfrac{{{\rho }^{\sigma \prime }}}{S_{7}^{1\sigma }{{m}^{\sigma }}}\dfrac{{{u}^{\sigma 2}}+{{u}^{2}}-2u_{x}^{\sigma }{{u}_{x}}-2u_{y}^{\sigma }{{u}_{y}}}{D+{{I}^{\sigma }}}+\dfrac{{{\rho }^{\sigma \prime }}}{S_{7}^{1\sigma }{{m}^{\sigma }}}{{\left( u_{y}^{\sigma }-{{u}_{y}} \right)}^{2}} \nonumber \\
	& +\dfrac{S_{4}^{2\sigma }{{\rho }^{\sigma }}}{S_{7}^{1\sigma }{{m}^{\sigma }}}\dfrac{{{u}^{\sigma 2}}-{{u}^{2}}}{D+{{I}^{\sigma }}}-\dfrac{S_{7}^{2\sigma }{{\rho }^{\sigma }}}{S_{7}^{1\sigma }{{m}^{\sigma }}}\left( u_{y}^{\sigma 2}-u_{y}^{2} \right)+\dfrac{\left( S_{4}^{2\sigma }-S_{7}^{2\sigma } \right){{\rho }^{\sigma }}}{S_{7}^{1\sigma }{{m}^{\sigma }}}\dfrac{{{T}^{\sigma }}-T}{{{m}^{\sigma }}} \nonumber \\
	& -\dfrac{S_{2}^{2\sigma }{{\rho }^{\sigma }}}{S_{7}^{1\sigma }{{m}^{\sigma }}}\dfrac{2u_{x}^{\sigma }\left( u_{x}^{\sigma }-{{u}_{x}} \right)}{D+{{I}^{\sigma }}}+\dfrac{S_{3}^{2\sigma }{{\rho }^{\sigma }}}{S_{7}^{1\sigma }{{m}^{\sigma }}}\dfrac{D+{{I}^{\sigma }}-1}{D+{{I}^{\sigma }}}2u_{y}^{\sigma }\left( u_{y}^{\sigma }-{{u}_{y}} \right)
	\label{hatf7_sneq}
	\tt{,}
\end{eqnarray}
at the NS level, which can be proved via the CE expansion.

Physically, both (${{\mathbf{\hat{f}}}^{\sigma neq}}$, ${{\mathbf{\hat{f}}}^{\sigma sneq}}$) and (${{\mathbf{S}}^{1 \sigma }}$, ${{\mathbf{S}}^{2 \sigma }}$) play roles in the thermodynamic and hydrodynamic behaviors, and various nonequilibrium modes are coupled in the relaxation process. Moreover, the nonequilibrium effects are important and traditional hydrodynamic models are not accurate in cases with small characteristic scales or large Knudsen numbers, particularly for multicomponent flows where various complex material and/or mechanical interfaces exist. For those complex nonequilibrium problems, the DBM provides a convenient tool to probe and analyze the nonequilibrium state and process.

\subsection{Nondimensionalization}

For numerical simulations and investigations, it is helpful to perform nondimensionalization. In this paper, ${{\Phi }_{d}}$ and ${{\Phi }_{n}}$ are designated as dimensional and nondimensional variables, respectively, and their ratio is ${{\Phi }_{r}}={{\Phi }_{d}}/{{\Phi }_{n}}$. The number density $n$, length $L$, flow speed $u=\left| \mathbf{u} \right|$, temperature $T$, and universal gas constant $R$ are adopted as references, see Table \ref{TableI}. Obviously, from these references, we can derive other ratios of dimensional to nondimensional variables, see Appendix \ref{APPENDIXC}. With the ratios and dimensional quantities, the nondimensional values are obtained in a straightforward way.

\begin{table}[h]
	\centering
	\tabcolsep 5mm
	\begin{tabular}{lccc}
		\hline\hline
		Variable & Dimension & Nondimension & Ratio \\
		\hline
		Number density & $n_{d}$ & $n_{n}$ & $n_{r}$ \\
		\hline
		Length & $L_{d}$ & $L_{n}$ & $L_{r}$ \\
		\hline
		Flow speed & $u_{d}$ & $u_{n}$ & $u_{r}$ \\
		\hline
		Temperature & $T_{d}$ & $T_{n}$ & $T_{r}$ \\
		\hline
		Universal gas constant & $R$ & $1$ & $R$ \\
		\hline\hline
	\end{tabular}
	\caption{References for nondimensionalization.}
	\label{TableI}
\end{table}

Additionally, various numerical schemes can be adopted to solve the discrete Boltzmann equations (\ref{DBEquation}). In this paper, we employ the third-order total variation diminishing Runge-Kutta \cite{TVD_RK1988} for handling the time derivative and the fifth-order weighted essentially non-oscillatory scheme for the space derivative \cite{WENO1996}. 
Consequently, in order to achieve good numerical stability, both the temporal and spatial steps must adhere to the Courant-Friedrichs-Lewy condition, and the temporal step should be smaller than the relaxation time. In the DBM with split collision, there are two sets of relaxation frequencies represented as ${{\mathbf{S}}^{1 \sigma }} = {\rm{diag}} \left( \begin{matrix} {{S}^{1 \sigma }_{1}} \ {{S}^{1 \sigma }_{2}} \ \cdots  \ {{S}^{1 \sigma }_{N}} \end{matrix} \right)$ and
${{\mathbf{S}}^{2 \sigma }} = {\rm{diag}}\left( \begin{matrix} {{S}^{2 \sigma }_{1}} \ {{S}^{2 \sigma }_{2}} \ \cdots  \ {{S}^{2 \sigma }_{N}} \end{matrix} \right)$, hence the temporal step $\Delta t$ should be smaller than the mimimum of $\left({{S}^{1 \sigma }_{i}} \right)^{-1}$ and $\left({{S}^{2 \sigma }_{i}} \right)^{-1}$. As for the DBM without split collision, there exists only one set of relaxation frequencies ${{\mathbf{S}}^{\sigma }} = {\rm{diag}} \left( \begin{matrix} {{S}^{\sigma }_{1}} \ {{S}^{\sigma }_{2}} \ \cdots  \ {{S}^{\sigma }_{N}} \end{matrix} \right)$, with the restriction $\Delta t < \left({{S}^{\sigma }_{i}} \right)^{-1}$. 

\section{Verification and validation} \label{SecIII}

To verify and validate the current model, six benchmarks are under consideration. First of all, multicomponent diffusion is adopted to confirm that the MRT DBM with the split collision term could provide a detailed relationship between the thermodynamic relaxation frequencies and the diffusivity of chemical species. 
Second, homogeneous mixture in a force field is used to test that the DBM can describe both thermal and isothermal systems where the acceleration and relaxtion frequencies are tunable. 
Third, the KH instability is simulated to demonstrate that the DBM could capture fluid flows with complex interfacial structures. 
Fourth, the laminar flame of a propane-air mixture is chosen to demonstrate that the DBM is capable of mimicking combustion. Fifth, simulations of the opposing reactions are performed to verify that the DBM can capture the chemical nonequilibrium process accurately. Finally, steady detonation is simulated to validate that the DBM has the ability to capture the supersonic reactive front with a strong compressible effect.

\subsection{Multicomponent diffusion}

Diffusion \cite{Cussler2000,Bird2002AMR} widely exists in fields of physics, chemistry, biology, etc. It plays a vital role in the combustion process, particularly where nonpremixed or partially mixed fuel and oxidant contact \cite{Law2006}. To mimic the diffusion in an accurate way is a prerequisite to simulating the combustion precisely. Actually, the diffusion (as well as viscosity and heat conduction) is a fundamental physical phenomenon in the thermodynamic nonequilibrium process. As the first benchmark, multicomponent diffusion is considered to validate our DBM for this kind of nonequilbrium process.

Give four chemical species $\sigma=A$, $B$, $C$, and $D$ with the same molar mass $m^{\sigma} = 1$ in the initial field as below
\begin{equation}
	\left\{
	\begin{array}{l}
		{{\left( {{n}^{A}}, {{n}^{B}}, {{n}^{C}}, {{n}^{D}} \right)}_{L}}=\left( 1, 2, 3, 4 \right) \tt{,} \\
		{{\left( {{n}^{A}}, {{n}^{B}}, {{n}^{C}}, {{n}^{D}} \right)}_{R}}=\left( 4, 3, 2, 1 \right) \tt{,}
	\end{array}
	\right.
\end{equation}
where the subscript $L$ indicates $0 \le x \le 0.05$, and $R$ indicates $0.05 < x \le 0.1$. The inflow and outflow boundary conditions are adopted in the $x$ direction, and the periodic boundary conditions in the $y$ direction. Moreover, the physical system is at rest ($\mathbf{u} = 0$) and can be regarded as isothermal ($T = 1$). Consequently, the exact solutions of the concentrations read \cite{Cussler2000,Bird2002AMR},
\begin{equation}
	{n}^{\sigma}=\dfrac{{n}^{\sigma}_{L}+{n}^{\sigma}_{R}}{2}-\dfrac{{n}^{\sigma}_{L}-{n}^{\sigma}_{R}}{2} {\rm Erf} \left(\dfrac{x-x_0}{\sqrt{4 {{\zeta}^{\sigma }} t}} \right)
	\label{Concentration_Exact}
	\tt{,}
\end{equation}
where ${\rm Erf}$ denotes the complementary error function, $x_0 = 0.05$ the interfacial position, and the diffusivity
\begin{equation}
	{{\zeta}^{\sigma }}=\dfrac{T}{{{m}^{\sigma }}S_{J }^{\sigma }}
	\label{Diffusion_expression}
	\tt{,}
\end{equation}
in terms of $S_{J }^{\sigma } = {{S}^{2 \sigma }_{2}} = {{S}^{2 \sigma }_{3}}$.

\begin{figure}[tbp]
	\begin{center}
		\includegraphics[width=0.5\textwidth]{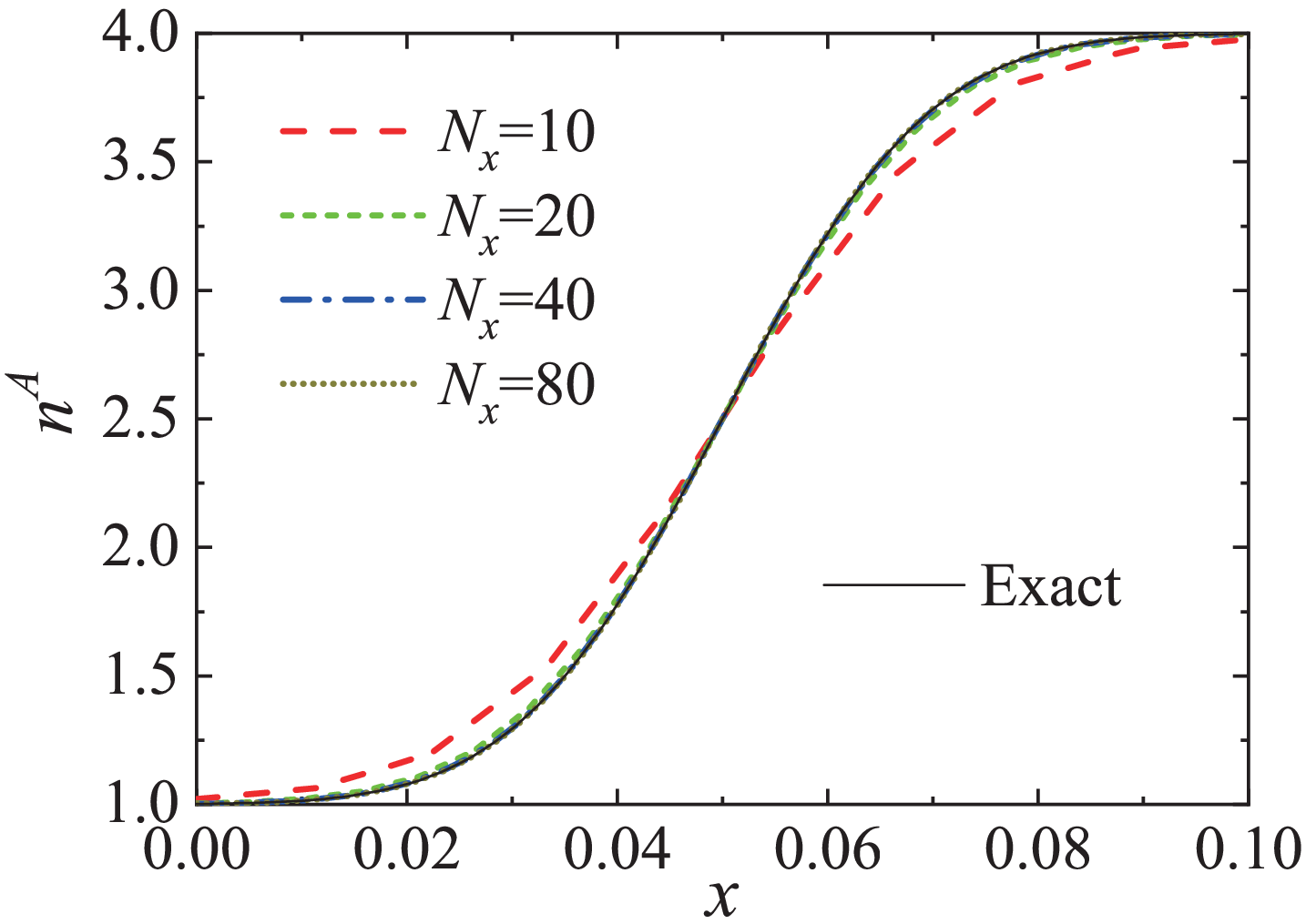}
	\end{center}
	\caption{Grid convergence analysis: the horizontal distribution of concentration $n^{A}$ at the time $t = 0.15$. The solid line represents the exact solution, and the other lines indicate the simulation results under different mesh grids.}
	\label{Fig02}
\end{figure}

To begin with, let us perform a grid convergence analysis, which is of great importance for numerical simulations. For this sake, four simulations are conducted under various mesh grids ${N_x} \times {N_y}$. To be specific, the mesh number is given as ${N_x}=10$, $20$, $40$, and $80$ in the horizontal direction and fixed as ${N_y}=1$ in the vertical direction. The relaxation frequencies are chosen as ${{S}^{1 \sigma }_{i}} = {{S}^{2 \sigma }_{i}} = 1250$, the temporal step $\Delta t = 2 \times 10^{-5}$, the extra degrees of freedom $I^{\sigma}=3$, and the parameters (${{v}_{a}^{\sigma}}$, ${{v}_{b}^{\sigma}}$, ${{v}_{c}^{\sigma}}$, ${{v}_{d}^{\sigma}}$, ${\eta }_{a }^{\sigma}$, ${\eta }_{b}^{\sigma}$, ${\eta }_{c}^{\sigma}$, ${\eta }_{d}^{\sigma}$) $=$ ($0.01$, $0.01$, $1.75$, $1.3$, $3$, $0$, $2.1$, $1$). Figure \ref{Fig02} delineates the concentration of species $A$. The long-dashed, short-dashed, dash-dotted, and short-dotted lines denote simulated results in the four cases, and the solid line stands for the analytical solution in Eq. \ref{Concentration_Exact}. As shown in Fig. \ref{Fig02}, with increasing resolution, the numerical results approach the exact solution. That is to say, the differences between these simulation results and theoretical solutions decrease as the spatial step reduces. In particular, numerical results for both ${N_x}=40$ and ${N_x}=80$ are quite close to the solution, which is satisfying.

\begin{figure}[tbp]
	\begin{center}
		\includegraphics[width=1.0\textwidth]{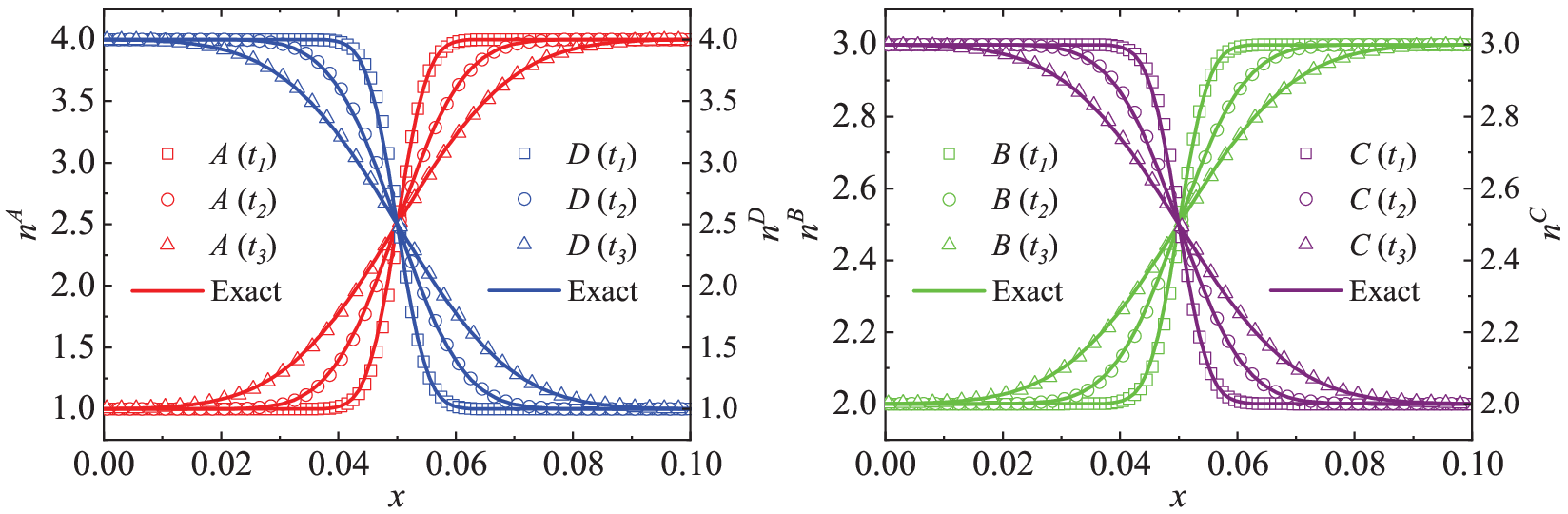}
	\end{center}
	\caption{Evolution of concentrations of four chemical species at time instants ${t_1} = 0.01$, ${t_2} = 0.05$, and ${t_3} = 0.15$, respectively. The left panel is for components $A$ and $D$, and the right panel for $B$ and $C$. Squares, circles, and triangles represent the simulation results at time instants ${t_1} = 0.01$, ${t_2} = 0.05$, and ${t_3} = 0.15$, respectively. Lines denote the corresponding exact solutions.}
	\label{Fig03}
\end{figure}

Next, a comparison is made between simulation results and exact solutions at various time instants during the evolution of multicomponent diffusion. In order to obtain simulation results as accurate as possible, the mesh grid is chosen as ${N_x} \times {N_y} = 200 \times 1$. Other parameters are the same as those used in Fig. \ref{Fig02}. Figure \ref{Fig03} displays the concentrations of chemical species $A$, $B$, $C$, and $D$ along the $x$ direction in the diffusion process. Symbols denote the DBM results, and solid lines indicate the corresponding exact solutions of Eq. (\ref{Concentration_Exact}). It can be observed that the simulated results match the exact solutions. Therefore, it is confirmed that the DBM can accurately describe the multicomponent diffusion.

\begin{figure}[tbp]
	\begin{center}
		\includegraphics[width=0.99\textwidth]{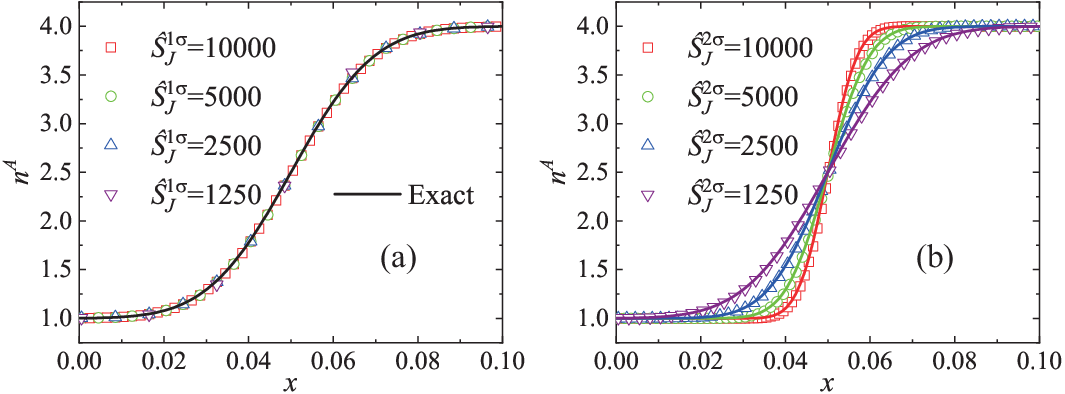}
	\end{center}
	\caption{Concentrations of species $A$ along the $x$ direction at a time instant $t = 0.15$ in the diffusion process: (a) For a fixed ${{S}^{2 \sigma }_{J}}$ and various ${{S}^{1 \sigma }_{J}}$; (b) For a fixed ${{S}^{1 \sigma }_{J}}$ and various ${{S}^{2 \sigma }_{J}} $. Symbols denote the DBM results, and lines represent the corresponding exact solutions.}
	\label{Fig04}
\end{figure}

Furthermore, as shown in Eq. (\ref{Diffusion_expression}), the diffusivity is a function of ${{S}^{2 \sigma }_{2}}$ and ${{S}^{2 \sigma }_{3}}$, but neither ${{S}^{1 \sigma }_{2}}$ nor ${{S}^{1 \sigma }_{3}}$. To verify it, a series of simulations are performed with different values of those parameters, i.e., ${{S}^{1 \sigma }_{2}} = {{S}^{1 \sigma }_{3}} = {{S}^{1 \sigma }_{J}}$, and ${{S}^{2 \sigma }_{2}} = {{S}^{2 \sigma }_{3}} = {{S}^{2 \sigma }_{J}}$. Figure \ref{Fig04} (a) illustrates concentrations of species $A$ versus $x$ for cases of a fixed ${{S}^{2 \sigma }_{J}} = 1250$ and various ${{S}^{1 \sigma }_{J}} = 10000$, $5000$, $2500$, and $1250$, respectively. Figure \ref{Fig04} (b) is for cases of a fixed ${{S}^{1 \sigma }_{J}} = 1250$ and various ${{S}^{2 \sigma }_{J}} = 10000$, $5000$, $2500$, and $1250$, respectively. Obviously, the parameters ${{S}^{1 \sigma }_{2}}$ and ${{S}^{1 \sigma }_{3}}$ have a negligible impact on the diffusion process. The diffusivity depends upon the values of ${{S}^{2 \sigma }_{2}}$ and ${{S}^{2 \sigma }_{3}}$. The numerical simulations are entirely consistent with the exact solutions. 

It should be mentioned that the two-step-relaxation DBM presents the same results as the one-step-relaxation model for ${{S}^{1 \sigma }_{i}} = {{S}^{2 \sigma }_{i}}$. It is verified that numerical results of the one-step-relaxation model are identical to those of the two-step-relaxation model when ${{S}^{1 \sigma }_{i}} = {{S}^{2 \sigma }_{i}}$ in Fig. \ref{Fig04} (a) or (b). Meanwhile, the simulations of the two-step-relaxation model in cases of ${{S}^{1 \sigma }_{i}} \ne {{S}^{2 \sigma }_{i}}$ are beyond the one-step-relaxation model. Compared to the latter one, the former model could present more details of nonequilibrium relaxation processes.

\begin{figure}[tbp]
	\begin{center}
		\includegraphics[width=1.0\textwidth]{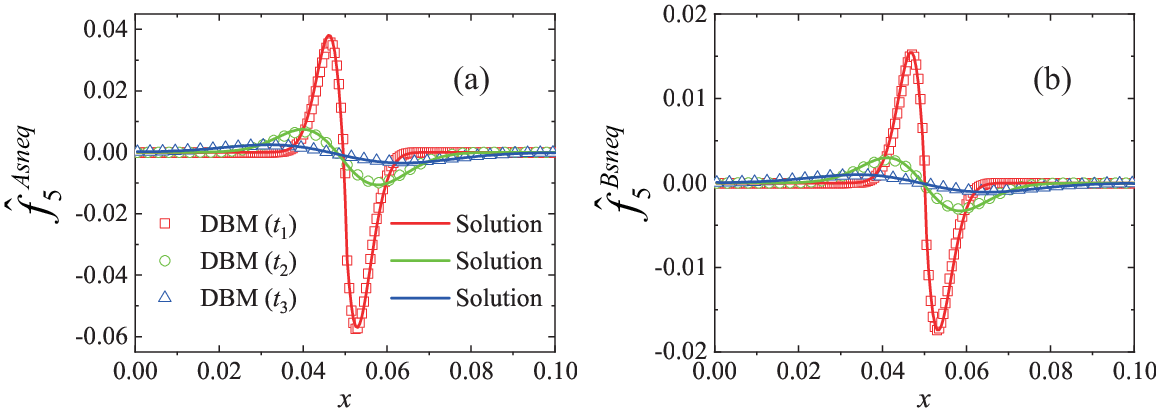}
	\end{center}
	\caption{Nonequilibrium quantities $\hat{f}_{5}^{\sigma sneq}$ in the diffusion process: (a) For species $\sigma = A$; (b) For species $\sigma = B$. Squares, circles, and triangles denote DBM results at time instants $t_1 = 0.01$, $t_2 = 0.05$, and $t_3 = 0.15$, respectively. Solid lines represent the corresponding analytical solutions.}
	\label{Fig05}
\end{figure}

In addition, to further validate that the DBM can be used to measure thermodynamic nonequilibrium manifestations, Fig. \ref{Fig05} displays nonequilibrium quantities $\hat{f}^{\sigma sneq}_{5}$ for species $\sigma = A$ and $B$ in the multicomponent diffusion process. Symbols stand for numerical results and lines for the corresponding analytical solutions in Eq. (\ref{hatf5_sneq}). It can be observed that the DBM results coincide with the analytical solutions. The profiles of species $C$ and $D$ are similar to those in Fig. \ref{Fig05} and are not plotted here for brevity. As a result, it is demonstrated that the DBM is capable of describing nonequilibrium behaviors.

\subsection{Mixture in the force field}

To confirm that chemical species with various accelerations and relaxation frequencies can be well described by using the current DBM, let us consider two situations, i.e., isothermal and thermal systems in force fields. Both the isothermal and thermal systems are homogeneous nonreactive mixtures that contain three components $\sigma = A$, $B$, and $C$, respectively. We choose the molar mass $\left( {{m}^{A}}, {{m}^{B}}, {{m}^{C}} \right)=\left( 2, 1.5, 1 \right)$, molar concentrations $\left( {{n}^{A}},{{n}^{B}},{{n}^{C}} \right) = \left( 1,4,2 \right)$, accelerations $\left( {{\mathbf{a}}^{A}},{{\mathbf{a}}^{B}},{{\mathbf{a}}^{C}} \right)=\left( -{{a}_{0}}, 0, {{a}_{0}} \right){{\mathbf{e}}_{x}}$, extra degrees of freedom $I^{\sigma} = 3$, parameters (${{v}_{a}^{\sigma}}$, ${{v}_{b}^{\sigma}}$, ${{v}_{c}^{\sigma}}$, ${{v}_{d}^{\sigma}}$, ${\eta }_{a }^{\sigma}$, ${\eta }_{b}^{\sigma}$, ${\eta }_{c}^{\sigma}$, ${\eta }_{d}^{\sigma}$)
$=$ ($0.5$, $1.5$, $2.2$, $3.5$, $0$, $5.2$, $3$, $0$).
Because the physical field is uniformly distributed, only one mesh grid ${N_x} \times {N_y} = 1 \times 1$ is used to have a high computing efficiency, and the periodic boundary conditions are adopted.

\begin{figure}[tbp]
	\begin{center}
		\includegraphics[bbllx=0pt,bblly=0pt,bburx=576pt,bbury=224pt,width=0.99\textwidth]{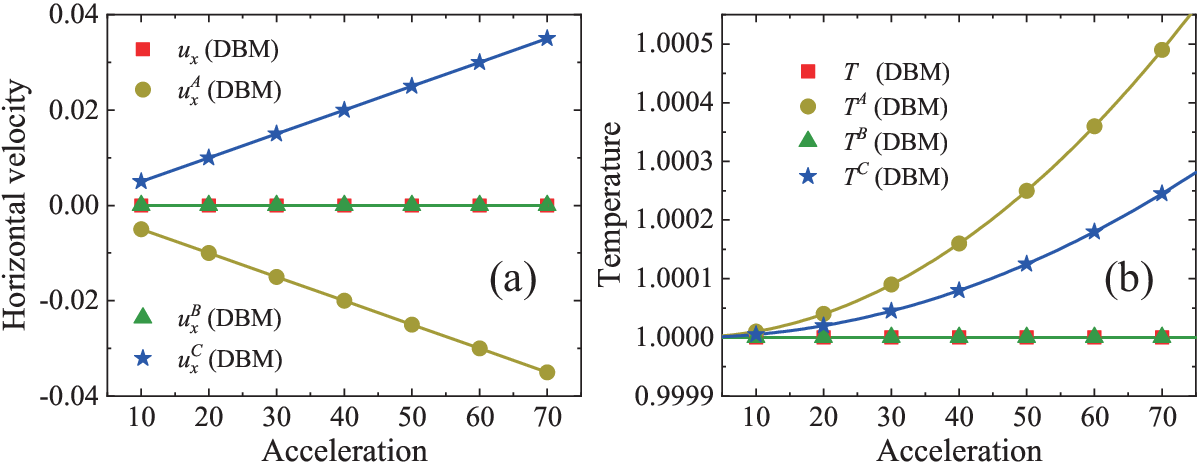}
	\end{center}
	\caption{Horizontal velocities (a) and temperatures (b) with different accelerations when the isothermal systems reach steady states. Symbols denote the DBM results, and solid lines represent the exact solutions.}
	\label{Fig06}
\end{figure}

As for the isothermal mixtures, the initial temperatures and velocities are given as $T^{\sigma} = 1$ and ${\mathbf{u}^{\sigma}} = 0$, respectively. Theoretically, the temperatures and velocities are expressed by Eqs. (\ref{Steady_u_force}) and (\ref{Steady_T_force}), when the systems reach steady states. To compare the DBM results with the theorical solutions, Fig. \ref{Fig06} displays the velocities and temperatures as the mixtures are imposed on various accelerations and the relaxation frequencies are chosen as ${{S}^{1 \sigma }_{2}} = {{S}^{2 \sigma }_{2}} = 2000$. The squares, circles, triangles, and diamonds indicate the simulation results of the mixing system, chemcial species $A$, $B$, and $C$, respectively. The solid lines stand for the corresponding exact results. Obviously, all DBM results coincide exactly with the theoretical solutions.

\begin{figure}[tbp]
	\begin{center}
		\includegraphics[bbllx=0pt,bblly=0pt,bburx=576pt,bbury=224pt,width=0.99\textwidth]{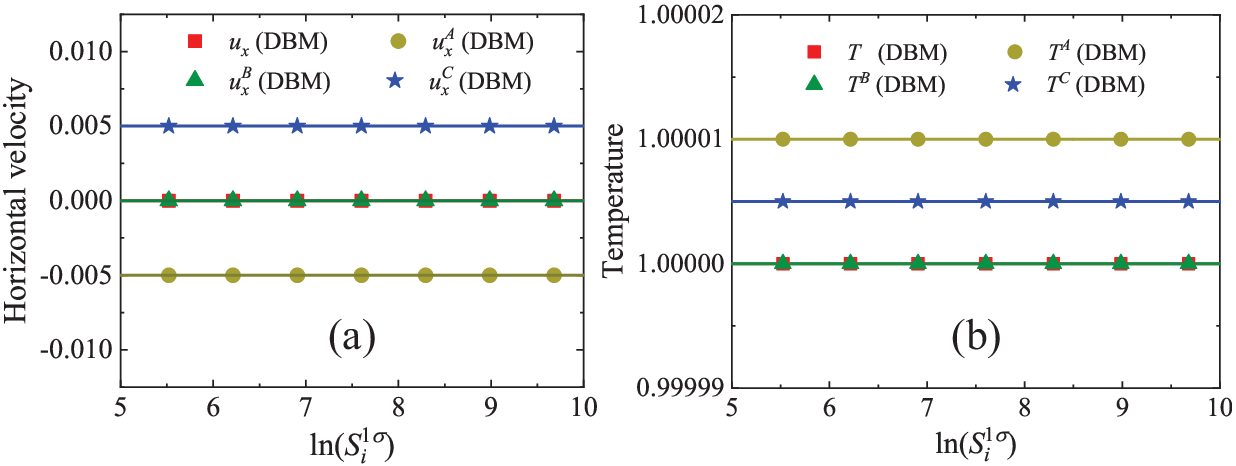}
	\end{center}
	\caption{Horizontal velocities (a) and temperatures (b) with variable ${{S}^{1 \sigma }_{i}}$ and fixed ${{S}^{2 \sigma }_{i}}$ in the stationary isothermal system. Symbols denote the DBM results, and solid lines represent the corresponding exact solutions.}
	\label{Fig07}
\end{figure}
\begin{figure}[tbp]
	\begin{center}
		\includegraphics[bbllx=0pt,bblly=0pt,bburx=576pt,bbury=224pt,width=0.99\textwidth]{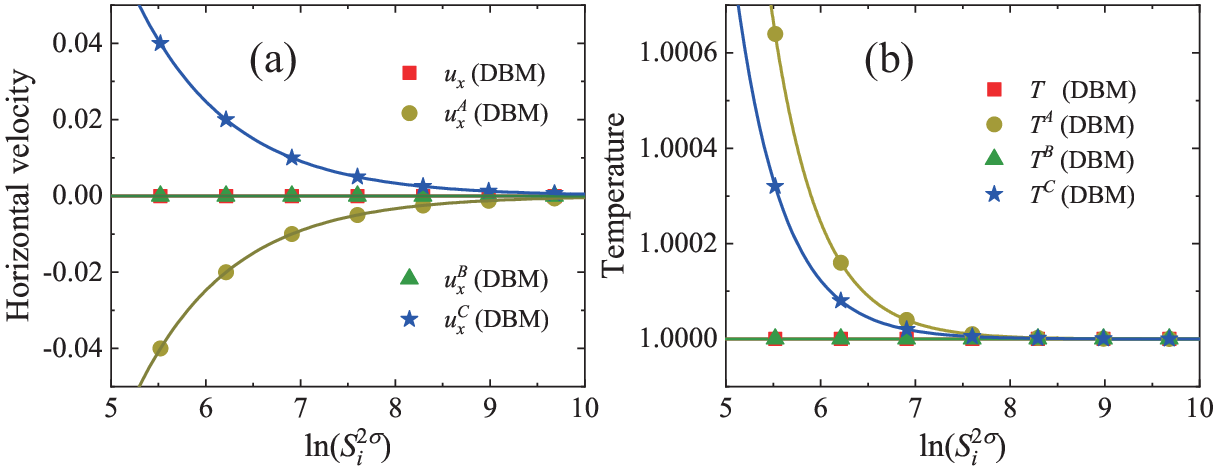}
	\end{center}
	\caption{Horizontal velocities (a) and temperatures (b) with fixed ${{S}^{1 \sigma }_{i}}$ and changeable ${{S}^{2 \sigma }_{i}}$ in the stationary isothermal system. Symbols denote the DBM results, and solid lines represent the corresponding exact solutions.}
	\label{Fig08}
\end{figure}

Let's now examine the impact of varying relaxation frequencies. 
In Fig. \ref{Fig07}, we observe the horizontal velocities and temperatures when employing variable ${{S}^{1 \sigma }_{i}}$ alongside a fixed ${{S}^{2 \sigma }_{i}} = 2000$ within the stationary isothermal system. Meanwhile, Fig. \ref{Fig08} showcases scenarios involving a constant ${{S}^{1 \sigma }_{i}} = 2000$ alongside a variable ${{S}^{2 \sigma }_{i}}$. Numerical results are denoted by symbols, while theoretical outcomes are represented by lines. 
It's apparent that the physical fields remain unaffected by ${{S}^{1 \sigma }_{i}}$ and are solely dependent on ${{S}^{2 \sigma }_{i}}$ within this homogeneous mixture under the specified body force. Notably, the DBM consistently yields exact results across all cases, which is quite satisfying.

\begin{figure}[tbp]
	\begin{center}
		\includegraphics[bbllx=0pt,bblly=0pt,bburx=664pt,bbury=503pt,width=0.5\textwidth]{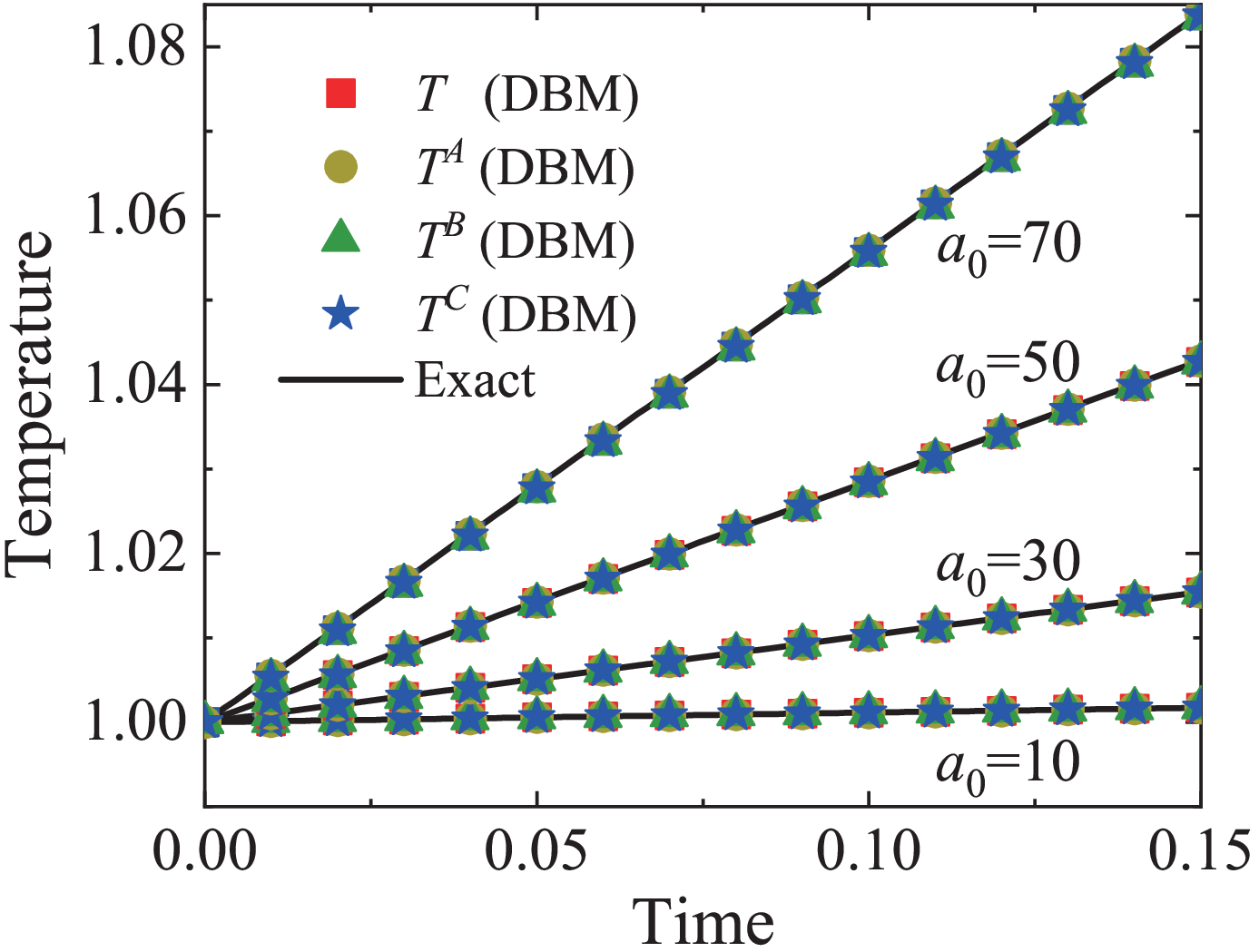}
	\end{center}
	\caption{Evolution of temperatures with different accelerations. Symbols denote the DBM results, and solid lines represent the exact solutions.}
	\label{Fig09}
\end{figure}

Next, the thermal mixtures are under consideration. The ultimate steady physical fields of the isothermal mixtures in Fig. \ref{Fig06} are set as the initial configurations of the thermal systems. Figure \ref{Fig09} plots the evolution of temperatures when the thermal mixtures are in force fields. Four values of accelerations $a_0 = 10$, $30$, $50$, and $70$ are under consideration. The squares, circles, triangles, and diamonds represent the simulated temperatures of the mixing system, chemcial species $A$, $B$, and $C$, respectively. The solid lines are for the exact results of the mixing temperatures. It is clear that the DBM results are in good agreement with the exact solutions.

\begin{figure}[tbp]
	\begin{center}
		\includegraphics[bbllx=0pt,bblly=0pt,bburx=576pt,bbury=224pt,width=0.99\textwidth]{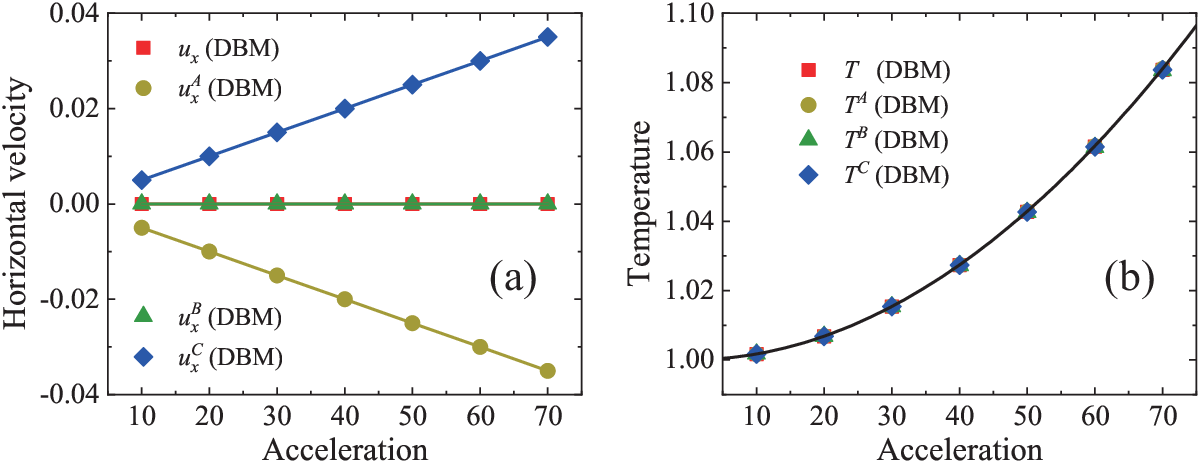}
	\end{center}
	\caption{Horizontal velocities (a) and temperatures (b) with different accelerations when the thermal systems are at the moment $t = 0.15$. Symbols denote the DBM results, and solid lines represent the exact solutions.}
	\label{Fig10}
\end{figure}

Figure \ref{Fig10} delineates the velocities and temperatures at the time $t = 0.15$ in the evolution of thermal systems  with various accelerations. Note that, the velocities of thermal mixtures in Fig. \ref{Fig10} (a) are exactly the same as those in Fig. \ref{Fig06} (a). Besides, the simulated mixing temperatures agree well with the exact solutions, and the individual and mixing temperatures are close to each other in the thermal systems, as shown in Fig. \ref{Fig10} (b). In fact, there are only minor differences among the individual and mixing temperatures in the isothermal systems as well, see Fig. \ref{Fig06}.

\begin{table}[h]
	\resizebox{\textwidth}{!}{
		\centering
		\begin{tabular}{lccccccc}
			\hline\hline
			Methods & $\ \ \rho \ \ $ & $\ u_x \ $ & $\ u_x^A \ $ & $\ u_x^B \ $ & $\ u_x^C \ $ & $\ \ \ \ \ \ \ T \ \ \ \ \ \ \ $ & Computing time \\
			\hline
			Method I & $10$ & $0$ & $-0.005$ & $0$ & $0.005$ & $1.01714857$ & $54$ s \\
			\hline
			Method II & $10$ & $0$ & $-0.005$ & $0$ & $0.005$ & $1.02572000$ & $60$ s \\
			\hline
			Method III \ & $10$ & $0$ & $-0.005$ & $0$ & $0.005$ & $1.01714857$ & $50$ s \\
			\hline
			Exact & $10$ & $0$ & $-0.005$ & $0$ & $0.005$ & $1.01714286$ & /  \\
			\hline\hline
	\end{tabular}}
	\caption{Simulation data about the three methods to calculate force terms.}
	\label{TableII}
\end{table}

It should be mentioned that the simulation results in Figs. \ref{Fig06} - \ref{Fig10} are calculated with Method I. Actually, Methods II and III give similar results. To compare Methods I, II, and III, Table \ref{TableII} lists the data in the simulation case of $a_0 = 10$ in Fig. \ref{Fig10}.
Simulations are performed on a personal computer with Intel(R) Core(TM) i9-9880H CPU @ 2.30GHz, RAM 64.0 GB, and a 64-bit version system for double precision floating point operations.
As shown in Table \ref{TableII}, the simulated density and velocities equal the corresponding exact solutions. It means that the conservation of mass and momentum is obeyed for all methods. Besides, the mixing temperatures simulated with Methods I, II, and III are $T = 1.01714857$, $1.02572000$, and $1.01714857$, respectively. Compared with the exact value $1.01714286$, the relative errors are $5.6 \times 10 ^{-6}$, $8.4 \times 10 ^{-3}$, and $5.6 \times 10 ^{-6}$, respectively. That is to say, Methods I and III give the same simulation results, both have a higher accuracy than Method II. Additionally, the computing time is $54$, $60$, and $50$ seconds for the three methods as the simulation runs $2 \times 10 ^{5}$ time steps. In other words, Methods I and II require $8 \%$ and $20 \%$ more computing time than Method III. Because it needs to calculate the equilibrium distribution functions ${f}^{\sigma seq}$ once per loop for Method I, twice per loop for Method II, while it does not demand ${f}^{\sigma seq}$ for Method III. Consequently, it takes less computational cost and running time for Method III.

\subsection{Kelvin-Helmholtz instability}

As a fundamental interfacial instability in fluid mechanics, the KH instability occurs when there is velocity shear across a wrinkled interface in a fluid system, and leads to the formation of vortices and turbulence \cite{Batchelor2000}. The KH instability is ubiquitous in nature and of considerable interest in scientific and engineering fields \cite{Batchelor2000,Liang2016PRE,Wang2017}. To show the capacity of our DBM in dealing with complex fluid systems, here we simulate the KH instability with a complicated interfacial dynamics.

\begin{figure}[tbp]
	\begin{center}
		\includegraphics[width=0.3\textwidth]{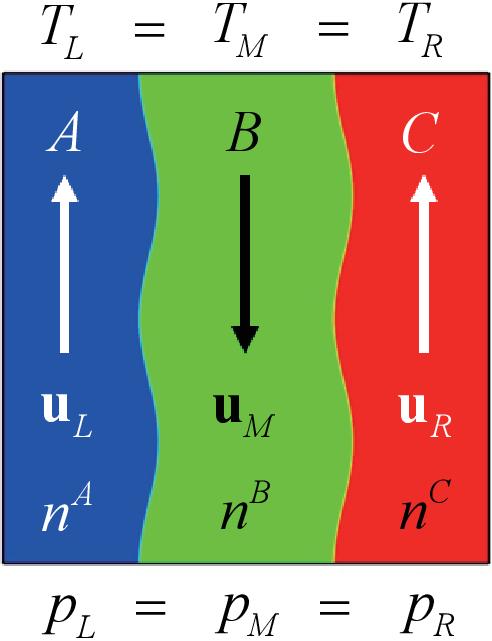}
	\end{center}
	\caption{Initial configuration of the KH instability.}
	\label{Fig11}
\end{figure}

Figure \ref{Fig11} portrays the sketch of the initial configuration of a three-component fluid. The length and height of the calculation domain are $L_{x} = L_{y} = 1$. The domain is divided into three parts, i.e., $0 < x \le x_{1}$, $x_{1} < x \le x_{2}$, and $x_{2} < x \le L_{x}$. Between the left and middle parts is the interface located at $x_{1} = 0.3 L_{x}$, between the middle and right parts is the interface located at $x_{2} = 0.7 L_{x}$. To trigger the KH instability rollup, both interfaces have an imposed cosinusoid perturbation, $w={{w}_{0}}\cos \left( {4\pi y}/{{L}_{y}} \right)$, with an amplitude ${w}_{0} = L_{x}/200$. Initially, the left (middle, right) part is occupied by species $A$ ($B$, $C$) moving upwards (downwards, upwards) with velocity $\mathbf{u}_{L} = u_{0} \mathbf{e}_{y}$ ($\mathbf{u}_{M} = - u_{0} \mathbf{e}_{y}$, $\mathbf{u}_{R} = u_{0} \mathbf{e}_{y}$). Both concentrations and temperatures in the three parts are equal, i.e., $n_{L} = n_{M} = n_{R}$ and $T_{L} = T_{M} = T_{R}$, hence the pressure is homogeneous across the two interfaces, $p_{L} = p_{M} = p_{R}$, due to the constitutive relation $p^{\sigma} = n^{\sigma} T^{\sigma}$. To be smooth across the interface, the initial profiles of the concentrations and velocities are given by
\[
{{n}^{A}}=\dfrac{{{n}_{L}}}{2}-\dfrac{{{n}_{L}}}{2}\tanh \left( \dfrac{x-{{x}_{1}}+w}{{{W}_{n}}} \right)
\tt{,}
\]
\[
{{n}^{C}}=\dfrac{{{n}_{R}}}{2}+\dfrac{{{n}_{R}}}{2}\tanh \left( \dfrac{x-{{x}_{2}}+w}{{{W}_{n}}} \right)
\tt{,}
\]
\[
{{n}^{B}}={{n}_{M}}-{{n}^{A}}-{{n}^{C}}
\tt{,}
\]
\[
{{\mathbf{u}}^{\sigma }}=
\left\{
\begin{array}{l}
	\dfrac{{{\mathbf{u}}_{L}}+{{\mathbf{u}}_{M}}}{2}-\dfrac{{{\mathbf{u}}_{L}}-{{\mathbf{u}}_{M}}}{2}\tanh \left( \dfrac{x-{{x}_{1}}+w}{{{W}_{\mathbf{u}}}} \right), {\rm{\ for \ }} 0 < x \le \dfrac{{L}_{x}}{2} \tt{,} \\
	\dfrac{{{\mathbf{u}}_{M}}+{{\mathbf{u}}_{R}}}{2}-\dfrac{{{\mathbf{u}}_{M}}-{{\mathbf{u}}_{R}}}{2}\tanh \left( \dfrac{x-{{x}_{2}}+w}{{{W}_{\mathbf{u}}}} \right), {\rm{\ for \ }} \dfrac{{L}_{x}}{2} <x\le {{L}_{x}} \tt{,}
\end{array}
\right.
\]
where ${W}_{n}$ (${W}_{\mathbf{u}}$) indicates the width of concentration (velocity) transition layer.

The boundary conditions are as follows: specular reflection boundary condition in the $x$ direction and periodic boundary condition in the $y$ direction. Simulation is carried out on a uniform mesh $N_x \times N_y = 2000 \times 2000$ with $\Delta x = \Delta y = 5 \times 10^{-4}$. The time step is set to be as small as $\Delta t = 2.5 \times 10^{-5}$ to keep the numerical dissipation negligible. The other parameters are $m^{A} = 1$, $m^{B} = 1.5$, $m^{C} = 2$, $I^{\sigma} = 3$, ${{S}^{1 \sigma }_{i}} = {{S}^{2 \sigma }_{i}} = 5\times 10^{3}$, and
(${{v}_{a}^{\sigma}}$, ${{v}_{b}^{\sigma}}$, ${{v}_{c}^{\sigma}}$, ${{v}_{d}^{\sigma}}$, ${\eta }_{a }^{\sigma}$, ${\eta }_{b}^{\sigma}$, ${\eta }_{c}^{\sigma}$, ${\eta }_{d}^{\sigma}$)
$=$
($2$, $1.414$, $3.9$, $2.758$, $1.5$, $0$, $5.5$, $0$).
In addition, as the number of mesh grids is large enough, the parallel programming with the message-passing interface is implemented in Fortran to improve the computing speed. Actually, because all information transfer is local in time and space in the evolution of the discrete Boltzmann equation, the DBM has natural parallelism with excellent scalability \cite{Lin2019CTP}.

\begin{figure}[tbp]
	\begin{center}
		\includegraphics[bbllx=0pt,bblly=0pt,bburx=519pt,bbury=345pt,width=0.9\textwidth]{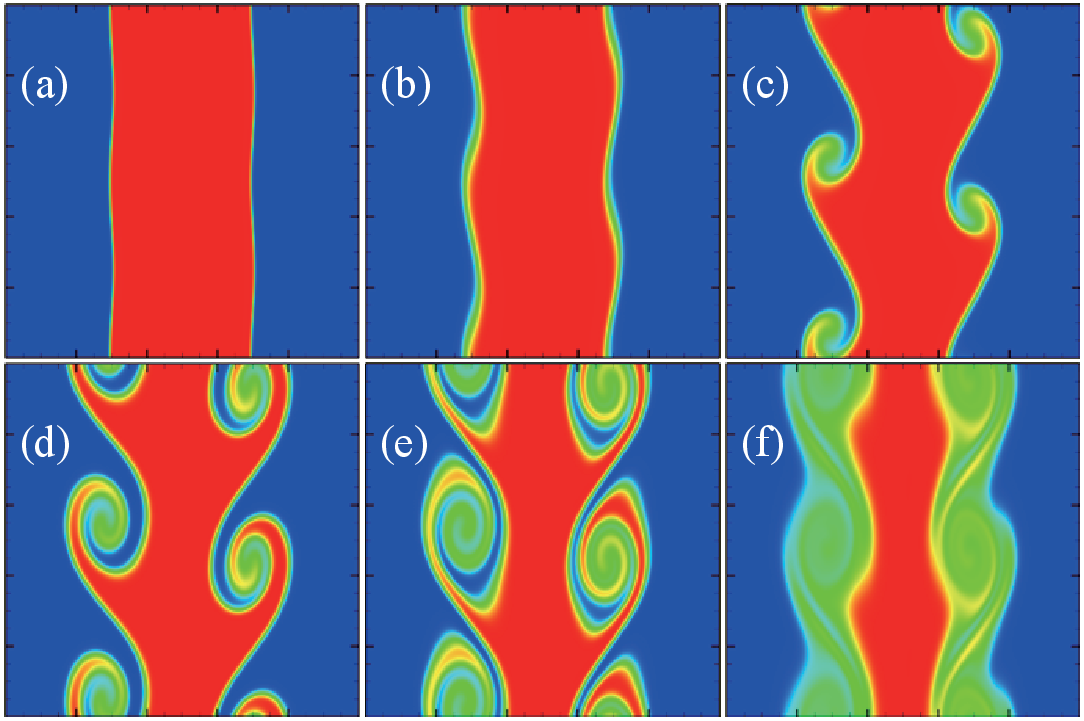}
	\end{center}
	\caption{Molar fraction $X^B$ at time instants $t = 0.0$, $0.5$, $1.0$, $1.5$, $2.0$, and $3.0$ in the evolution of the KH instability. The color from blue to red  corresponds to the value from $0$ to $1$.}
	\label{Fig12}
\end{figure}

Figure \ref{Fig12} depicts contours of the molar fraction of species $B$ at representative times in the evolution of the miscible KH instability. It can be observed in Figs. \ref{Fig12} (a)-(b) that the amplitude of perturbation $w$ increases due to the shear effect and the width of concentration transition layer ${W}_{n}$ increases because of diffusion. The fluid interface begins wiggling and its shape changes from regular to irregular gradually. It can be found in Fig. \ref{Fig12} (c) that, as time advances, several pairs of vortices appear, and the middle fluid penetrates into the left and right ones. Figures \ref{Fig12} (d)-(e) show that the continuous growth of vortices leads to formation of billows, and nonregular interfaces become more complex at the later stage. As shown in Fig. \ref{Fig12} (f), fluid structures are chaotic and the KH instability promotes the mixture between different fluid species. The above dynamic process of the KH instability obtained by our model is basically consistent with the scenarios in previous studies \cite{Wang2017,Liang2016PRE}.

\begin{figure}[tbp]
	\begin{center}
		\includegraphics[width=0.5\textwidth]{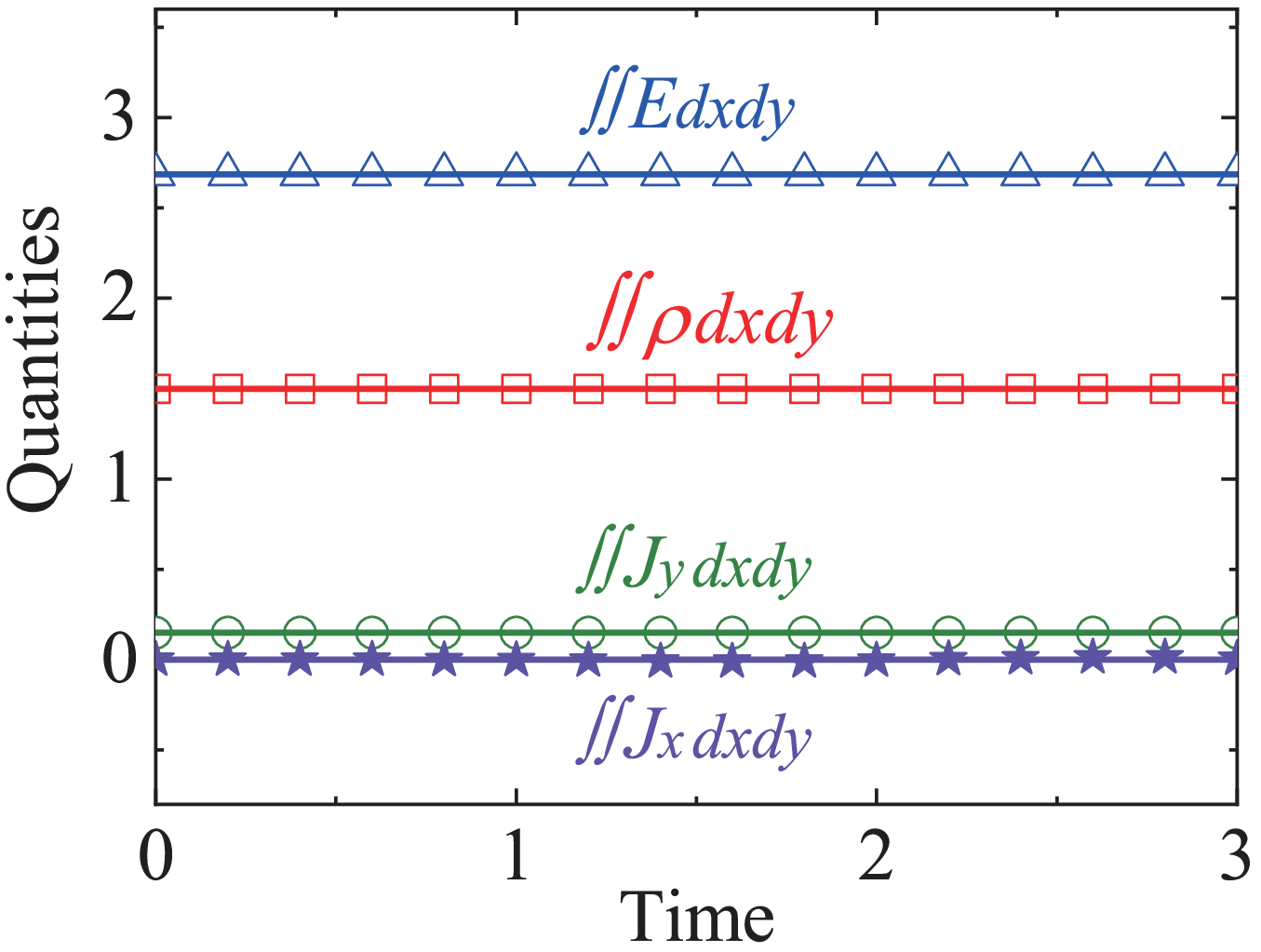}
	\end{center}
	\caption{Conserved quantities ($\iint{\rho dxdy}$, $\iint{J_{x} dxdy}$, $\iint{J_{y} dxdy}$, $\iint{E dxdy}$) in the evolution of the KH instability. Symbols and lines indicate numerical and exact results, respectively.}
	\label{Fig13}
\end{figure}

To further verify our model, we measure conserved quantities in the process of KH instability, see Fig. \ref{Fig13}. Squares, pentagrams, circles, and triangles indicate numerical results of the mass $\iint{\rho dxdy}$, momentum in the $x$ direction $\iint{J_{x} dxdy}$, momentum in the $y$ direction $\iint{J_{y} dxdy}$), and energy $\iint{E dxdy}$, respectively. Solid lines are for the corresponding exact solutions $\iint{\rho dxdy}=1.5$, $\iint{J_{x} dxdy}=0$, $\iint{J_{y} dxdy}=0.15$, and $\iint{E dxdy}=2.685$, respectively. It is clear in Fig. \ref{Fig13} that our computed results agree well with these exact solutions. Moreover, it is found that the numerical results of $\iint{\rho^{\sigma} dxdy}$ are exactly equal to their corresponding theoretical solutions $0.3$, $0.6$, and $0.6$ for species $\sigma = A$, $B$, and $C$, respectively. The results are very good and satisfactory.

\subsection{Laminar flame}

In this subsection, the objective is to demonstrate that the DBM is suitable for combustion. For this purpose, we simulate a laminar flame of propane-air mixture. The combustion is controlled by the one-step overall reaction,
\begin{equation}
	{{\rm{C}}_{3}}{{\rm{H}}_{8}}+5{{\rm{O}}_{2}}\to 3\rm{C}{{\rm{O}}_{2}}+4{{\rm{H}}_{2}}\rm{O}
	\tt{,}
\end{equation}
\begin{equation}
	{{\omega }^{\sigma }}={{s}^{\sigma }}\cdot {{{m}}^{\sigma }}\cdot {{\omega }_{\rm{ov}}}
	\tt{,}
\end{equation}
\begin{equation}
	\omega_{\rm{ov}}={{k}_{\rm{ov}}}{{n}^{{{\rm{C}}_{3}}{{\rm{H}}_{8}}}}{{n}^{{{\rm{O}}_{2}}}}\exp \left( -{{E}_{a}}/RT \right)
	\tt{,}
\end{equation}
where ${{\rm{C}}_{3}}{{\rm{H}}_{8}}$, ${{\rm{O}}_{2}}$, ${\rm{C}{{\rm{O}}_{2}}}$, and ${{\rm{H}}_{2}}\rm{O}$ stand for propane, oxygen, carbon dioxide, and water, respectively. Nitrogen ${\rm{N}}_{2}$ is assumed to be inert. The stoichiometric coefficients are [$s^{{{\rm{C}}_{3}}{{\rm{H}}_{8}}}$, 
$s^{{{\rm{O}}_{2}}}$, 
$s^{{\rm{C}{{\rm{O}}_{2}}}}$, 
$s^{{{\rm{H}}_{2}}\rm{O}}$, 
$s^{{\rm{N}}_{2}}$] $=$ [$-1$, $-5$, $3$, $4$, $0$], the molar mass [$m^{{{\rm{C}}_{3}}{{\rm{H}}_{8}}}$, $m^{{{\rm{O}}_{2}}}$, $m^{{\rm{C}{{\rm{O}}_{2}}}}$, $m^{{{\rm{H}}_{2}}\rm{O}}$, $m^{{\rm{N}}_{2}}$] $=$ [$4.4$, $3.2$, $4.4$, $1.8$, $2.8$] $\times {{10}^{-2}}$ [$\rm{kg} / \rm{mol}$], the reaction coefficient ${{k}_{\rm{ov}}}=9.9\times {{10}^{7}}\left[ {{\rm{m}}^{3}}\cdot \rm{mo}{{\rm{l}}^{-1}}\cdot {{\rm{s}}^{-1}} \right]$, the universal gas constant $R=8.315\left[ \rm{J}\cdot \rm{mo}{{\rm{l}}^{-1}}\cdot {{\rm{K}}^{-1}} \right]$, the effective activation energy ${{E}_{a}}=1.26\times {{10}^{5}}\left[ \rm{J}\cdot \rm{mo}{{\rm{l}}^{-1}} \right]$, the chemical heat of overall reaction $Q=2.05\times {{10}^{6}}\left[ \rm{J}\cdot \rm{mo}{{\rm{l}}^{-1}} \right]$, the overall reaction rate $\omega_{\rm{ov}}$, the mass change rate of species ${{\omega }^{\sigma }}$, and the parameters (${{v}_{a}^{\sigma}}$, ${{v}_{b}^{\sigma}}$, ${{v}_{c}^{\sigma}}$, ${{v}_{d}^{\sigma}}$, ${\eta }_{a }^{\sigma}$, ${\eta }_{b}^{\sigma}$, ${\eta }_{c}^{\sigma}$, ${\eta }_{d}^{\sigma}$) $=$ ($0.7$, $0.7$, $3.9$, $2.758$, $1.5$, $0$, $5.5$, $0$).
Initially, a channel with length $L=4$ $\rm{cm}$ is filled with the propane-air mixture with an equivalence ratio $0.6$. The molar concentration is $44.6$ ${\rm{mol}}\cdot {{\rm{m}}^{-3}}$, the temperature $300$ $\rm{K}$, the pressure $1$ $\rm{atm}$. After ignition in the left part of the channel $0 \le x \le L/64$, the flame starts to move downstream. The periodic boundary conditions are used at the upper and lower walls, the specular reflection (outflow) boundary condition at the left (right) side. The grid is chosen as $N_x \times N_y = 1600 \times 1$, the spatial step $\Delta x = \Delta y = 2.5 \times 10 ^{-5}$ $\rm{m}$, and the temporal step $\Delta t = 1.25 \times 10 ^{-10}$ $\rm{s}$.
\begin{figure}[tbp]
	\begin{center}
		\includegraphics[width=0.5\textwidth]{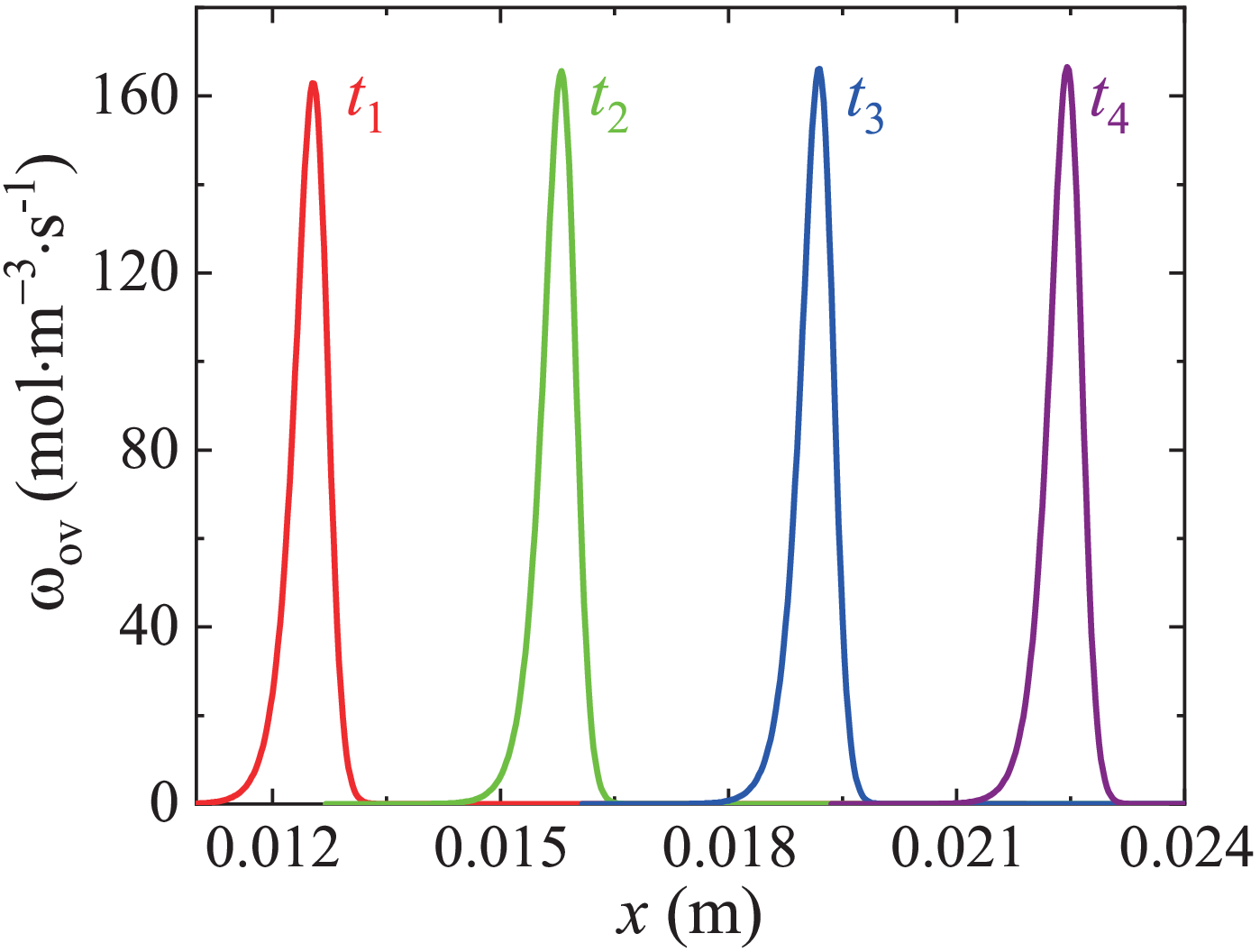}
	\end{center}
	\caption{The overall reaction rate in the evolution of the laminar flame at time instants $t_{1} = 3.125 \times 10 ^{-2}$ s, $t_{2} = 3.75 \times 10 ^{-2}$ s, $t_{3} = 4.375 \times 10 ^{-2}$ s, and $t_{4} = 5 \times 10 ^{-2}$ s from left to right.}
	\label{Fig14}
\end{figure}
\begin{figure}[tbp]
	\begin{center}
		\includegraphics[width=0.5\textwidth]{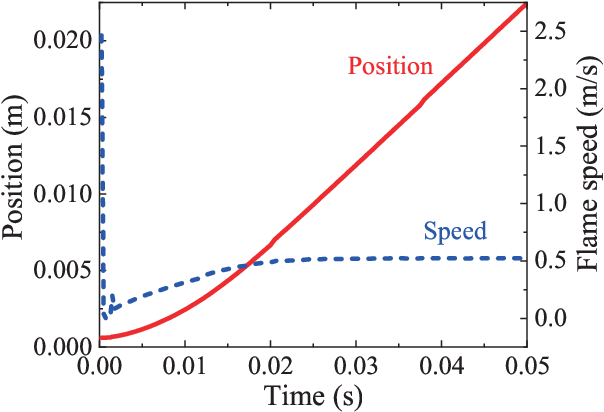}
	\end{center}
	\caption{Temporal evolution of the flame position and speed presented by the solid and dashed lines, respectively.}
	\label{Fig15}
\end{figure}

Figure \ref{Fig14} delineates the overall reaction rate at various time instants in the evolution of the laminar flame. Clearly, the overall reaction rate first increases then decreases and forms a peak at the combustion front as the flame propagates forwards. The profile of reaction rate gradually becomes steady as time goes on. Let us define the flame position as the location where the reaction takes its maximum. Then we can obtain the temporal evolution of the flame position and its velocity, as shown in Fig. \ref{Fig15}. It is clear that the flame moves rightwards after the initial ignition stage, and its speed tends to be a constant in the later period. To be specific, the flame speed is around ${U_f} = 0.522$ $\rm{m/s}$ at $t = 0.05$ s. Meanwhile, the value of flow velocity is about ${U_o} = 0.413$ $\rm{m/s}$ ahead of the moving flame. Note that the burning velocity can be estimated by the relation, ${U_f}={U_o}+{U_b}$, hence the resultant burning velocity is ${U_b} = 0.109$ $\rm{m/s}$.

\begin{figure}[tbp]
	\begin{center}
		\includegraphics[width=0.5\textwidth]{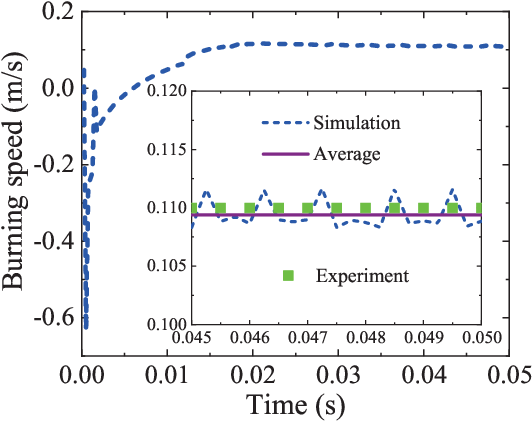}
	\end{center}
	\caption{The burning speed in the evolution of the laminar flame. The insert shows the zoom-in view within the range $0.045 \ {\rm{s}} \le t \le 0.05 \ {\rm{s}}$. The dashed and solid lines denote simulation results and their average, and the squares represent experimental data.}
	\label{Fig16}
\end{figure}
\begin{figure}[tbp]
	\begin{center}
		\includegraphics[width=0.5\textwidth]{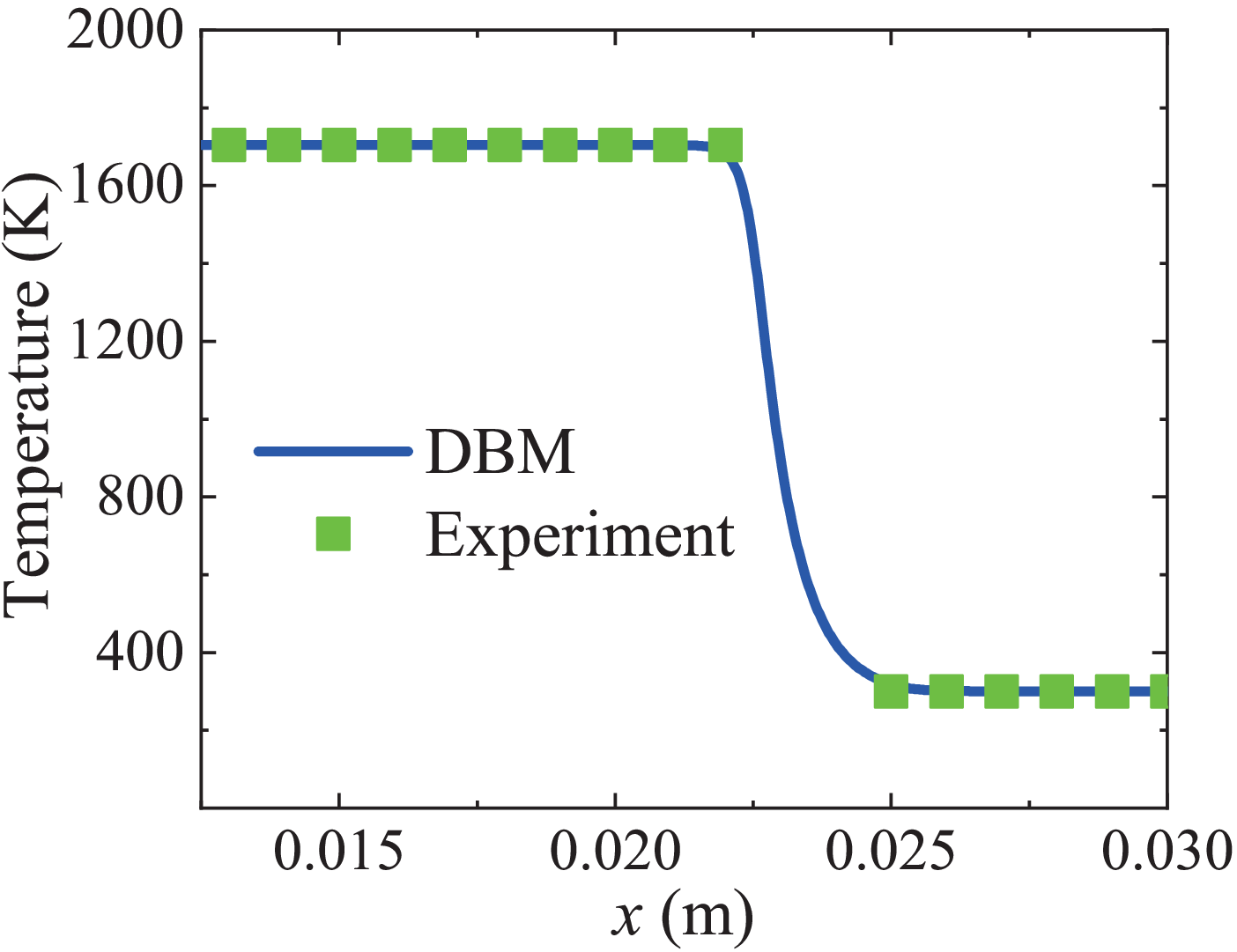}
	\end{center}
	\caption{Temperature profiles in the evolution of the laminar flame at time instant $t = 0.05$ s. The line and squares stand for the DBM results and experimental data \cite{Law2006}, respectively.}
	\label{Fig17}
\end{figure}

Figure \ref{Fig16} exhibits the evolution of the burning speed. To give a clearer depiction, an insert is attached for the enlargement within the period $0.045 \ {\rm{s}} \le t \le 0.05 \ {\rm{s}}$, during which the flame speed is almost steady with only small perturbations. It is easy to calculate the average burning speed, ${U_b} = 0.10941$, within the time range $0.045 \ {\rm{s}} \le t \le 0.05 \ {\rm{s}}$. Clearly, the numerical result approaches the experimental datum $0.11$ $\rm{m/s}$ in Ref. \cite{Yamaoka1985SC}. Furthermore, Fig. \ref{Fig17} displays the temperature profiles at $t = 0.05$ s. Specifically, the temperatures are $1704$ K and $1705$ K in the simulation and experiment \cite{Law2006}, respectively. The satisfying result of the DBM is due to its outstanding advantages:
(i) Physical properties such as the extra degrees of freedom and the specific heat ratio are flexible in the DBM;
(ii) The physical fields of concentration, velocity, temperature, and pressure are naturally coupled together in the DBM;
(iii) The DBM is suitable for both low-speed incompressible and high-speed compressible fluid flows.

\subsection{Opposing reaction}

To demonstrate that our DBM is suitable for a chemical nonequilibrium system, the opposing reactions $A\rightleftharpoons B$ are used as a typical benchmark. The forward and reverse reaction coefficients are $k_{1}$ and $k_{-1}$, respectively. Hence the overall reaction rate reads ${{\omega }_{\rm{ov}}} = {{k}_{1}}{{n}^{A}}-{{k}_{-1}}{{n}^{B}}$. Initially, the concentrations are (${{n}^{A}}$, ${{n}^{B}}$) = ($n_{0}$, $0$) = ($1$, $0$), the velocity $\mathbf{u} = 0$, the temperature $T=1$. The concentrations are
\[{{n}^{A}}=n_{e}^{A}+n_{e}^{B}\exp \left( -\dfrac{{{k}_{1}}{{n}_{0}}}{n_{e}^{B}}t \right) \tt{,} \]
\[{{n}^{B}}=n_{e}^{B}-n_{e}^{B}\exp \left( -\dfrac{{{k}_{1}}{{n}_{0}}}{n_{e}^{B}}t \right) \tt{,} \]
during the nonequilibrium reaction process, and
\[n_{e}^{A}=\dfrac{{{k}_{-1}}{{n}_{0}}}{{{k}_{1}}+{{k}_{-1}}} \tt{,} \]
\[n_{e}^{B}=\dfrac{{{k}_{1}}{{n}_{0}}}{{{k}_{1}}+{{k}_{-1}}} \tt{,} \]
when the chemical reaction reaches equilibrium. Given the heat release $Q = 10$ and extra degrees of freedom $I^{A} = I^{B} =I = 3$, the temperature reads
\[T={{T}_{0}}+\dfrac{2{{n}^{B}}Q}{{{n}_{0}}\left( D+I \right)} \tt{,} \]
as the chemical reaction takes place.

\begin{figure}[tbp]
	\begin{center}
		\includegraphics[width=0.5\textwidth]{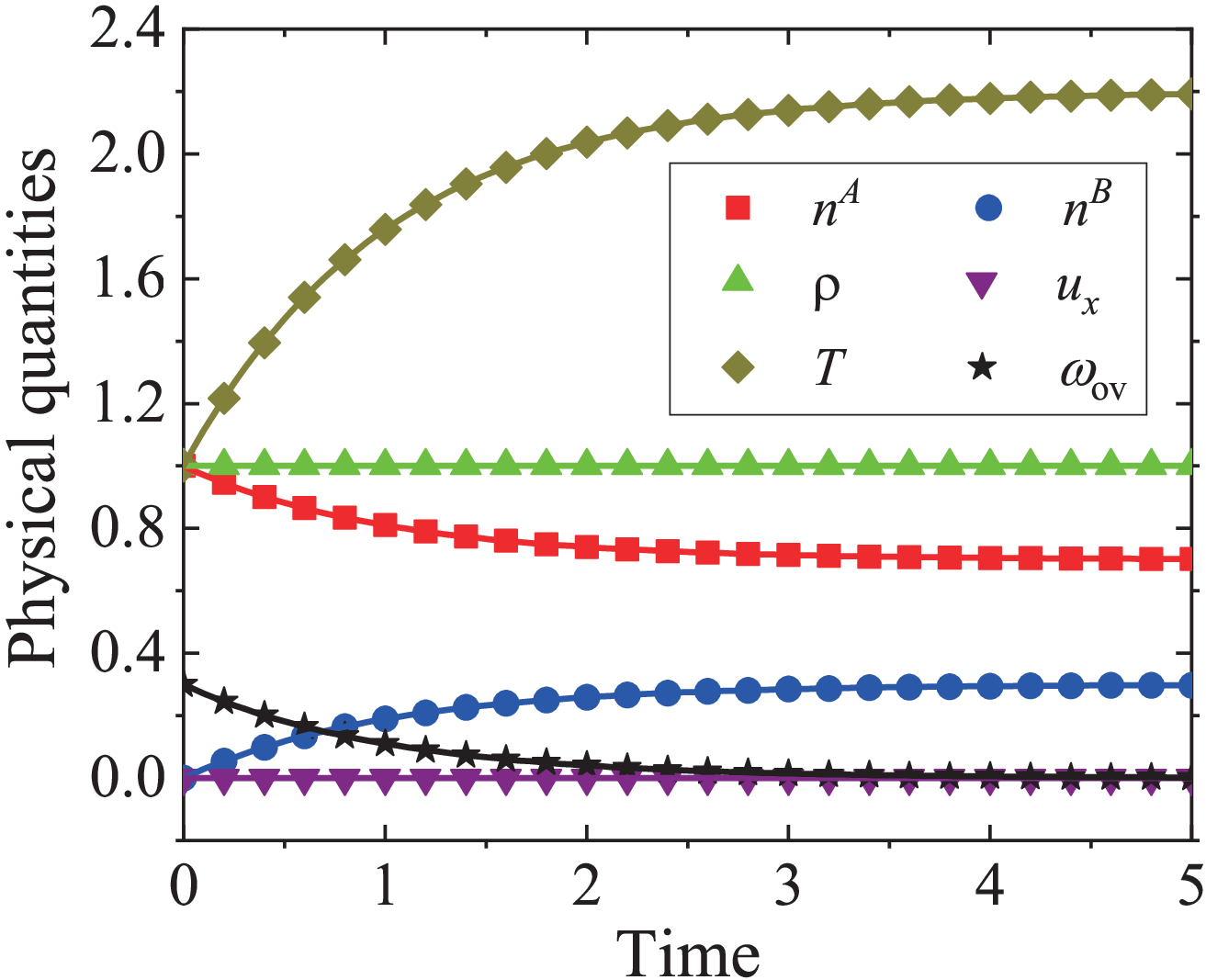}
	\end{center}
	\caption{Physical quantities versus time in the evolution of opposing reaction. Symbols denote DBM results in the legend, and lines denote corresponding exact solutions.}
	\label{Fig18}
\end{figure}
Figure \ref{Fig18} illustrates physical quantities (${{n}^{A}}$, ${{n}^{B}}$, $\rho$, $u_{x}$, $T$, ${{\omega }_{\rm{ov}}}$) versus time in the case of ($k_{1}$, $k_{-1}$) $=$ ($0.3$, $0.7$). The relaxation frequencies are choosen as ${{S}^{1 \sigma }_{i}} = {{S}^{2 \sigma }_{i}} = 10^{4}$, and the parameters (${{v}_{a}^{\sigma}}$, ${{v}_{b}^{\sigma}}$, ${{v}_{c}^{\sigma}}$, ${{v}_{d}^{\sigma}}$, ${\eta }_{a }^{\sigma}$, ${\eta }_{b}^{\sigma}$, ${\eta }_{c}^{\sigma}$, ${\eta }_{d}^{\sigma}$) $=$ ($0.7$, $0.7$, $2.7$, $2.7$, $0$, $5.3$, $5$, $0$).
To be specific, concentration of species $A$ ($B$) decreases (increases) gradually as time goes on. The mixture density remains $\rho = 1$, which equals the exact solution $\rho = m^{A} n^{A} + m^{B} n^{B}$ with $m^{A} = m^{B} = 1$. The system remains motionless, i.e., $u_x = u_y = 0$, in line with the momentum conservation. The temperature rises and the reaction rate reduces over time. On the whole, the DBM results coincide well with the exact solutions in the evolution of chemical reaction. All quantities become constant when the opposing reaction reaches equilibrium. At $t=20$, the numerical results (${{n}^{A}}$, ${{n}^{B}}$, $\rho$, $u_{x}$, $T$, ${{\omega }_{\rm{ov}}}$) $=$ ($0.7$, $0.3$, $1$, $0$, $2.2$, $0$) match the exact solutions. Consequently, it is verified that the DBM is capable of both chemical equilibrium and nonequilibrium processes.

\begin{figure}[tbp]
	\begin{center}
		\includegraphics[width=0.5\textwidth]{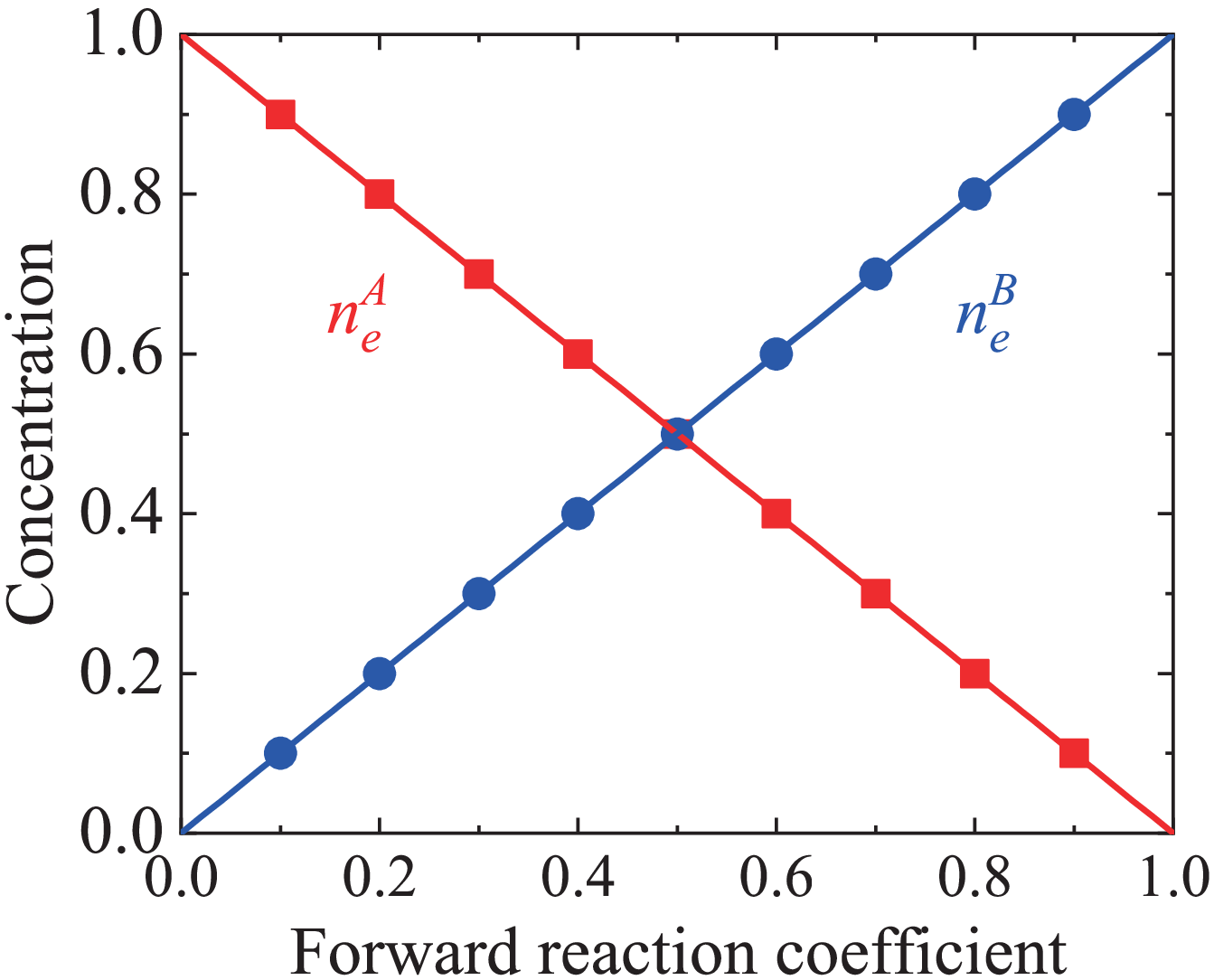}
	\end{center}
	\caption{Concentrations of species $A$ and $B$ versus the reaction coefficient. Symbols represent DBM results in the legend, and lines represent corresponding exact solutions.}
	\label{Fig19}
\end{figure}

Let us consider more cases of opposing reactions with various forward reaction coefficients $k_1$. For simplicity, the corresponding reverse reaction coefficients are set as ${k_{-1} }= 1 - {k_1}$. Figure \ref{Fig19} displays the concentrations of species $A$ and $B$ after the chemical reaction reaches equilibrium. It is clear that the concentration of species $A$ ($B$) reduces (increases) linearly with the increasing forward reaction coefficient. There is a satisfying agreement between the simulation results and exact solutions.

\subsection{Detonation wave}

Detonation is a particular type of combustion with violent chemical heat release around a supersonic exothermic front accelerating through a medium. The physical fields have strong temporal and spatial changes near the detonation wave, which poses a great challenge to the numerical robustness and physical accuracy of computational fluid dynamics. In this subsection, we demonstrate that the DBM has the capability of capturing the detonation wave travelling at a supersonic speed. As the detonation wave passes in the $x$ direction, the chemical reactant $A$ changes into the product $B$, i.e., $A\to B$, and the chemical energy is released. 
In theory, the speed of the steady detonation front is a function of the chemical heat release of reactant per unit mass, $q$, i.e.,
\begin{equation}
	D=\sqrt{\frac{(\gamma^2-1)q}{2}+\gamma{T_0}}+\sqrt{\frac{(\gamma^2-1)q}{2}} \text{,}
\end{equation}
where ${T_0}$ denotes the temperature in front of the detonation wave and $\gamma$ represents the specific heat ratio. Then, the Mach number ${\rm{Ma}}$ is calculated by
\begin{equation}
	{\rm{Ma}} = \frac{D}{\sqrt{\gamma {T_0}}} \text{.}
\end{equation}
Here we focus on a specific case where $q=1$, yielding a corresponding Mach number of ${\rm{Ma}}=1.74436$. It is worth noting that the DBM excels in simulating detonations under a high Mach number, employing a robust numerical scheme. 

The reaction rate is controlled by
\begin{equation}
	\omega_{\rm{ov}}={{k}_{\rm{ov}}}{{n}^{A}}\exp \left( -\frac{{E}_{a}}{RT} \right)
	\tt{,}
\end{equation}
in terms of $k_{\rm{ov}}=5 \times {{10}^{5}}$ and ${{E}_{a}}=10$. 
The initial configuration is set as
\begin{equation}
	\left\{
	\begin{array}{l}
		{{\left( {{n}^{A}}, {{n}^{B}}, {{u}_{x}}, T \right)}_{L}}=\left( 0, 1.38837, 0.57735, 1.57856 \right) \tt{,} \\
		{{\left( {{n}^{A}}, {{n}^{B}}, {{u}_{x}}, T \right)}_{R}}=\left( 1, 0, 0, 1 \right) \tt{,}
	\end{array}
	\right.
\end{equation}
where the subscript $L$ indicates $0\le x\le 0.04$, and $R$ indicates $0.04<x\le 0.4$. The quantities in the left and right parts satisfy the Hugoniot relationship for detonation wave.
The parameters are $m^{\sigma} = 1$, $\gamma^{\sigma} =1.5$, ${{S}^{1 \sigma }_{i}} = {{S}^{2 \sigma }_{i}} = 2 \times 10^{4}$, and (${{v}_{a}^{\sigma}}$, ${{v}_{b}^{\sigma}}$, ${{v}_{c}^{\sigma}}$, ${{v}_{d}^{\sigma}}$, ${\eta }_{a }^{\sigma}$, ${\eta }_{b}^{\sigma}$, ${\eta }_{c}^{\sigma}$, ${\eta }_{d}^{\sigma}$) $=$ ($0.5$, $1.5$, $2.2$, $3.5$, $0$, $5.2$, $3$, $0$).
In addition, inflow and outflow boundary conditions are adopted in the $x$ direction, and the periodic boundary conditions in the $y$ direction.

\begin{figure}[tbp]
	\begin{center}
		\includegraphics[width=0.5\textwidth]{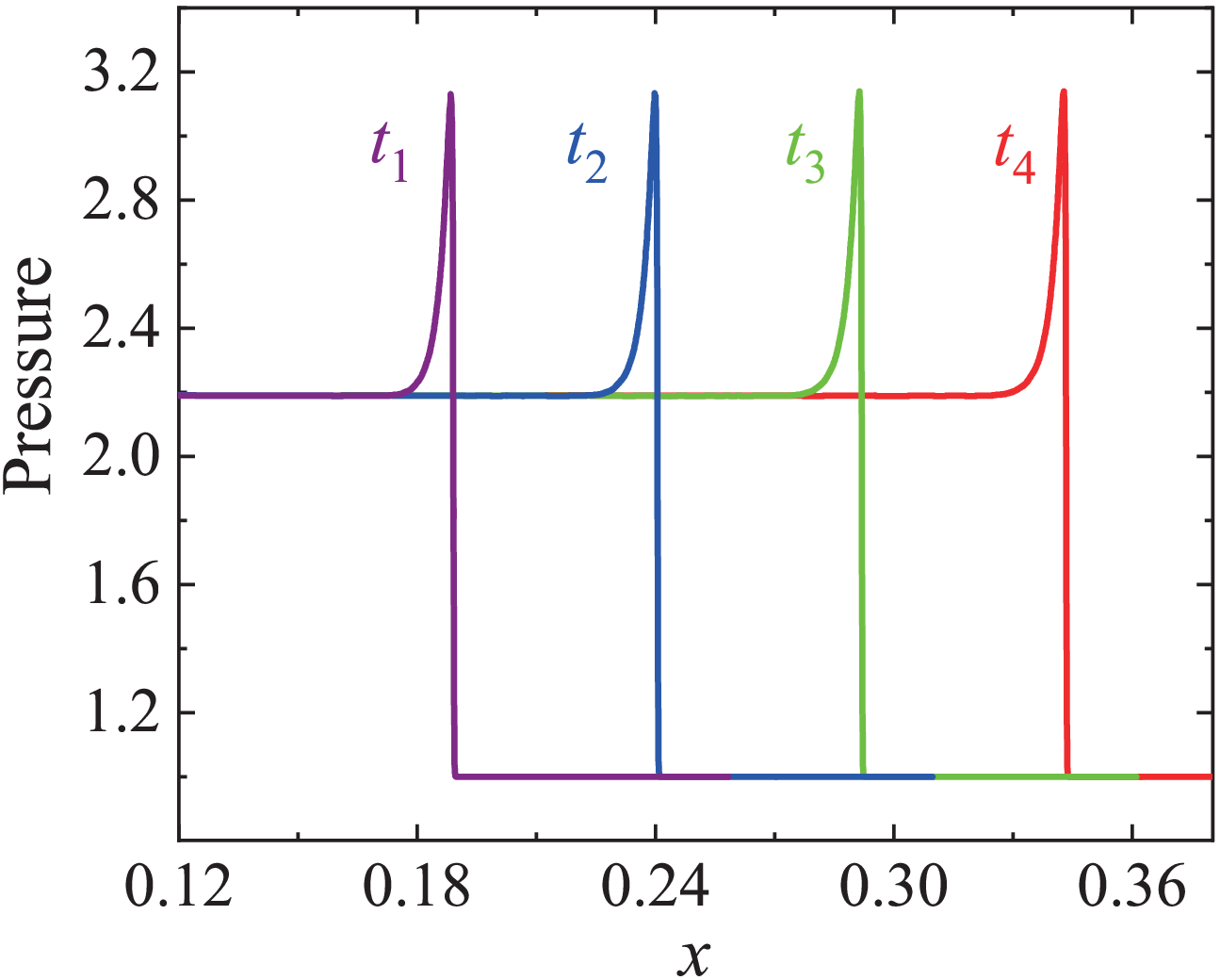}
	\end{center}
	\caption{Pressure profiles at time instants ${t_1} = 0.075$, ${t_2} = 0.1$,  ${t_3} = 0.125$, and ${t_4} = 0.15$ in the evolution of detonation.}
	\label{Fig20}
\end{figure}
Figure \ref{Fig20} depicts the evolution of pressure profiles around the detonation wave that propagates forwards. It is clear that the spatial distribution of the pressure field is quite similar to each other at the four time instants ${t} = 0.075$, $0.1$, $0.125$, and $0.15$. It indicates that the detonation wave moves forwards in a steady state. Then the speed of the steady detonation front can be calculated, $v_d = 2.058$. Compared with the analytic solution $2.06395$, the relative error is about $0.0029$. The simulation results are satisfactory.

\begin{figure}[tbp]
	\begin{center}
		\includegraphics[bbllx=0pt,bblly=0pt,bburx=528pt,bbury=363pt,width=0.9\textwidth]{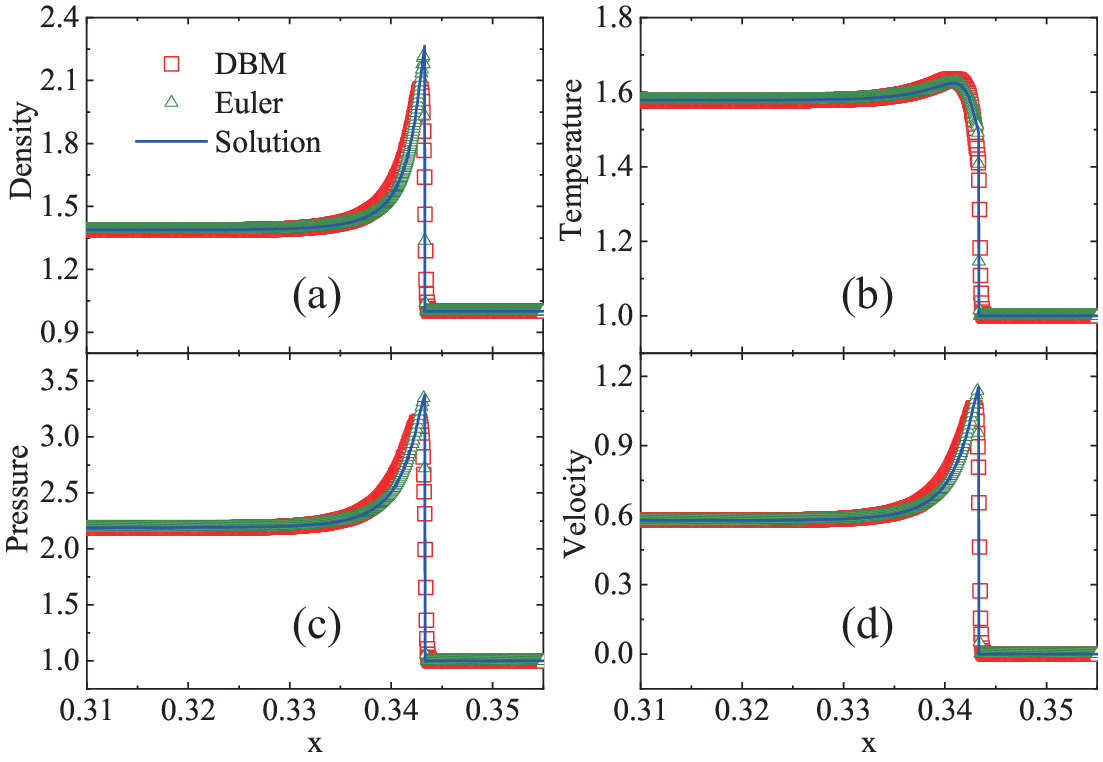}
	\end{center}
	\caption{Physical quantities around the detonation wave. The squares, triangles, and lines represent the results obatined from the DBM, Euler solver, and ZND theory \cite{Law2006}, respectively.}
	\label{Fig21}
\end{figure}
Figure \ref{Fig21} gives the density (a), temperature (b), pressure (c), and horizontal velocity (d) at $t = 0.15$ in the detonation process. The squares represent the DBM results, the triangles denote the numerical outcomes obtained from an Euler solver, and the lines stand for the analytical solutions of the Zel'dovich-Neumann-D\"oring (ZND) results theory \cite{Law2006}. Obviously, there is a satisfying agreement among the three models in regions distant from the detonation wave. To be specific, the DBM results are ($\rho$, $T$, $p$, ${{u}_{x}}$) $=$ ($1.38748$, $1.57724$, $2.18838$, $0.57892$) after the detonation wave, resulting in relative errors of ($0.0006$, $0.0008$, $0.0015$, $0.0027$) compared to the ZND results \cite{Law2006}. 
However, slight differences between them emerge at the von-Neumann-peak. This disparity arises from the fact that both the ZND theory and Euler solver neglect the viscosity and heat conduction, assuming a sharp discontinuity at the von-Neumann-peak. 
In contrast, the DBM takes into account the viscosity, heat conduction, and other thermodynamic nonequilibrium effects. As a result, the physical fields simulated by the DBM exhibit smoothness across the detonation wave, aligning more closely with real-world conditions. 

It should be mentioned that previous research on detonation mechanisms has primarily relied on traditional computational fluid dynamics methods, which differ substantially from the current DBM. Physically, this DBM can be likened to a modified continuous fluid model augmented with a coarse-grained representation of significant thermodynamic nonequilibrium effects. Consequently, the DBM exhibits the capability to accurately capture detonation phenomena with nonequilibrium effects, encompassing diffusion, viscosity, and thermal conduction. In fact, the DBM is suitable for both steady and unsteady detonation scenarios, although the latter is not explicitly demonstrated in this manuscript due to space constraints. 

\section{Conclusion}\label{SecIV}

An MRT DBM with split collision is presented for both subsonic and supersonic reactive flows. The external forces, chemical reactions, and multi-physical fields are coupled naturally through the collision, force and reaction terms on the right-hand side of the  discrete Boltzmann equations that describe the evolution of reactive mixture. Through the CE expansion, it can be proved that the DBM is consistent with the reactive NS equations with external forces, the Fick's law and Stefan-Maxwell diffusion equation in the hydrodynamic limit. Each chemical species owns individual adjustable molar mass, concentration, velocity, acceleration, temperature, pressure, diffusivity, dynamic viscosity, thermal conductivity, specific heat ratio, Prandtl number, Reynolds number, and Schmidt number, etc. 

Compared to the one-step-relaxation MRT or BGK model \cite{Lin2017SR,Lin2019PRE}, the DBM with the splitting technique, a two-step-relaxation model, offers greater flexibility in parameters and is applicable to a broader range of physicochemical systems. 
(i) The relaxation frequencies ${{\mathbf{S}}^{1 \sigma }}$ and ${{\mathbf{S}}^{2 \sigma }}$ govern the thermodynamic nonequilibrium process, guiding the approach toward temporary individual equilibrium and ultimate mixing equilibrium. 
(ii) The relaxation frequencies in the split collision term influence both self- and cross-collisions, thereby impacting the evolution of the mixture.
(iii) Specific relationships can be established between the relaxation frequencies and other physical quantities, such as thermodynamic nonequilibrium quantities, diffusivity, dynamic viscosity, and thermal conductivity.
(iv) Classical dimensionless numbers in fluid mechanics, such as the Reynolds number, Prandtl number, and Schmidt number of each species, can be adjusted. 
Consequently, the DBM with the split collision term presents a more detailed relationship between the thermodynamic relaxation process and nonequilibrium effects. Moreover, the two-step-relaxation DBM can reduce to the one-step-relaxation or BGK model under special conditions. 

It should be stressed that the hydrodynamic, thermodynamic, and chemical nonequilibrium effects can be captured and measured by the versatile kinetic DBM dynamically. Physically, the DBM is more general than traditional NS solvers since it contains more detailed thermodynamic nonequilibrium information. Mathematically, a set of uniform discrete Boltzmann equations is used to describe the reactive mixtures, and the algorithm is easy to code due to the linearization of evolution equations. Computationally, it can be implemented on very large parallel clusters with exceptional scalability because all information transfer in DBM is local in time and space, which is similar to other LBMs.

In addition, three methods to calculate the source terms (including the force and reaction terms) are introduced into the multicomponent DBM.
As a traditional idea, Method I uses the discretization form of the formula of source terms in the velocity space directly.
Method II expresses the source terms as the change of discrete distribution functions due to the source influences over a small time interval.
Method III gives the expression of source terms by using the matrix inversion method.
Methods I and III own higher accuracy than Method II which possesses only the first-order accuracy. Besides, Method III has the highest computational efficiency because it requires to calculate the equilibrium distribution functions zero, one and two times per loop for Methods III, I and II, respectively.

Finally, several canonical systems, including the multicomponent diffusion, mixture in the force field, KH instability, laminar flame of propane-air mixture, opposing reactions, and detonation wave are simulated to validate this model. The first three benchmarks are physical systems without chemical reaction, and the last three benchmarks have chemical reactions, with the last one containing rather violent chemical heat release. It is demonstrated that the current DBM is suitable for multicomponent mixtures with or without the chemical reaction. The interplay among different chemical species can be described accurately. The complex interfacial structures could be captured dynamically. The essential nonequilibrium and compressible effects can be quantified. In the near future, this DBM will be employed to investigate more practical combustion problems with significant nonequilibrium and compressible effects.

\appendix

\section{}\label{APPENDIXA}

The square matrix ${{\mathbf{M}}^{\sigma }}$ takes the form,
\begin{equation}
	{{\mathbf{M}}^{\sigma }}=
	\left(
	\begin{array}{cccc}
		{M_{11}^{\sigma }} & M_{12}^{\sigma } & \cdots  & M_{1N}^{\sigma }  \\
		{M_{21}^{\sigma }} & M_{22}^{\sigma } & \cdots  & M_{2N}^{\sigma }  \\
		\vdots  & \vdots  & \ddots  & \vdots   \\
		{M_{N1}^{\sigma }} & M_{N2}^{\sigma } & \cdots  & M_{NN}^{\sigma }
	\end{array}
	\right)
	\tt{,}
\end{equation}
with ${{M}^{\sigma}_{1i}}=1$, $M_{2i}^{\sigma }=v_{ix}^{\sigma }$, $M_{3i}^{\sigma }=v_{iy}^{\sigma }$, $M_{4i}^{\sigma }=v_{i}^{\sigma 2}+\eta _{i}^{\sigma 2}$, $M_{5i}^{\sigma }=v_{ix}^{\sigma 2}$,
$M_{6i}^{\sigma }=v_{ix}^{\sigma }v_{iy}^{\sigma }$, ${{M}^{\sigma}_{7i}}=v_{iy}^{2}$, $M_{8i}^{\sigma }=\left( v_{i}^{\sigma 2}+\eta _{i}^{\sigma 2} \right)v_{ix}^{\sigma }$, $M_{9i}^{\sigma }=\left( v_{i}^{\sigma 2}+\eta _{i}^{\sigma 2} \right)v_{iy}^{\sigma }$,
$M_{10i}^{\sigma }=v_{ix}^{\sigma 3}$, $M_{11i}^{\sigma }=v_{ix}^{\sigma 2}v_{iy}^{\sigma }$, $M_{12i}^{\sigma }=v_{ix}^{\sigma }v_{iy}^{\sigma 2}$, $M_{13i}^{\sigma }=v_{iy}^{\sigma 3}$,
$M_{14i}^{\sigma }=\left( v_{i}^{\sigma 2}+\eta _{i}^{\sigma 2} \right)v_{ix}^{\sigma 2}$, $M_{15i}^{\sigma }=\left( v_{i}^{\sigma 2}+\eta _{i}^{\sigma 2} \right)v_{ix}^{\sigma }v_{iy}^{\sigma }$, and $M_{16i}^{\sigma }=\left( v_{i}^{\sigma 2}+\eta _{i}^{\sigma 2} \right)v_{iy}^{\sigma 2}$. Its inverse $({{\mathbf{M}}^{\sigma }})^{-1}$ could be obtained by using a calculation software, such as Matlab or Mathematica.

The column matrix ${{\mathbf{\hat{f}}}^{\sigma eq}}$ is given by
\begin{equation}
	{{\mathbf{\hat{f}}}^{\sigma eq}}={{\left( \begin{matrix} {\hat{f}_{1}^{\sigma eq}} \ \hat{f}_{2}^{\sigma eq} \ \cdots  \ \hat{f}_{N}^{\sigma eq} \end{matrix} \right)}^{\rm{T}}}
	\label{Expression_hatfeq}
	\tt{,}
\end{equation}
in terms of $\hat{f}_{1}^{\sigma eq}={{n}^{\sigma }}$, $\hat{f}_{2}^{\sigma eq}={{n}^{\sigma }}{{u}_{x}}$, $\hat{f}_{3}^{\sigma eq}={{n}^{\sigma }}{{u}_{y}}$, $\hat{f}_{4}^{\sigma eq}={{n}^{\sigma }}[ ( D+{{I}^{\sigma }} ){T}/{{{m}^{\sigma }}}+{{u}^{2}} ]$,
$\hat{f}_{5}^{\sigma eq}={{n}^{\sigma }}( {T}/{{{m}^{\sigma }}}+u_{x}^{2} )$, $\hat{f}_{6}^{\sigma eq}={{n}^{\sigma }}{{u}_{x}}{{u}_{y}}$, $\hat{f}_{7}^{\sigma eq}={{n}^{\sigma }}( {T}/{{{m}^{\sigma }}}+u_{y}^{2} )$,
$\hat{f}_{8}^{\sigma eq}={{n}^{\sigma }}{{\xi }^{\sigma }}{{u}_{x}}$, $\hat{f}_{9}^{\sigma eq}={{n}^{\sigma }}{{\xi }^{\sigma }}{{u}_{y}}$,
$\hat{f}_{10}^{\sigma eq}=3{{n}^{\sigma }}{{u}_{x}}{T}/{{{m}^{\sigma }}}+{{n}^{\sigma }}u_{x}^{3}$, $\hat{f}_{11}^{\sigma eq}={{n}^{\sigma }}{{u}_{y}}{T}/{{{m}^{\sigma }}}+{{n}^{\sigma }}u_{x}^{2}{{u}_{y}}$,
$\hat{f}_{12}^{\sigma eq}={{n}^{\sigma }}{{u}_{x}}{T}/{{{m}^{\sigma }}}+{{n}^{\sigma }}{{u}_{x}}u_{y}^{2}$, $\hat{f}_{13}^{\sigma eq}=3{{n}^{\sigma }}{{u}_{y}}{T}/{{{m}^{\sigma }}}+{{n}^{\sigma }}u_{y}^{3}$,
$\hat{f}_{14}^{\sigma eq}={{n}^{\sigma }}{{\xi }^{\sigma }}{T}/{{{m}^{\sigma }}}+{{n}^{\sigma }}u_{x}^{2}( {{\xi }^{\sigma }}+2{T}/{{{m}^{\sigma }}} )$, $\hat{f}_{15}^{\sigma eq}={{n}^{\sigma }}{{u}_{x}}{{u}_{y}}( {{\xi }^{\sigma }}+2{T}/{{{m}^{\sigma }}} )$, and
$\hat{f}_{16}^{\sigma eq}={{n}^{\sigma }}{{\xi }^{\sigma }}{T}/{{{m}^{\sigma }}} + {{n}^{\sigma }}u_{y}^{2}( {{\xi }^{\sigma }}+2{T}/{{{m}^{\sigma }}} )$, with ${{\xi }^{\sigma }}=( D+{{I}^{\sigma }}+2 ){T}/{{{m}^{\sigma }}}+{{u}^{2}}$.

The column matrix ${{\mathbf{\hat{f}}}^{\sigma seq}}$ is expressed by
\begin{equation}
	{{\mathbf{\hat{f}}}^{\sigma seq}}={{\left( \begin{matrix} {\hat{f}_{1}^{\sigma seq}} \ \hat{f}_{2}^{\sigma seq} \ \cdots  \ \hat{f}_{N}^{\sigma seq} \end{matrix} \right)}^{\rm{T}}}
	\label{Expression_hatfseq}
	\tt{,}
\end{equation}
whose elements are similar to those in Eq. (\ref{Expression_hatfeq}). The elements $\hat{f}_{i}^{\sigma seq}$ are functions of (${n}^{\sigma }$, ${u}^{\sigma }_{\alpha}$, $T^{\sigma }$), and $\hat{f}_{i}^{\sigma eq}$ are functions of (${n}^{\sigma }$, ${u}_{\alpha}$, $T$). The former are given by substituting (${n}^{\sigma }$, ${u}^{\sigma }_{\alpha}$, $T^{\sigma }$) for (${n}^{\sigma }$, ${u}_{\alpha}$, $T$) in the latter, respectively.

Moreover, the column matrix ${{\mathbf{\hat{F}}}^{\sigma }}$ takes the form
\begin{equation}
	{{\mathbf{\hat{F}}}^{\sigma }}={{\left( \begin{matrix} {\hat{F}_{1}^{\sigma }} \ \hat{F}_{2}^{\sigma } \ \cdots  \ \hat{F}_{N}^{\sigma } \end{matrix} \right)}^{\rm{T}}}
	\tt{,}
\end{equation}
where $\hat{F}_{1}^{\sigma }=0$,
$\hat{F}_{2}^{\sigma }={{n}^{\sigma }}{{a}_{x}^{\sigma }}$,
$\hat{F}_{3}^{\sigma }={{n}^{\sigma }}{{a}_{y}^{\sigma }}$,
$\hat{F}_{4}^{\sigma }=2{{n}^{\sigma }}\left( u_{x}^{\sigma }{{a}_{x}^{\sigma }}+u_{y}^{\sigma }{{a}_{y}^{\sigma }} \right)$,
$\hat{F}_{5}^{\sigma }=2{{n}^{\sigma }}u_{x}^{\sigma }{{a}_{x}^{\sigma }}$,
$\hat{F}_{6}^{\sigma }={{n}^{\sigma }}\left( u_{x}^{\sigma }{{a}_{y}^{\sigma }}+u_{y}^{\sigma }{{a}_{x}^{\sigma }} \right)$,
$\hat{F}_{7}^{\sigma }=2{{n}^{\sigma }}u_{y}^{\sigma }{{a}_{y}^{\sigma }}$,
$\hat{F}_{8}^{\sigma }=2{{n}^{\sigma }}u_{x}^{\sigma }\left( u_{x}^{\sigma }{{a}_{x}^{\sigma}}+u_{y}^{\sigma }{{a}_{y}^{\sigma}} \right)+{{n}^{\sigma }}{{a}_{x}^{\sigma}}\left[ {{u}^{\sigma 2}}+\left( D+{{I}^{\sigma }}+2 \right){{{T}^{\sigma }}}/{{{m}^{\sigma }}}\; \right]$,
$\hat{F}_{9}^{\sigma }=2{{n}^{\sigma }}u_{y}^{\sigma }\left( u_{x}^{\sigma }{{a}_{x}^{\sigma}}+u_{y}^{\sigma }{{a}_{y}^{\sigma}} \right) + {{n}^{\sigma }}{{a}_{y}^{\sigma}} [ {{u}^{\sigma 2}}+\left( D+{{I}^{\sigma }}+2 \right){{{T}^{\sigma }}}/{{{m}^{\sigma }}} ]$,
$\hat{F}_{10}^{\sigma }=3{{n}^{\sigma }}{{a}_{x}^{\sigma }}\left( u_{x}^{\sigma 2}+{{{T}^{\sigma }}}/{{{m}^{\sigma }}}\; \right)$,
$\hat{F}_{11}^{\sigma }=2{{n}^{\sigma }}{{a}_{x}^{\sigma }}u_{x}^{\sigma }u_{y}^{\sigma } + {{n}^{\sigma }}{{a}_{y}^{\sigma }} (u_{x}^{\sigma 2} + {{{T}^{\sigma }}}/{{m}^{\sigma }} )$,
$\hat{F}_{12}^{\sigma }={{n}^{\sigma }}{{a}_{x}^{\sigma }}\left( u_{y}^{\sigma 2}+{{{T}^{\sigma }}}/{{{m}^{\sigma }}}\; \right)+2{{n}^{\sigma }}{{a}_{y}^{\sigma }}u_{x}^{\sigma }u_{y}^{\sigma }$,
$\hat{F}_{13}^{\sigma }=3{{n}^{\sigma }}{{a}_{y}^{\sigma }}\left( u_{y}^{\sigma 2}+{{{T}^{\sigma }}}/{{{m}^{\sigma }}}\; \right)$,
$\hat{F}_{14}^{\sigma }=2{{n}^{\sigma }}{{a}_{x}^{\sigma }}u_{x}^{\sigma } \left[ 2u_{x}^{\sigma 2}+u_{y}^{\sigma 2}+\left( D+{{I}^{\sigma }}+5 \right){{{T}^{\sigma }}}/{{{m}^{\sigma }}}\; \right]+2{{n}^{\sigma }}{{a}_{y}^{\sigma }}u_{y}^{\sigma }\left( u_{x}^{\sigma 2}+{{{T}^{\sigma }}}/{{{m}^{\sigma }}}\; \right)$,
${{\hat{F}}_{15}}={{n}^{\sigma }}{{a}_{x}^{\sigma }}u_{y}^{\sigma }[ 3u_{x}^{\sigma 2}+u_{y}^{\sigma 2} + \left( D+{{I}^{\sigma }}+4 \right){{T}^{\sigma }}/{{m}^{\sigma }} ]+{{n}^{\sigma }}{{a}_{y}^{\sigma }}u_{x}^{\sigma }[ u_{x}^{\sigma 2}+3u_{y}^{\sigma 2}+( D+{{I}^{\sigma }}+4){{T}^{\sigma }}/{{m}^{\sigma }} ]$, and
$\hat{F}_{16}^{\sigma }=2{{n}^{\sigma }}{{a}_{x}^{\sigma }}u_{x}^{\sigma }\left( u_{y}^{\sigma 2}+{{{T}^{\sigma }}}/{{{m}^{\sigma }}} \right)+2{{n}^{\sigma }}{{a}_{y}^{\sigma }}u_{y}^{\sigma }[ u_{x}^{\sigma 2}+2u_{y}^{\sigma 2}+( D+{{I}^{\sigma }}+5){{{T}^{\sigma }}}/{{{m}^{\sigma }}}]$.

Additionally, the column matrix ${{\mathbf{\hat{R}}}^{\sigma }}$ is calculated by
\begin{equation}
	{{\mathbf{\hat{R}}}^{\sigma }}={{\left( \begin{matrix} {\hat{R}_{1}^{\sigma }} \ \hat{R}_{2}^{\sigma } \ \cdots  \ \hat{R}_{N}^{\sigma } \end{matrix} \right)}^{\rm{T}}}
	\tt{,}
\end{equation}
with
$\hat{R}_{1}^{\sigma }={{n}^{\sigma }}^{\prime }$,
$\hat{R}_{2}^{\sigma }={{n}^{\sigma }}^{\prime }{{u}_{x}}$,
$\hat{R}_{3}^{\sigma }={{n}^{\sigma }}^{\prime }{{u}_{y}}$,
$\hat{R}_{4}^{\sigma }={{n}^{\sigma }}^{\prime }[ ( D+{{I}^{\sigma }} ){T}/{{{m}^{\sigma }}}+{{u}^{2}} ]+( D+{{I}^{\sigma }} ){{n}^{\sigma }}{{{T}'}}/{{{m}^{\sigma }}}$,
$\hat{R}_{5}^{\sigma }={{n}^{\sigma }}^{\prime }( {T}/{{{m}^{\sigma }}}+u_{x}^{2} )+{{n}^{\sigma }}{{{T}'}}/{{{m}^{\sigma }}}$,
$\hat{R}_{6}^{\sigma }={{n}^{\sigma }}^{\prime }{{u}_{x}}{{u}_{y}}$,
$\hat{R}_{7}^{\sigma }={{n}^{\sigma }}^{\prime }( {T}/{{{m}^{\sigma }}}+u_{y}^{2} )+{{n}^{\sigma }}{{{T}'}}/{{{m}^{\sigma }}}$,
$\hat{R}_{8}^{\sigma }={{n}^{\sigma }}^{\prime }{{u}_{x}}{{\xi }^{\sigma }}+( D+{{I}^{\sigma }}+2 ){{n}^{\sigma }}{{u}_{x}}{{{T}'}}/{{{m}^{\sigma }}}$,
$\hat{R}_{9}^{\sigma }={{n}^{\sigma }}^{\prime }{{u}_{y}}{{\xi }^{\sigma }}+( D+{{I}^{\sigma }}+2 ){{n}^{\sigma }}{{u}_{y}}{{{T}'}}/{{{m}^{\sigma }}}$,
$\hat{R}_{10}^{\sigma }=3{{n}^{\sigma }}^{\prime }{{u}_{x}}{T}/{{{m}^{\sigma }}}+{{n}^{\sigma }}^{\prime }u_{x}^{3}+3{{n}^{\sigma }}{{u}_{x}}{{{T}'}}/{{{m}^{\sigma }}}$,
$\hat{R}_{11}^{\sigma }={{n}^{\sigma }}^{\prime }{{u}_{y}}{T}/{{{m}^{\sigma }}}+{{n}^{\sigma }}^{\prime }u_{x}^{2}{{u}_{y}}+{{n}^{\sigma }}{{u}_{y}}{{{T}'}}/{{{m}^{\sigma }}}$,
$\hat{R}_{12}^{\sigma }={{n}^{\sigma }}^{\prime }{{u}_{x}}{T}/{{{m}^{\sigma }}}+{{n}^{\sigma }}^{\prime }{{u}_{x}}u_{y}^{2}+{{n}^{\sigma }}{{u}_{x}}{{{T}'}}/{{{m}^{\sigma }}}$,
$\hat{R}_{13}^{\sigma }=3{{n}^{\sigma }}^{\prime }{{u}_{y}}{T}/{{{m}^{\sigma }}}+{{n}^{\sigma }}^{\prime }u_{y}^{3}+3{{n}^{\sigma }}{{u}_{y}}{{{T}'}}/{{{m}^{\sigma }}}$,
$\hat{R}_{14}^{\sigma }={{n}^{\sigma }}^{\prime }{{\xi }^{\sigma }}{T}/{{{m}^{\sigma }}}+{{n}^{\sigma }}^{\prime }u_{x}^{2}( {{\xi }^{\sigma }}+2{T}/{{{m}^{\sigma }}} )+{{n}^{\sigma }}[ 2( D+{{I}^{\sigma }}+2 ){T}/{{{m}^{\sigma }}}+{{u}^{2}}+( D+{{I}^{\sigma }}+4 )u_{x}^{2} ]{{{T}'}}/{{{m}^{\sigma }}}$,
$\hat{R}_{15}^{\sigma }={{n}^{\sigma }}^{\prime }{{u}_{x}}{{u}_{y}}( {{\xi }^{\sigma }}+2{T}/{{{m}^{\sigma }}} )+{{n}^{\sigma }}[ ( D+{{I}^{\sigma }}+4 ){{u}_{x}}{{u}_{y}} ]{{{T}'}}/{{{m}^{\sigma }}}$, and
$\hat{R}_{16}^{\sigma }={{n}^{\sigma }}^{\prime }{{\xi }^{\sigma }}{T}/{{{m}^{\sigma }}}+{{n}^{\sigma }}^{\prime }u_{y}^{2}( {{\xi }^{\sigma }}+2{T}/{{{m}^{\sigma }}} )+{{n}^{\sigma }}[ 2( D+{{I}^{\sigma }}+2 ){T}/{{{m}^{\sigma }}}+{{u}^{2}}+( D+{{I}^{\sigma }}+4 )u_{y}^{2} ]{{{T}'}}/{{{m}^{\sigma }}}$.

\section{}\label{APPENDIXB}

Via the CE expansion, it can be found that the current DBM is consistent with the reactive NS equations in the hydrodynamic limit. Based on the Einstein summation convention, the NS equations of individual species read,
\begin{equation}
	\dfrac{\partial {{\rho }^{\sigma }}}{\partial t}+\dfrac{\partial }{\partial \alpha }\left( {{\rho }^{\sigma }}u_{\alpha }^{\sigma } \right)={{\rho }^{\sigma }}^{\prime }
	\label{NSequation1}
	\tt{,}
\end{equation}
\begin{eqnarray}
	& \dfrac{\partial }{\partial t}\left( {{\rho }^{\sigma }}u_{\alpha }^{\sigma } \right)+\dfrac{\partial }{\partial \beta }\left( {{\delta }_{\alpha \beta }}{{p}^{\sigma }}+{{\rho }^{\sigma }}u_{\alpha }^{\sigma }u_{\beta }^{\sigma }+P_{\alpha \beta }^{\sigma }+U_{\alpha \beta }^{\sigma }+V_{\alpha \beta }^{\sigma } \right) \nonumber \\
	& =S_{J\alpha }^{\sigma }{{\rho }^{\sigma }}\left( {{u}_{\alpha }}-u_{\alpha }^{\sigma } \right)+{{\rho }^{\sigma }}a_{\alpha }^{\sigma }+{{\rho }^{\sigma \prime }}{{u}_{\alpha }}
	\label{NSequation2}
	\tt{,}
\end{eqnarray}
\begin{eqnarray}
	& \dfrac{\partial {{E}^{\sigma }}}{\partial t}+\dfrac{\partial }{\partial \alpha }\left( {{E}^{\sigma }}u_{\alpha }^{\sigma }+{{p}^{\sigma }}u_{\alpha }^{\sigma }-\kappa _{\alpha }^{\sigma }\dfrac{\partial {{T}^{\sigma }}}{\partial \alpha }+u_{\beta }^{\sigma }P_{\alpha \beta }^{\sigma }+X_{\alpha }^{\sigma }+Y_{\alpha }^{\sigma } \right)  \nonumber \\
	& =\dfrac{1}{2}S_{4}^{2 \sigma }{{\rho }^{\sigma }}\left[ \left( D+{{I}^{\sigma }} \right)\dfrac{T-{{T}^{\sigma }}}{{{m}^{\sigma }}}+{{u}^{2}}-{{u}^{\sigma 2}} \right]
	+{{\rho }^{\sigma }}u_{\alpha }^{\sigma }a_{\alpha }^{\sigma }+{{E}^{\sigma }}^{\prime }
	\label{NSequation3}
	\tt{,}
\end{eqnarray}
in terms of
\begin{equation}
	P_{\alpha \beta }^{\sigma }=\mu _{\alpha \beta }^{\sigma }\left( \dfrac{2{{\delta }_{\alpha \beta }}}{D+{{I}^{\sigma }}}\dfrac{\partial u_{\chi }^{\sigma }}{\partial \chi }-\dfrac{\partial u_{\alpha }^{\sigma }}{\partial \beta }-\dfrac{\partial u_{\beta }^{\sigma }}{\partial \alpha } \right)
	\label{P_ab}
	\tt{,}
\end{equation}
\begin{eqnarray}
	& U_{\alpha \beta }^{\sigma }=\dfrac{S_{4}^{2\sigma }}{S_{\alpha \beta }^{1\sigma }}\dfrac{{{\delta }_{\alpha \beta }}{{\rho }^{\sigma }}}{D+{{I}^{\sigma }}}\left( {{u}^{\sigma 2}}-{{u}^{2}} \right)-\dfrac{S_{J\chi }^{\sigma }}{S_{\alpha \beta }^{1\sigma }}\dfrac{2{{\delta }_{\alpha \beta }}{{\rho }^{\sigma }}}{D+{{I}^{\sigma }}}u_{\chi }^{\sigma }\left( u_{\chi }^{\sigma }-{{u}_{\chi }} \right) \nonumber \\
	& -\dfrac{S_{\alpha \beta }^{2\sigma }}{S_{\alpha \beta }^{1\sigma }}{{\rho }^{\sigma }}\left( u_{\alpha }^{\sigma }u_{\beta }^{\sigma }-{{u}_{\alpha }}{{u}_{\beta }} \right)+\dfrac{S_{J\alpha }^{\sigma }}{S_{\alpha \beta }^{1\sigma }}{{\rho }^{\sigma }}\left( u_{\alpha }^{\sigma }-{{u}_{\alpha }} \right)u_{\beta }^{\sigma } \nonumber \\
	& +\dfrac{S_{J\beta }^{\sigma }}{S_{\alpha \beta }^{1\sigma }}{{\rho }^{\sigma }}u_{\alpha }^{\sigma }\left( u_{\beta }^{\sigma }-{{u}_{\beta }} \right)+{{\delta }_{\alpha \beta }}\dfrac{S_{4}^{2\sigma }-S_{\alpha \beta }^{2\sigma }}{S_{\alpha \beta }^{1\sigma }}{{\rho }^{\sigma }}\dfrac{{{T}^{\sigma }}-T}{{{m}^{\sigma }}}
	\tt{,}
\end{eqnarray}
\begin{equation}
	V_{\alpha \beta }^{\sigma }=\dfrac{{{\rho }^{\sigma \prime }}}{S_{\alpha \beta }^{1 \sigma }}\left( {{u}_{\alpha }}{{u}_{\beta }}+u_{\alpha }^{\sigma }u_{\beta }^{\sigma }-{{u}_{\alpha }}u_{\beta }^{\sigma }-u_{\alpha }^{\sigma }{{u}_{\beta }}-{{\delta }_{\alpha \beta }}\dfrac{{{u}^{\sigma 2}}+{{u}^{2}}-2u_{\chi }^{\sigma }{{u}_{\chi }}}{D+{{I}^{\sigma }}} \right)
	\tt{,}
\end{equation}
\begin{eqnarray}
	& X_{\alpha }^{\sigma }=\dfrac{S_{4}^{2\sigma }}{S_{\kappa \alpha }^{1\sigma }}\dfrac{{{\rho }^{\sigma }}u_{\alpha }^{\sigma }}{D+{{I}^{\sigma }}}\left( {{u}^{\sigma 2}}-{{u}^{2}} \right)-\dfrac{2S_{J\beta }^{\sigma }}{S_{\kappa \alpha }^{1\sigma }}\dfrac{{{\rho }^{\sigma }}u_{\alpha }^{\sigma }}{D+{{I}^{\sigma }}}u_{\beta }^{\sigma }\left( u_{\beta }^{\sigma }-{{u}_{\beta }} \right) \nonumber \\
	& +\dfrac{S_{J\alpha }^{\sigma }-S_{J\beta }^{\sigma }}{S_{\kappa \alpha }^{1\sigma }}{{\rho }^{\sigma }}u_{\alpha }^{\sigma }u_{\beta }^{\sigma }\left( u_{\beta }^{\sigma }-{{u}_{\beta }} \right) \nonumber \\
	& +\dfrac{S_{4}^{2\sigma }u_{\alpha }^{\sigma }-S_{\kappa \alpha }^{2\sigma }{{u}_{\alpha }}}{2S_{\kappa \alpha }^{1\sigma }}{{\rho }^{\sigma }}\left[ \left( D+{{I}^{\sigma }}+2 \right)\dfrac{{{T}^{\sigma }}-T}{{{m}^{\sigma }}}+{{u}^{\sigma 2}}-{{u}^{2}} \right] \nonumber \\
	& +\dfrac{S_{J\alpha }^{\sigma }-S_{\kappa \alpha }^{2\sigma }}{2S_{\kappa \alpha }^{1\sigma }}{{\rho }^{\sigma }}\left( u_{\alpha }^{\sigma }-{{u}_{\alpha }} \right)\left[ \left( D+{{I}^{\sigma }}+2 \right)\dfrac{{{T}^{\sigma }}}{{{m}^{\sigma }}}+{{u}^{\sigma 2}} \right]
	\tt{,}
\end{eqnarray}
\begin{eqnarray}
	& Y_{\alpha }^{\sigma }=\dfrac{{{\rho }^{\sigma }}^{\prime }}{2S_{\kappa \alpha }^{1\sigma }}\left( u_{\alpha }^{\sigma }-{{u}_{\alpha }} \right)\left[ \left( D+{{I}^{\sigma }}+2 \right)\dfrac{{{T}^{\sigma }}-T}{{{m}^{\sigma }}}+{{u}^{\sigma 2}}-{{u}^{2}} \right] \nonumber \\
	& -\dfrac{1}{S_{\kappa \alpha }^{1\sigma }}\dfrac{{{\rho }^{\sigma \prime }}u_{\alpha }^{\sigma }}{D+{{I}^{\sigma }}}{{\left( u_{\beta }^{\sigma }-{{u}_{\beta }} \right)}^{2}}+\left( D+{{I}^{\sigma }}+2 \right)\dfrac{{{\rho }^{\sigma }}{T}'}{2S_{\kappa \alpha }^{1\sigma }{{m}^{\sigma }}}\left( {{u}_{\alpha }}-u_{\alpha }^{\sigma } \right)
	\tt{,}
\end{eqnarray}
with the change rate of individual energy due to the chemical reaction
\begin{equation}
	{{E}^{\sigma }}^{\prime }={{\rho }^{\sigma }}^{\prime }\left( \frac{D+{{I}^{\sigma }}}{2}\frac{T}{{{m}^{\sigma }}}+\frac{{{u}^{2}}}{2} \right)+\frac{D+{{I}^{\sigma }}}{2}\frac{{{\rho }^{\sigma }}}{{{m}^{\sigma }}}{T}'=\frac{{{m}^{\sigma }}{{{\hat{R}}}_{4}}}{2}
	\label{Thermal_conductivity}
	\tt{,}
\end{equation}
the thermal conductivity
\begin{equation}
	\kappa _{\alpha }^{\sigma }=\dfrac{D+{{I}^{\sigma }}+2}{2S_{ \kappa \alpha }^{1 \sigma }}\dfrac{{{\rho }^{\sigma }}{{T}^{\sigma }}}{{{m}^{\sigma 2}}}
	\label{Thermal_conductivity}
	\tt{,}
\end{equation}
the dynamic viscosity
\begin{equation}
	\mu _{\alpha \beta }^{\sigma }=\dfrac{{{p}^{\sigma }}}{S_{\alpha \beta }^{1 \sigma }}
	\label{Dynamic_viscosity}
	\tt{,}
\end{equation}
and the parameters
($S_{Jx}^{\sigma }$, $S_{Jy}^{\sigma }$,
$S_{xx}^{1 \sigma }$, $S_{xy}^{1 \sigma }$, $S_{yy}^{1 \sigma }$,
$S_{xx}^{2 \sigma }$, $S_{xy}^{2 \sigma }$, $S_{yy}^{2 \sigma }$,
$S_{\kappa x}^{1 \sigma }$, $S_{\kappa y}^{1 \sigma }$
$S_{\kappa x}^{2 \sigma }$, $S_{\kappa y}^{2 \sigma }$)
$=$
($S_{2}^{2 \sigma }$, $S_{3}^{2 \sigma }$,
$S_{5}^{1 \sigma }$, $S_{6}^{1 \sigma }$, $S_{7}^{1 \sigma }$,
$S_{5}^{2 \sigma }$, $S_{6}^{2 \sigma }$, $S_{7}^{2 \sigma }$,
$S_{8}^{1 \sigma }$, $S_{9}^{1 \sigma }$,
$S_{8}^{2 \sigma }$, $S_{9}^{2 \sigma }$).

In the case of $S_{5}^{1\sigma }=S_{6}^{1\sigma }=S_{7}^{1\sigma }=S_{\mu }^{\sigma }$ and $S_{8}^{1\sigma }=S_{9}^{1\sigma }=S_{\kappa }^{\sigma }$, the Prandtl number is
\begin{equation}
	{{\Pr }^{\sigma }}=\dfrac{S_{\kappa }^{\sigma }}{S_{\mu }^{\sigma }}
	\tt{.}
\end{equation}

Moreover, the specific heat at constant pressure and volume are $c_{p}^{\sigma }=( D+{{I}^{\sigma }}+2 ) /( 2{{m}^{\sigma }} )$ and $c_{v}^{\sigma }={\left( D+{{I}^{\sigma }} \right)}/{\left( 2{{m}^{\sigma }} \right)}$, respectively. Consequently, the specific heat ratio is
\begin{equation}
	{{\gamma }^{\sigma }}=\dfrac{c_{p}^{\sigma }}{c_{v}^{\sigma }}=\dfrac{D+{{I}^{\sigma }}+2}{D+{{I}^{\sigma }}}
	\tt{.}
\end{equation}

Additionally, summing Eqs. (\ref{NSequation1}) - (\ref{NSequation3}) over all species $\sigma$, we get the following reactive NS equations,
\begin{equation}
	\dfrac{\partial \rho }{\partial t}+\dfrac{\partial }{\partial \alpha }\left( \rho {{u}_{\alpha }} \right)=0
	\tt{,}
\end{equation}
\begin{equation}
	\dfrac{\partial }{\partial t}\left( \rho {{u}_{\alpha }} \right)+\dfrac{\partial }{\partial \beta }\sum\nolimits_{\sigma }{\left( {{\delta }_{\alpha \beta }}{{p}^{\sigma }}+{{\rho }^{\sigma }}u_{\alpha }^{\sigma }u_{\beta }^{\sigma }+P_{\alpha \beta }^{\sigma }+U_{\alpha \beta }^{\sigma }+V_{\alpha \beta }^{\sigma } \right)}=\rho {{a}_{\alpha }}
	\tt{,}
\end{equation}
\begin{equation}
	\dfrac{\partial E}{\partial t}+\dfrac{\partial }{\partial \alpha }\sum\nolimits_{\sigma }{\left( {{E}^{\sigma }}u_{\alpha }^{\sigma }+{{p}^{\sigma }}u_{\alpha }^{\sigma }-\kappa _{\alpha }^{\sigma }\dfrac{\partial {{T}^{\sigma }}}{\partial \alpha }+u_{\beta }^{\sigma }P_{\alpha \beta }^{\sigma }+X_{\alpha }^{\sigma }+Y_{\alpha }^{\sigma } \right)}=\rho {{u}_{\alpha }}{{a}_{\alpha }}+{E}'
	\tt{,}
	\label{NSequation_Mixing3}
\end{equation}
in the case of $a_{\alpha }^{\sigma }={{a}_{\alpha }}$, where ${E}'=\sum\nolimits_{\sigma }{{{E}^{\sigma }}^{\prime }}$ denotes the change rate of mixing energy due to the chemical reaction.

Actually, under some corresponding conditions, we could obtain the Fick's laws of diffusion and Maxwell-Stefan diffusion equation from Eqs. (\ref{NSequation1}) and (\ref{NSequation2}) as well \cite{Lin2016CNF}. More discussion is beyond this work.

\section{}\label{APPENDIXC}

Now, let us demonstrate some important parameters in the current DBM.

(i) Time. The dimensional time is ${t_{d}} = {L_{d}}/{u_{d}}$, the nondimensional time ${t_{n}} = {L_{n}}/{u_{n}}$, and the time ratio
\begin{equation}
	{t_{r}} = \dfrac{{L_{d}} u_{n}}{{L_{n}} {u_{d}}}
	\tt{.}
\end{equation}

(ii) Energy. The internal energies with dimension and nondimension are ${E_{id}} = (D+I) {n_{d}} R {T_{d}}/2$ and ${E_{in}} = (D+I) {n_{n}} {T_{n}}/2$, respectively, and the energy ratio reads
\begin{equation}
	{E_{r}} = \frac{{E_{id}}}{{E_{in}}} = {n_{r}} R {T_{r}}
	\label{Er1}
	\tt{.}
\end{equation}

(iii) Mass. Given the dimensional and nondimensional molar mass $m_{d}$ and $m_{n}$, the mass densities are ${\rho}_{d} = m_{d} n_{d}$ and ${\rho}_{n} = m_{n} n_{n}$ in dimensional and nondimensional forms, the kinetic energies are ${E_{kd}} = {{\rho}_{d}} {u_{d}^{2}} / 2$ and ${E_{kn}} = {{\rho}_{n}} {u_{n}^{2}} / 2$ in the two forms, so the energy ratio is
\begin{equation}
	{E_{r}} = \frac{{E_{kd}}}{{E_{kn}}}= {m_{r}} {n_{r}} {u_{r}^{2}}
	\label{Er2}
	\tt{.}
\end{equation}
From Eqs. (\ref{Er1}) and (\ref{Er2}), we obtain the mass ratio
\begin{equation}
	m_{r} = \dfrac{R {T_r}}{u_{r}^2}
	\tt{,}
\end{equation}
which leads to the mass density ratio ${\rho}_{r} = m_{r} n_{r}$.

(iv) Viscosity. The Reynolds number is defined as
\begin{equation}
	{\rm{Re}}=\dfrac{{{\rho}_{d}}{{u}_{d}}{{L}_{d}}}{{{\mu }_{d}}}=\dfrac{{{\rho}_{n}}{{u}_{n}}{{L}_{n}}}{{{\mu }_{n}}}
	\tt{,}
\end{equation}
where ${\mu }_{d}$ and ${\mu }_{n}$ denote the dynamic viscosity in dimensional and nondimensional forms, respectively. The ratio of dimensional to nondimensional dynamic viscosity takes the form
\begin{equation}
	{{\mu }_{r}}=\dfrac{n_{r} L_{r} R T_{r}}{u_{r}}
	\tt{,}
\end{equation}
from which we could get the kinematic viscosity ratio ${{\nu}_{r}}={{\mu}_{r}}/{{\rho}_{r}}$.

(v) Thermal diffusivity. The Prandtl number is a dimensionless number defined as the ratio of momentum diffusivity (i.e., kinematic viscosity) to thermal diffusivity,
\begin{equation}
	\Pr =\dfrac{{{\nu}_{d}}}{{{K}_{d}}}=\dfrac{{{\nu}_{n}}}{{{K}_{n}}}
	\tt{,}
\end{equation}
where (${{\nu}_{d}}$, ${{\nu}_{n}}$) and (${{K}_{d}}$, ${{K}_{n}}$) denote the momentum diffusivity and thermal diffusivity in dimensional and nondimensional forms, respectively. Consequently, the ratio of dimensional to nondimensional thermal diffusivity is ${K}_{r}={\nu}_{r}$.

(vi) Mass diffusivity. The Schmidt number is a dimensionless number defined as the ratio of momentum diffusivity (kinematic viscosity) to mass diffusivity,
\begin{equation}
	{\rm{Sc}}=\dfrac{{{\nu }_{d}}}{{{\zeta}_{d}}}=\dfrac{{{\nu }_{n}}}{{{\zeta}_{n}}}
	\tt{,}
\end{equation}
where ${\zeta}_{d}$ and ${\zeta}_{n}$ stand for the dimensional and nondimensional mass diffusivity, respectively. Hence, the ratio of dimensional to nondimensional mass diffusivity is ${\zeta}_{r}={\nu}_{r}$.

\section{}\label{APPENDIXD}

When a homogeneous nonreactive mixture reaches its steady state in the force field, the time and space derivatives equal zero, hence Eq. (\ref{NSequation2}) leads to
\begin{equation}
	S_{J\alpha }^{\sigma }{{\rho }^{\sigma }}\left( {{u}_{\alpha }}-u_{\alpha }^{\sigma } \right)+{{\rho }^{\sigma }}a_{\alpha }^{\sigma }=0
	\tt{.}
\end{equation}
Namely, the individual velocities read
\begin{equation}
	u_{\alpha }^{\sigma }={{u}_{\alpha }}+\left(S{_{J\alpha }^{\sigma }}\right)^{-1}a_{\alpha }^{\sigma }
	\tt{.}
	\label{Steady_u_force}
\end{equation}

Moreover, if the homogeneous nonreactive mixture is an isothermal system in the steady state,  Eq. (\ref{NSequation3}) gives
\begin{equation}
	\frac{1}{2}S_{4}^{2\sigma }{{\rho }^{\sigma }}\left[ \left( D+{{I}^{\sigma }} \right)\frac{T-{{T}^{\sigma }}}{{{m}^{\sigma }}}+{{u}^{2}}-{{u}^{\sigma 2}} \right]+{{\rho }^{\sigma }}u_{\alpha }^{\sigma }a_{\alpha }^{\sigma }=0
	\tt{.}
\end{equation}
Consequently, the individual temperatures take the form
\begin{equation}
	{{T}^{\sigma }}=T+\frac{{{m}^{\sigma }}}{D+{{I}^{\sigma }}}\left[ {{\left| \mathbf{u} \right|}^{2}}-{{\left| {{\mathbf{u}}^{\sigma }} \right|}^{2}}+2\big(S{{_{4}^{2\sigma }}\big)^{-1}}{{\mathbf{u}}^{\sigma }}\cdot {{\mathbf{a}}^{\sigma }} \right]
	\tt{.}
	\label{Steady_T_force}
\end{equation}

On the contrary, for a thermal homogeneous mixture in the force field, if the individual accelerations are not equal, the chemical species propagate collectively with different velocities and collide randomly with each other, which leads to the change of energies and temperatures. That is to say, the work done by external forces transforms into the kinetic and internal energies of the thermal system, i.e,
\begin{equation}
	\frac{\partial E}{\partial t}=\sum\nolimits_{\sigma }{{{\rho }^{\sigma }}{{\mathbf{a}}^{\sigma }}\cdot {{\mathbf{u}}^{\sigma }}}
	\tt{,}
	\label{E_force}
\end{equation}
which can be derived from Eq. (\ref{NSequation_Mixing3}).
Let us consider a special case where the mixing velocity keeps constant $\mathbf{u} = 0$ in the force field, then Eq. (\ref{E_force}) changes into
\begin{equation}
	\frac{\partial {{E}_{\operatorname{int}}}}{\partial t}=\sum\nolimits_{\sigma }{\big(S{{_{J}^{\sigma }}\big)^{-1}}{{\rho }^{\sigma }}{{\mathbf{a}}^{\sigma }}\cdot {{\mathbf{a}}^{\sigma }}}
	\tt{.}
	\label{Eint_force}
\end{equation}
From Eqs. (\ref{T_expression}) and (\ref{Eint_force}), we get
\begin{equation}
	{{T}^{\sigma }}\approx T=\frac{\sum\nolimits_{\sigma }{\left( D+{{I}^{\sigma }} \right){{n}^{\sigma }}{{T}_{0}}}+2{{\sum\nolimits_{\sigma }{(S{{_{J}^{\sigma }})^{-1}}{{\rho }^{\sigma }}\left| {{\mathbf{a}}^{\sigma }} \right|}}^{2}}t}{\sum\nolimits_{\sigma }{\left( D+{{I}^{\sigma }} \right){{n}^{\sigma }}}}
	\tt{,}
	\label{Change_T_force}
\end{equation}
where $T_0$ is the initial mixing temperature.

\vspace*{2mm}

\vspace*{-1mm}
\begin{small}\baselineskip=10pt\itemsep-2pt
\bibliography{References}
\end{small}
\end{CJK*}
\end{document}